\documentclass[letterpaper]{mn2e}

\pdfoutput=1

\usepackage{subfigure}
\usepackage{graphicx}
\usepackage{longtable}
\usepackage{float}
\usepackage{captcont}
\usepackage{endfloat}
\usepackage{morefloats}
\usepackage{mathrsfs}
\usepackage{alphalph}

\title[Galaxies with Background QSOs, II]{Galaxies with Background QSOs: II. An Automated Search for Multiple Galaxy Emission Lines}

\author[Lorrie A. Straka]
{Lorrie A. Straka,$^{1,2}$ Zakary L. Whichard,$^3$ Varsha P. Kulkarni,$^2$ Michael Bishof,$^{1,4}$ 
\newauthor
David Bowen,$^5$  Pushpa Khare,$^{6}$ Donald G. York$^{1,7}$\\
$^1$Department of Astronomy and Astrophysics, University of Chicago, Chicago, IL 60637\\
$^2$Department of Physics and Astronomy, University of South Carolina, Columbia, SC 29208\\
$^3$Department of Physics and Astronomy, Northwestern University, Evanston, IL 60208 \\ 
$^4$Department of Physics, University of Colorado, Boulder, Co 80309 \\
$^5$Department of Physics and Astronomy, Princeton University, Princeton, NJ, 08544\\
$^{6}$CSIR Emeritus Scientist, IUCAA, Ganeshkhind, Pune 411007, India \\
$^{7}$Enrico Fermi Institute, University of Chicago, Chicago, IL 60637, USA \\
}

\date{}

\pagerange{\pageref{firstpage}--\pageref{lastpage}} \pubyear{}

\def\LaTeX{L\kern-.36em\raise.3ex\hbox{a}\kern-.15em
    T\kern-.1667em\lower.7ex\hbox{E}\kern-.125emX}

\begin{document}

\maketitle

\label{firstpage}

\begin{abstract}

We have improved upon our previous search technique of systematically searching QSO spectra for narrow galactic H$\alpha$ emission, which indicates a foreground galaxy within the Sloan Digital Sky Survey (SDSS) spectral fiber. We now search for H$\alpha$ plus eight other galactic emission lines in the same manner. { We have scanned the SDSS DR7 QSO catalog spectra searching for these emission lines.} Here we present our sample which focuses on the redshift range $z<0.401$ where galactic H$\alpha$ is detectable in the SDSS spectra. This has revealed 27 unique galaxies on top of QSOs (GOTOQs). We have deblended the QSOs from the respective galaxies and determined the photometric properties of these systems. We find upon deblending that most of the galaxies are primarily blue, late-type galaxies with colors in the range -0.71$<$(u-r)$<$2.07. We find a slight anti-correlation between reddening and impact parameter (E(B-V)$_{(g-i)}$ vs. b). The galaxies have average star formation rates of 0.01 to 1 M$_{\sun}$ yr$^{-1}$, with an average of 0.6 M$_{\sun}$ yr$^{-1}$. They range in z from 0 to 0.4 and in stellar luminosity from about 0.01 L$^{\ast}$ to 3.0 L$^{\ast}$. They are foreground to QSOs of brightness 17.4 to 20.4 magnitudes (r-band) with impact parameters of 1 to 10 kpc. They represent a fair sample of typical galaxies { for which it should be possible to determine accurately various quantities (e.g. abundances, dust extinction, Faraday rotation) using follow-up analysis of the background QSOs}. 

Where present, Ca II $\lambda3934$ (K)  and Na I $\lambda 5892$ (D2) absorption lines were also measured in the QSO spectrum. We find 15 systems with Ca II K absorption and 6 with Na I D2 absorption. We find no trends relating the equivalent widths of these lines to impact parameter or reddening of the background QSO. Eight of our fields show significant reddening (E(B-V)$_{(g-i)}>0.20$), which are targeted for followup studies of interstellar clouds. We find three systems (Q0059-0009, Q1033+2059, and Q2356+0029) with detectable galactic spectral emission, but no visible galaxy in imaging and no detectable absorption features. We speculate on the nature of these galaxies, which are strong candidates for dark galaxies and dwarf halos.  

\end{abstract}


\section{Introduction}\label{section-introduction}

{ Galaxies on Top of QSOs (GOTOQs)}, defined roughly as foreground galaxies lying either directly along the line of sight to a background QSO or within a few arcseconds of the line of sight, offer unique insight into the contents of these galaxies through their imprint on QSO spectra. While QSO absorbers have been known since shortly after QSOs were discovered (e.g. Sandage 1965; Burbidge, Lynds, \& Burbidge 1966; Arp, Bolton, \& Kinman 1967), little is known about the environments in which these systems arise. It is, however, well known that the evolutionary processes and histories of gas and stars are closely linked, and therefore studies of the properties of these systems can lead us to a better understanding of the environments in which they reside. Critical to determining the characteristics of these environments is detecting these absorber galaxies in emission, both spectroscopically and visually in imaging. Such detections will allow us to ascertain the galaxy morphologies, star formation rates (SFRs), reddening values via Balmer decrement and color excess, and luminosities. The QSO impact parameter can be accurately determined if the emitting galaxy is well-imaged. One can also study how the stellar properties of the emission galaxy correlate with the properties of the interstellar material revealed through absorption lines. 

{ The detection of absorption host galaxies in emission at high redshifts has proven difficult. Recent studies have made some progress in this regime, but the success rate is still low (e.g., Moller et al. 2004; Fynbo et al. 2010, 2011, 2013; P\'eroux et al. 2011, 2012; Bouch\'e et al. 2012; Noterdaeme et al. 2012; Kulkarni et al. 2012).  The issue is that surface brightness decreases rapidly with increasing distance, and therefore higher-$z$ galaxies showing absorption in QSO spectra are much more difficult to detect in emission. We have circumvented this issue by choosing a sample of low-$z$ galaxies intervening with background QSOs in the Sloan Digital Sky Survey (SDSS). These GOTOQs  are easier to detect in imaging and in spectral emission due to their low-$z$ nature. This sample is critical to understanding the links between absorbers and galaxies because it is the only way to study galaxies in both absorption and emission, whereas high-$z$ GOTOQs may be only successfully observed in absorption. It is important to note that both low- and high-$z$ studies are necessary to track the evolution properties with cosmic time in GOTOQs. }

A variety of techniques exist for detecting GOTOQs, including matching catalogs of QSOs and galaxies searching for overlapping or near overlapping positions (Hewitt \& Burbidge 1989, Womble et al. 1993), seeking absorption features in QSO spectra (Bergeron \& Boisse 1991, Yanny \& York 1992, Zych et al. 2007),  finding galaxies with matching redshifts, and searching QSO spectra for overlying narrow galactic emission lines (Noterdaeme et al. 2010; York et al. 2012, hereafter referred to as Paper I). A more thorough discussion of these techniques can be found in Paper I. 

Paper I utilized the last of these techniques to consider a sample of GOTOQs systematically found in the SDSS by their H$\alpha$ emission line signature overlaid in QSO spectra (York et al. 2000). We have herein refined this search technique to search for nine galactic emission lines sharing a common redshift, including H$\alpha$, similarly overlaid on QSO spectra. Such a refinement promises to be much more complete, allowing us to detect GOTOQs with fewer false positives and to increase the sample size of pairs to be followed up with absorption line spectroscopy to obtain the physical conditions in a number of galaxies. Additionally, the increased sample size allows us to place finer constraints on the relationships between each of the properties we have measured. 

Section~\ref{section-sample} details our sample selection and addresses any bias in the sample. Section~\ref{section-analysis} details our data analysis. Section~\ref{section-results} discusses our results. Finally, section~\ref{section-conclusion} summarizes our conclusions. Throughout we have assumed the concordance cosmology (H$_{0}=70$ km s$^{-1}$ Mpc$^{-1}$, $\Omega_{m}=0.30$, $\Omega_{\Lambda}=0.70$). 

\section{Sample Selection}\label{section-sample}

GOTOQs were detected in the SDSS DR7 (Schneider et al. 2010) by systematically searching QSO spectra for narrow galaxy emission lines. The process involved utilizing code originally written to detect absorption lines in SDSS DR7 QSO spectra. { Absorption line systems returned from this original search were compiled in a catalog, analyzed, and used in the results of, e.g., York et al. (2006), Vanden Berk et al. (2008), and Khare et al. (2012).}  Rather than rewrite the code to detect emission lines, the input spectra were inverted, thereby turning emission lines into ``absorption'' lines capable of being detected by this code (Lundgren et al. 2009; Vikas et al. 2013; York et al., in preparation). The output is then a list of identified wavelengths for detected emission lines. Nine emission lines were searched for: H$\alpha$ ($\lambda$6564.61), H$\beta$ ($\lambda$4862.72), [O II]$\lambda$3727, [O III]$\lambda\lambda$4960.30, 5008.24,  [N II]$\lambda\lambda$6549.86, 6585.27, and [S~II]~$\lambda\lambda$6718.29, 6732.67 (all cited wavelengths are vacuum wavelengths). A minimum of two emission lines are required to determine an accurate redshift for any foreground objects to which the emission lines belong, however, any system with fewer than three detected emission lines is ignored. A further requirement for this sample was the presence of the H$\alpha$ line, therefore restricting the redshift range to $z<$0.401. At redshifts above this value, the H$\alpha$ line is out of SDSS spectrograph range (3800 \AA ~to 9200 \AA).  

The line identification code was run on the database of QSO spectra in DR7. { The code searched 104,374 QSO spectra, returning 635 systems with at least three emission lines, 85 of which had at least four emission lines.} All the systems chosen for this study had at least 3 lines with significance level (SL) $>4$ (that is, the {measured flux} is at least 4 times stronger than the 1$\sigma$ detection limit for the relevant region of the spectrum). Even though the QSO spectra search is automated, each detection must be confirmed by eye to eliminate false positives that may arise. We present here { the results of this search}. Of these 85 objects found, 50 were found to be legitimate pairs after visually examining both the spectra and the images contained in the SDSS Catalogue Archive Server (CAS) database. Within these legitimate pairs, 23 were previously treated in Paper I and so are not relisted here, leaving us with 27 unique targets new to this paper. False positives were caused mainly by poor sky line subtraction and misidentification of the H$\alpha$ line due to noise. Table~\ref{tbl-qgp} lists the subsequent systems found graded B or higher (see below); the index number by which they'll be referred to throughout the paper; the SDSS ID of the QSO; the plate, fiber, and MJD numbers;  the redshifts of both the QSO and the foreground galaxy producing the emission lines; and the impact parameter of the galaxy in both arcseconds and kpc. Those graded B have at least 3 emission lines detected with the code with SL$>4$, while those graded A have four or more emission lines detected with the code at SL$>4$. 

{ The remaining 550 QSO spectra in DR7 that were flagged as low quality systems, and therefore unlikely to be real cases, are currently being evaluated by eye. About 1 in every 164 spectra was flagged by the program as being a potential GOTOQ (104,374 total with 635 flagged), with a false positive rate of about 10 to 1. Therefore, we expect to find another approximately 20 systems in the low grade regime, estimated from the above numbers and taking into account the lower quality of the spectra which are noisier than our grade A and B systems by about a factor of 2.} 

\begin{figure}
\begin{minipage}{1.0\linewidth}
\begin{center}

\subfigure[3]{
\includegraphics[trim = 0mm 0mm 0mm 0mm, clip, width=0.3\textwidth]{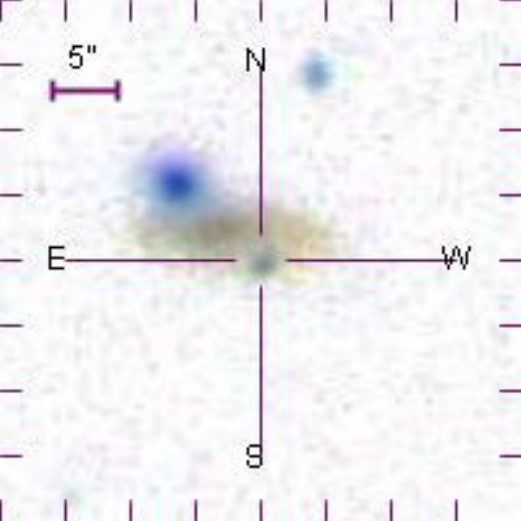}
}
\subfigure[$z_{gal}=0.0279$]{
\includegraphics[trim = 4.5mm 2.5mm 5mm 2.5mm, clip, width=0.6\textwidth]{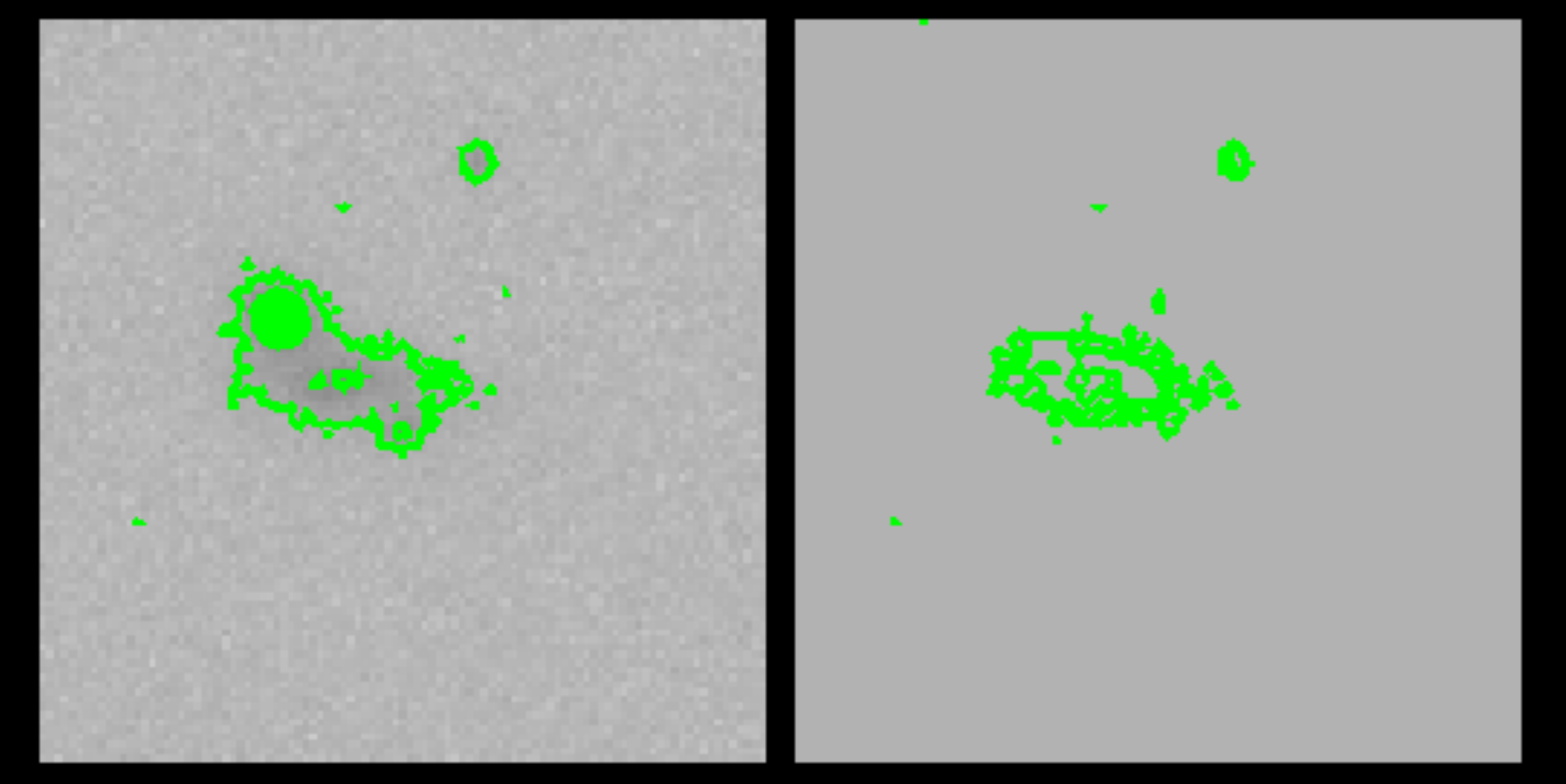}
}
\subfigure[4]{
\includegraphics[trim = 0mm 0mm 0mm 0mm, clip, width=0.3\textwidth]{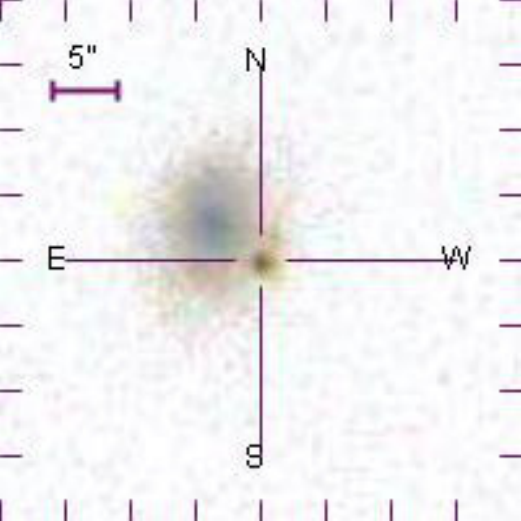}
}
\subfigure[$z_{gal}=0.0430$]{
\includegraphics[trim = 4.5mm 2.5mm 5mm 2.5mm, clip, width=0.6\textwidth]{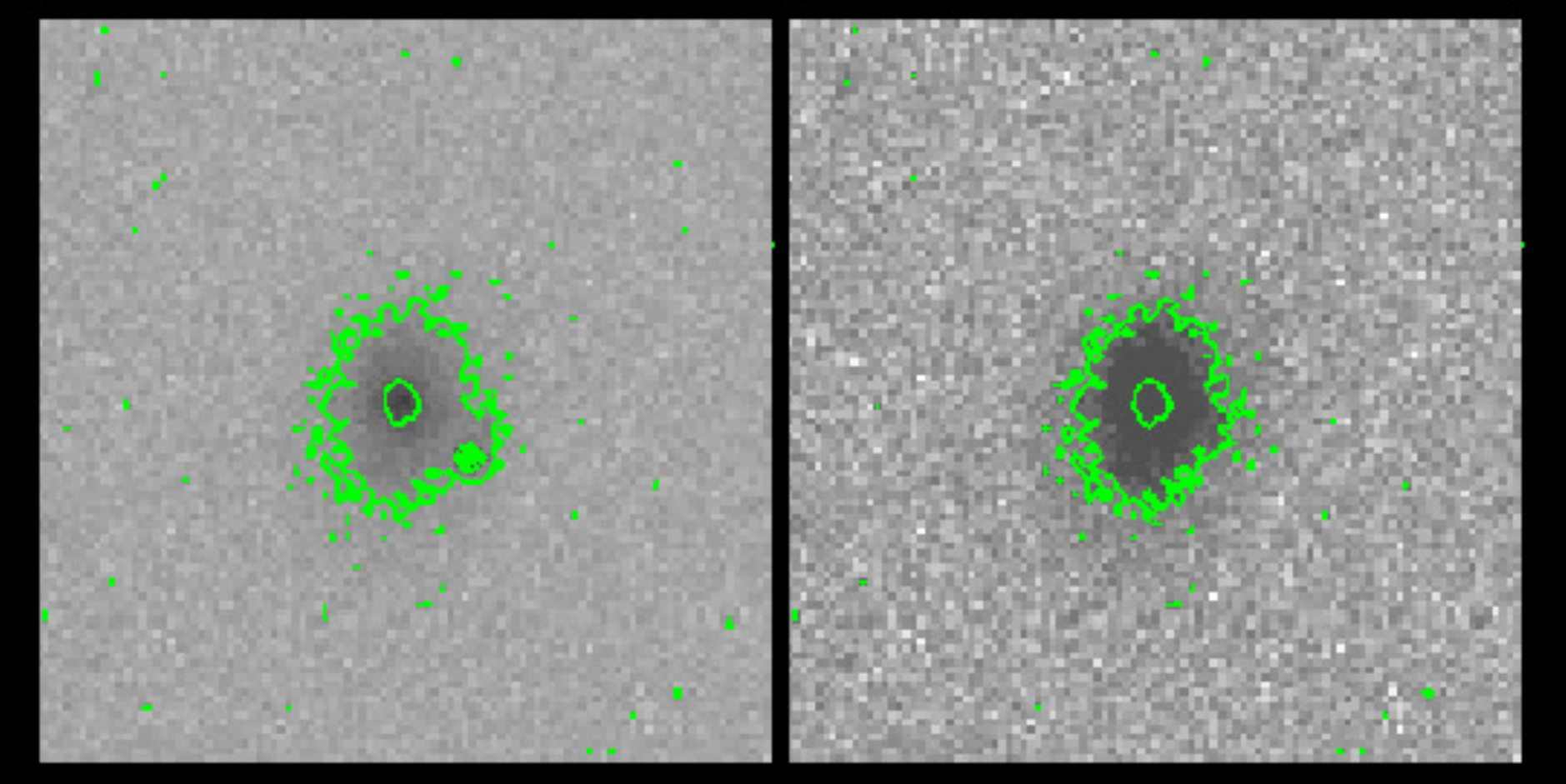}
}
\subfigure[5]{
\includegraphics[trim = 0mm 0mm 0mm 0mm, clip, width=0.3\textwidth]{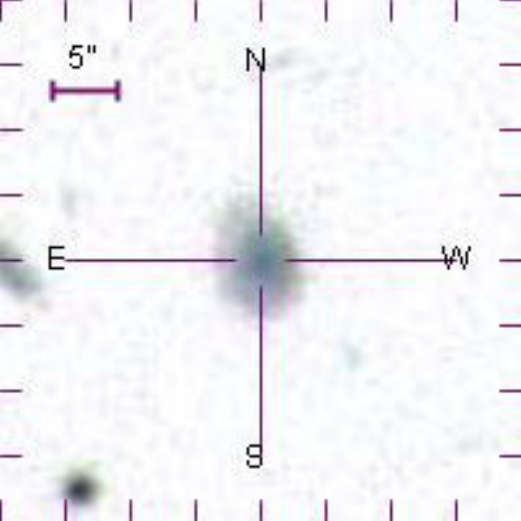}
}
\subfigure[$z_{gal}=0.1147$]{
\includegraphics[trim = 4.5mm 24.5mm 5mm 3mm, clip, width=0.6\textwidth]{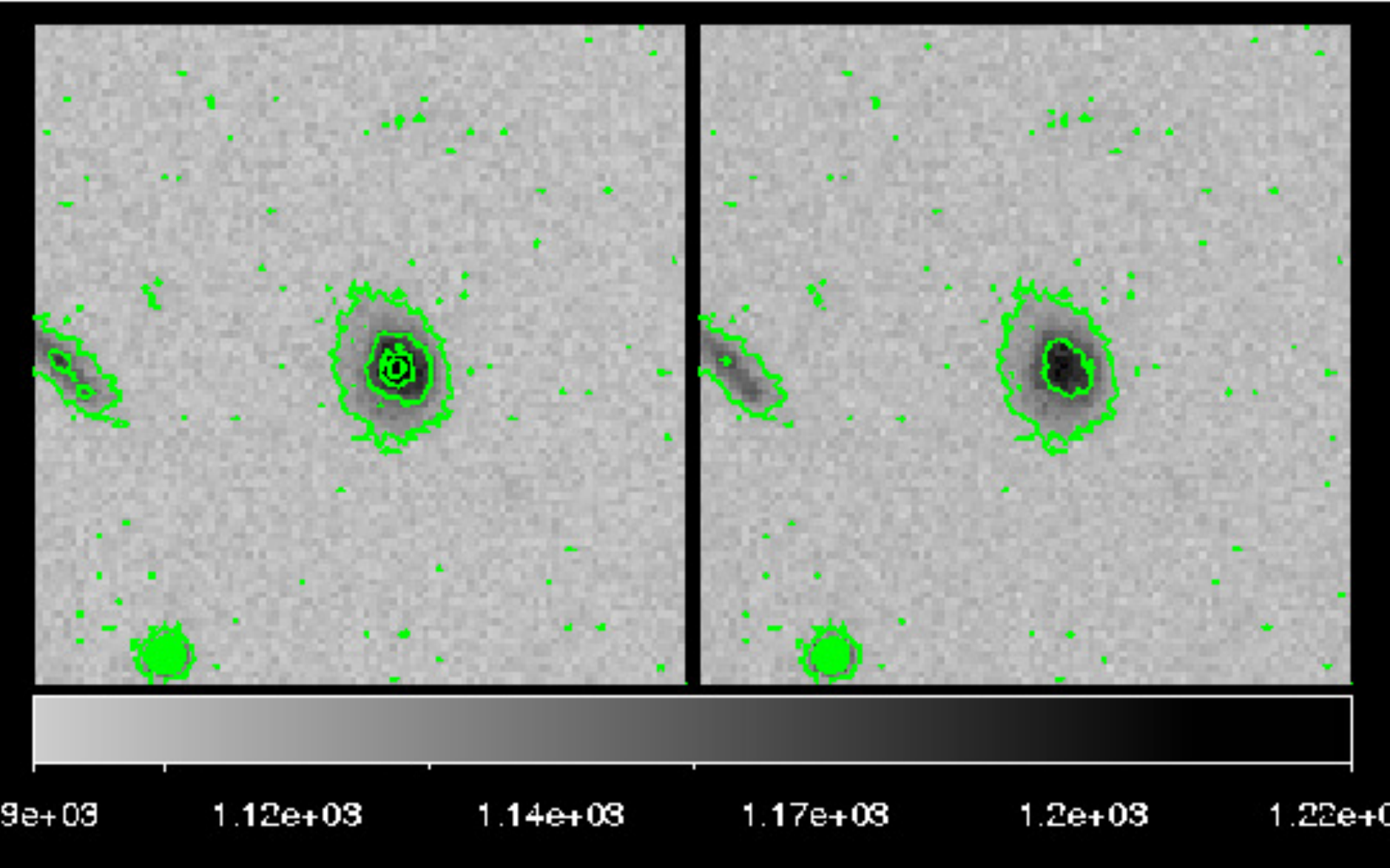}
}

\end{center}
\captcont{SDSS multicolor images of fields with a QSO intercepting a detected visible low-z galaxy, ordered by the increasing RA and increasing numerical index. The scale of the images is indicated in the upper left hand corner. The orientation is north up and east left. The numbers under the composite color images refer to the index numbers in Table 1. The numbering is not continuous: Figure 2 includes images for the missing objects, that is, the ones with no visible galaxies in the SDSS images. the central image is a contour map representation with contours representing (arbitrary) steps in flux levels in the SDSS r-band image. the right most contour image has all stars and QSOs that overlap with the galaxy image removed by procedures described in the text.  }\label{fig-thumb}
\end{minipage}
\end{figure}

\begin{figure}
\begin{minipage}{1.0\linewidth}
\begin{center}

\subfigure[1]{
\includegraphics[trim = 0mm 0mm 0mm 0mm, clip, width=0.3\textwidth]{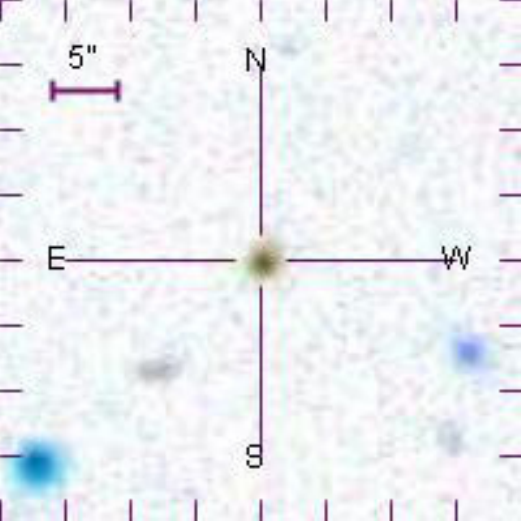}
}
\subfigure[2]{
\includegraphics[trim = 0mm 0mm 0mm 0mm, clip, width=0.3\textwidth]{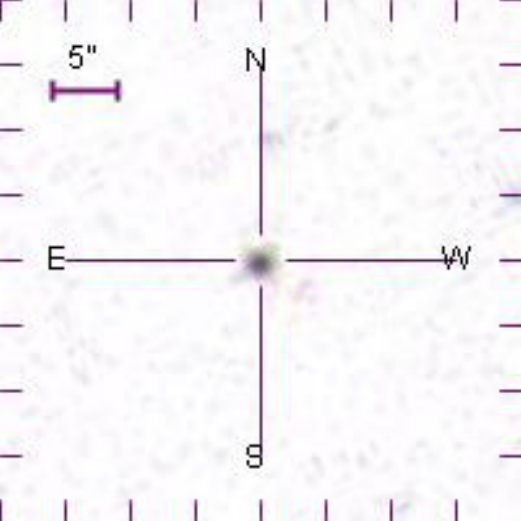}
}
\subfigure[6]{
\includegraphics[trim = 0mm 0mm 0mm 0mm, clip, width=0.3\textwidth]{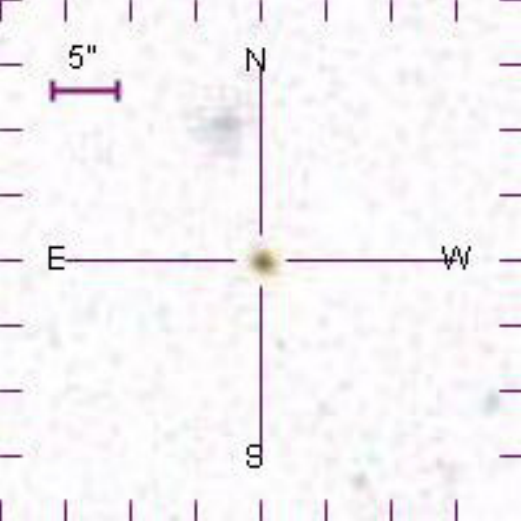}
}
\subfigure[13]{
\includegraphics[trim = 0mm 0mm 0mm 0mm, clip, width=0.3\textwidth]{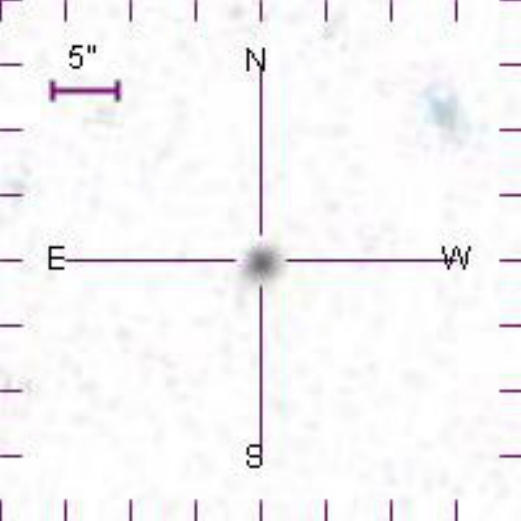}
}
\subfigure[17]{
\includegraphics[trim = 0mm 0mm 0mm 0mm, clip, width=0.3\textwidth]{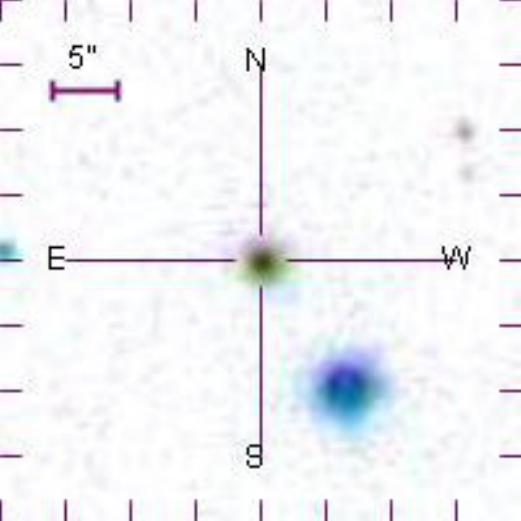}
}
\subfigure[21]{
\includegraphics[trim = 0mm 0mm 0mm 0mm, clip, width=0.3\textwidth]{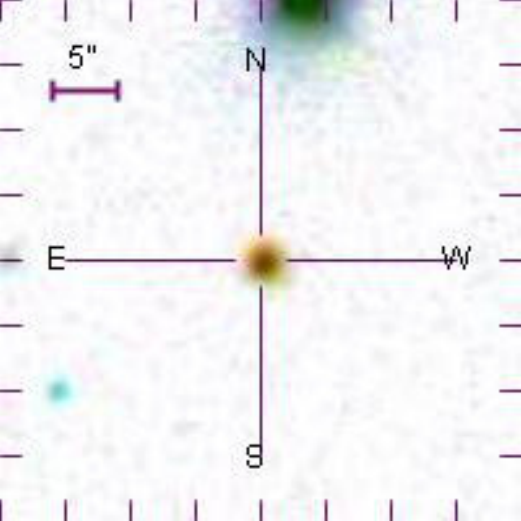}
}
\subfigure[27]{
\includegraphics[trim = 0mm 0mm 0mm 0mm, clip, width=0.3\textwidth]{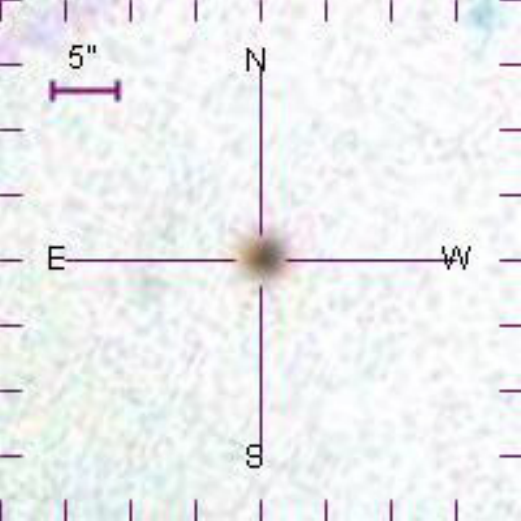}
}

\end{center}
\caption{SDSS multicolor images of fields with no detected visible galaxy with a QSO intercepting a low-z galaxy, ordered by the increasing RA and increasing numerical index (1, 2, 6, 13, 17, 21, and 27). The scale of the images is indicated in the upper left hand corner. The orientation is north up and east left. Six of these seven objects are in the highest quartile by redshift, so are selectively not visible because of the QSO brightness. \label{fig-thumb-2}}
\end{minipage}
\end{figure}

\begin{figure}
\begin{minipage}{1.0\linewidth}
\begin{center}

\includegraphics[trim = 0mm 0mm 0mm 0mm, clip, width=0.7\textwidth]{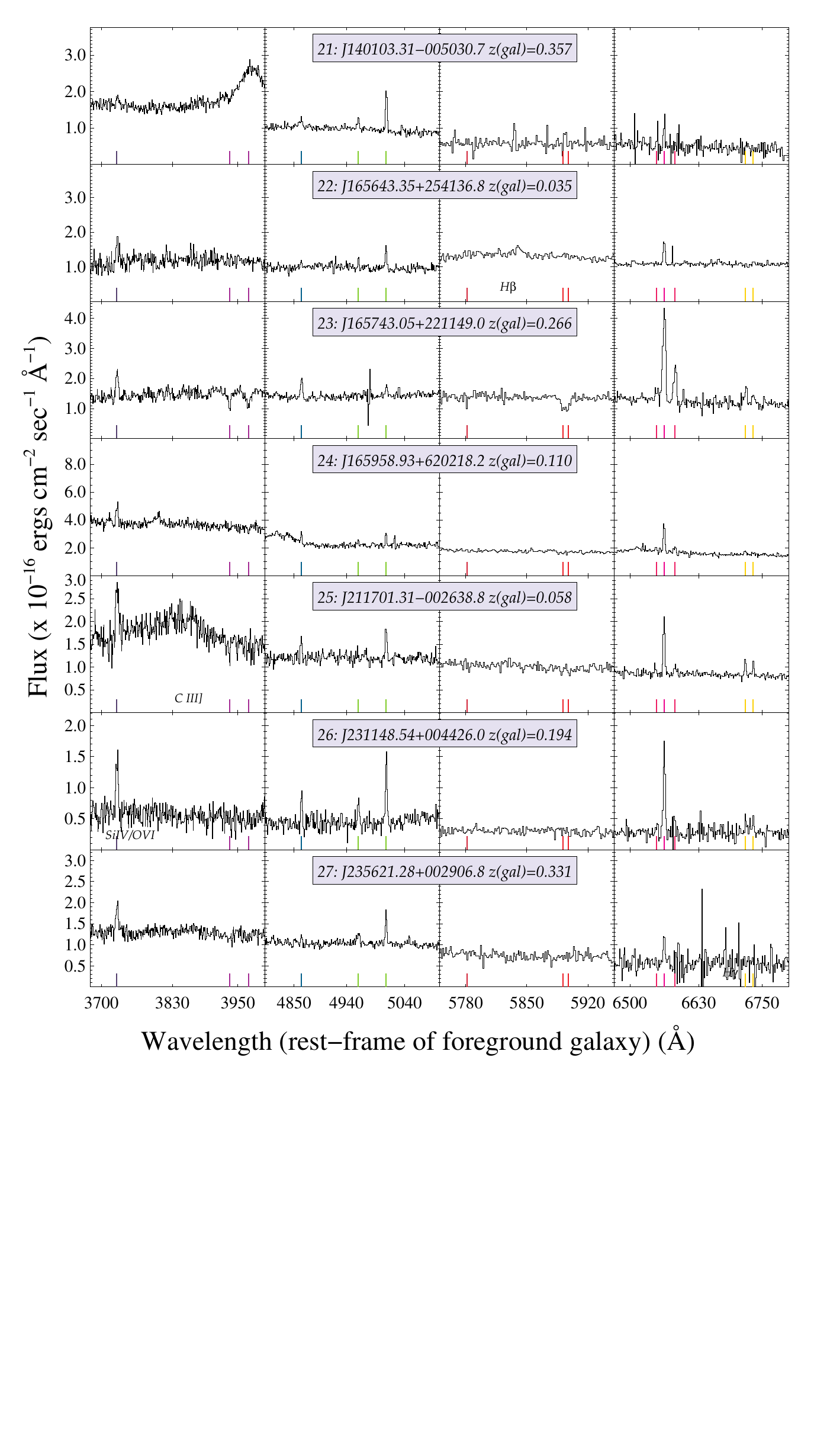}

\end{center}
\caption{Spectra of targets 21-27 of our sample.\label{fig-spectra}}
\end{minipage}
\end{figure}

Figure~\ref{fig-thumb} shows the thumbnail tricolor composite images (bands $g$, $r$, and $i$) from the SDSS images for each of our 27 objects. { This figure also presents the before and after PSF subtracted data images. Figure~\ref{fig-thumb-2} shows the thumbnail tricolor composite images for those systems with no galaxy detected in imaging.} They are arranged in order of increasing RA and index number in their respective subcategories. This technique of searching for multiple galactic emission lines also successfully returned all 23 of our detected systems from Paper I of this series, in which we only used the presence of the H$\alpha$ line to detect visual foreground galaxies. { These 23 are not repeated in this sample or this paper, but can be found in Paper I. They are included in any calculation referring to the combined sample. The remaining systems were either false positives or did not have 3 emission lines with SL$>4$, and so are not presented here. Figure~\ref{fig-spectra} shows the SDSS QSO spectra for each target, broken into four areas. These areas highlight the regions of [O II]$\lambda 3727$ and Ca II H and K; H$\beta$ and [O III]$\lambda\lambda$4960.30, 5008.24; Na I; and [N II]$\lambda\lambda$6549.86, 6585.27, H$\alpha$, and [S II]$\lambda\lambda$6718.29, 6732.67 respectively.}

{ Sample I was found by searching SDSS DR5 (MJD $< 53552)$, whereas Sample II was found by searching SDSS DR7 (MJD $<54663$). As stated above, we have found 27 unique targets in Sample II beyond those presented in Sample I. Of these 27, 8 were not present in DR5 and so could not be found by our initial search for Sample I (these are marked in Table~\ref{tbl-qgp}). However, the remaining 19 in Sample II were also present in DR5. These were not detected because they did not meet either the flux or FWHM requirement for Sample I. Additionally, the search employed for Sample I concentrated on one strong emission line, which may have been lost in noise. However, the current multi-emission line search may have first picked up other strong lines in low-noise regions, and subsequent lines were confirmed by eye.}

Throughout the paper, we will be referring to several different samples of GOTOQs. The H$\alpha$ selected sample from York et al. (2012; Paper I) will be referred to as Sample I, while the sample new to this paper { (which excludes the targets from Paper I)} will be Sample II. Additionally, the sample found via [O III] detection published by Noterdaeme et al. (2010) will be referred to as ND10. Note that Sample I was found from SDSS DR5, while Sample II was found from SDSS DR7. { Table~\ref{tbl-sample} details the criteria of each of these samples.}

Between Samples I and II, we have a total of 50 targets. Of these 50, 23 were also found independently by ND10. This leaves 27 unique results to our total sample. Of those 23 found in ND10 that we did not find in either Sample I or II, 17 had redshifts higher than $z=0.401$, and so we could not have detected them with our search due to our H$\alpha$ emission requirement. The other 6 targets not detected in our search but detected in ND10 are within our redshift range, however, only Q1329+6304 ($z=0.366$) has undetectable H$\alpha$. These systems were not detectable because their emission lines were too few or not significant enough (not SL$>4$) by our criteria, or fell within noisy spectra which obscured the signal. { We note that the comparison between these two samples indicates our search is only less than 90\% complete, estimated from the targets found by ND10 that we have missed.} We will discuss these [O III] detected objects in a subsequent paper. Anywhere Sample I or II is referred to, we have included those objects from ND10 that overlap with these samples because they were detected through independent means by our group. 

Sources of sample bias are sky line residuals and other strong emission lines that may make it difficult to detect the galactic emission lines we seek. When strong emission lines occur near galactic emission lines, they can sometimes be misidentified as a second occurrence of the detected galactic line. This occurs because our emission template allows for 5 \AA ~of uncertainty in the position of each line, a window that, although small, can sometimes include more than one line. These misidentifications occur primarily with the [O~III] lines, most often in the absence of H$\alpha$, and are easily identified by eye for separate consideration. We expect the same unbiased nature in this QSO sample as in Paper I, as the selection criteria are similar, but with additional emission lines. That is, we expect no bias in detecting QSOs due to reddening. In principle, it is harder to detect emission lines near bright QSOs. However, very bright QSOs (m$_{i}<17$) are very rare. Additionally, the detection limit in flux is not as high when the QSOs are brighter, but strong galaxy emission lines can still be seen, allowing detections towards these objects (Noterdaeme et al. 2010). There are a few cases with emission lines too weak or too few to be flagged by our program as positive detections, but upon investigation by eye we find them to be real. Discussion of these systems is deferred to a later paper.

\section{Data Analysis}\label{section-analysis}

\subsection{Broad-band flux measurements}

SDSS images have already been reduced. Bias subtraction, dark correction, and flat fielding have all been applied through the SDSS reduction pipeline as presented on the website\footnote{http://www.sdss.org/dr7/} at the time of each data release, so no further image reduction steps were applied by the authors. However, we have performed PSF subtraction on the QSO of interest in each of the fields, { which is not performed by SDSS prior to SDSS measurements if the galaxy and QSO are not resolved (this is the case for nearly all of our targets)}. The removal of the QSO allows us to measure the photometric properties of the removed QSO and the remaining emission galaxy separately, without overlapping flux from either. Appropriate stars were chosen from within the same field as the QSO to act as our PSF stars. The software program IDP3 (Image Display Paradigm 3; Stobie \& Ferro 2006) was used to calculate the PSF subtraction by minimizing the variance. { In some cases, the PSF subtraction was imperfect, leaving some residuals at the position of the QSO. Analysis of these residuals indicates that they are not significant, being $ < 1\sigma$ in the noise. } Table~\ref{tbl-qgp} details the offsets and impact parameter from the QSO. 

We then used the IRAF package PHOT to perform the photometric measurements on the galaxy. { All measurements have been made on the deconvolved images.} Measurements of the QSO were accomplished by knowing the flux scale between the QSO and the PSF defining stars and the total flux of the PSF star itself. Reported apparent magnitudes are the asinh magnitudes as defined by Stoughton et al. (2002) and as presented on the SDSS website. Table~\ref{tbl-phot-QSOs} details the magnitudes calculated for the QSOs in filters $u$, $g$, $r$, and $i$. Table~\ref{tbl-phot-galaxies} details the magnitudes and luminosities for the galaxies in these same four filters. Also presented in Table~\ref{tbl-phot-QSOs} is the $r$-band PSF magnitude model fit to the QSO { by SDSS. However, the SDSS pipeline does not distinguish between the galaxy and QSO in nearly all cases, and so does not remove the galaxy light prior to photometric measurements of the QSO (thus skewing the measurements of the QSO). We include the PSF magnitude as a comparison to our own PSF subtracted magnitudes to show the difference and improvement}. Additionally, the Petrosian radius for each QSO is provided. The Petrosian (angular) radius is defined such that the ratio of the local surface brightness averaged over an annulus between 0.8$\mathcal{R}_{P}$ and 1.25$\mathcal{R_{P}}$ to the mean surface brightness within $\mathcal{R_{P}}$ equals 0.2.  Values of $\mathcal{R_{P}}$ larger than about 1.4$\arcsec$ indicate profiles that deviate from a point source. This could be an indicator of a galaxy's presence even if such a galaxy could not be seen and measured. All fields in our sample but those with indices 1, 2, 6, 13, 17, 21, and 27 had Petrosian radii $\ge1.4\arcsec$. Of these, only 1, 13, and 27 are also lacking absorption feature detections.

We have also calculated the dust reddening estimates for the QSOs and galaxies. We have used Milky Way extinction corrected QSO and galaxy colors. The observer-frame color excess ($\Delta$(g-i)) is calculated by taking the color difference (g-i) for the QSO and comparing it to the median (g-i) for QSOs at the same redshift, taken from { Schneider et al. (2007). These median values are measured from the Schneider SDSS DR5 QSO catalogue.} This value allows us to estimate the reddening of the QSO due to dust in the intervening galaxy.  Additionally, from this $\Delta$(g-i) value, we have calculated the E(B-V) value, which is the absorber rest-frame color excess. From York et al. (2006):

\begin{equation}
E(B-V)_{(g-i )} = \frac{\Delta(g-i)(1 + z_{abs})^{-1.2}}{1.506} 
\end{equation}
Table~\ref{tbl-phot-QSOs} details the color and reddening estimates for each of the pairs. 

\subsection{Emission line measurements}

Each of the nine previously mentioned emission lines found in the spectra were measured with the SPLOT task within IRAF (where detected). Determined flux values can be found in Table~\ref{tbl-emission} measured in 10$^{-17}$ erg  cm$^{-2}$ s$^{-1}$. We use vacuum wavelengths given in Angstroms. Error values are determined by measuring the standard deviation in a featureless region of the continuum near the emission line, and scaling that by the square root of the number of pixels occupied by the emission line. 

Additional spectral inferences include star formation rate (SFR), Balmer decrement reddening correction, and Balmer decrement reddening values. Star formation rates were calculated using the prescription of Kennicutt (1998): 

\begin{equation}
SFR_{H\alpha}~(M_{\sun} ~yr^{-1}) = 4.9\times 10^{-41}~L(H\alpha) ~(erg ~s^{-1})
\end{equation}

\begin{equation}
SFR_{[O II]} ~(M_{\sun} ~yr^{-1}) = 1.4\times10^{-41} ~L([O II]) ~(erg ~s^{-1})
\end{equation}
From the H$\alpha$/H$\beta$ flux values (also found in Table~\ref{tbl-SFR}), we can calculate the reddening as for the Small Magellanic Cloud (SMC) extinction curve:

\begin{equation}
E(B-V)_{H\alpha/H\beta} = \frac{1.086}{k(H\beta)-k(H\alpha)}ln\left(\frac{H\alpha}{2.88H\beta}\right)
\end{equation}
with k(H$\alpha$) and k(H$\beta$) taken from Pei (1992).

We report the Balmer decrement corrected and uncorrected SFRs in Table~\ref{tbl-SFR}. The Balmer decrement corrected and uncorrected SFRs calculated from these equations and listed in the table can be taken as a lower limit, as it is possible a part of the galaxy area falls outside the spectral fiber.  
A full discussion of these measurements can be found below. 
 
\section{Results}\label{section-results}

\subsection{Photometry}

For the combined Sample I$+$II, the median galaxy (u-r) value is 1.30, whereas the mean is 1.21. For Sample II (those objects reported in this paper), we find the median to be 1.05 with a mean of 1.07. It can be seen that our sample falls directly in the range of late-type galaxies (u-r$<$2.22), which is perhaps unsurprising given that we have selected our targets based on emission. The largest values for the combined sample are 2.25 and 2.64 (targets number 4 and 5), the only two values to fall within the early-type range (above 2.22; Strateva et al. 2001). These targets could be good candidates for green valley blue ellipticals or red spirals (Bell et al. 2004, Jaffe et al. 2011, Masters et al. 2011).   Table \ref{tbl-small-II} summarizes the small number statistics for each sample individually and the combined sample for all our measurement categories reported in the text. That is, (u-r), $\Delta$(g-i), E(B-V)$_{(g-i)}$, H$\alpha$/H$\beta$, E(B-V)$_{H\alpha/H\beta}$, and SFR (H$\alpha$).  

{ Many of our galaxies are very faint, with m$_{r} > 18.5$. Those systems that do not have imaging detections have $r$-band limits of m$_{r}>24$ (similarly for m$_{i}$). While many targeted galaxy surveys go deep enough to detect objects of these magnitudes (reaching limits of I$\sim27$ in some cases; i.e., COSMOS, GEMS, and GOODS), many of the galaxies presented here are small enough in size to be blended with the background QSO (17/23 in Sample I and 18/27 in Sample II -- over half of the entire sample). Thus many surveys would classify these systems as QSOs and stars. Therefore, their presence would be missed except for their spectral emission present in QSO spectra.  Additionally, surveys such as the SDSS cannot resolve and individually measure overlapping targets such as these (unless the galaxy is at a higher impact parameter). Therefore, automated detections of galaxies in imaging can be difficult, even if they are luminous, and detections and measurements by eye prohibitively time consuming. 

The detection of these galaxies, which might otherwise be completely overlooked, by emission in QSO spectra is a unique and robust method. However, their presence being now known, they are detectable by deep targeted observations. Traditionally, imaging surveys are biased towards the largest and brightest galaxies. QSOALS are unique in that they avoid this bias. Our sample selection is not based on imaging, but on spectral emission, which can also trace those galaxies that are otherwise difficult to detect in imaging (with constraints on redshift). Table~\ref{tbl-phot-galaxies} lists the magnitudes, magnitude limits, luminosities in four optical filters, and (u-r) values for all of our galaxy targets. } 

\subsection{Dust measurements} 

\begin{figure}
\begin{minipage}{1.0\linewidth}
\begin{center}

\includegraphics[trim = 0mm 0mm 0mm 0mm, clip, width=1.0\textwidth]{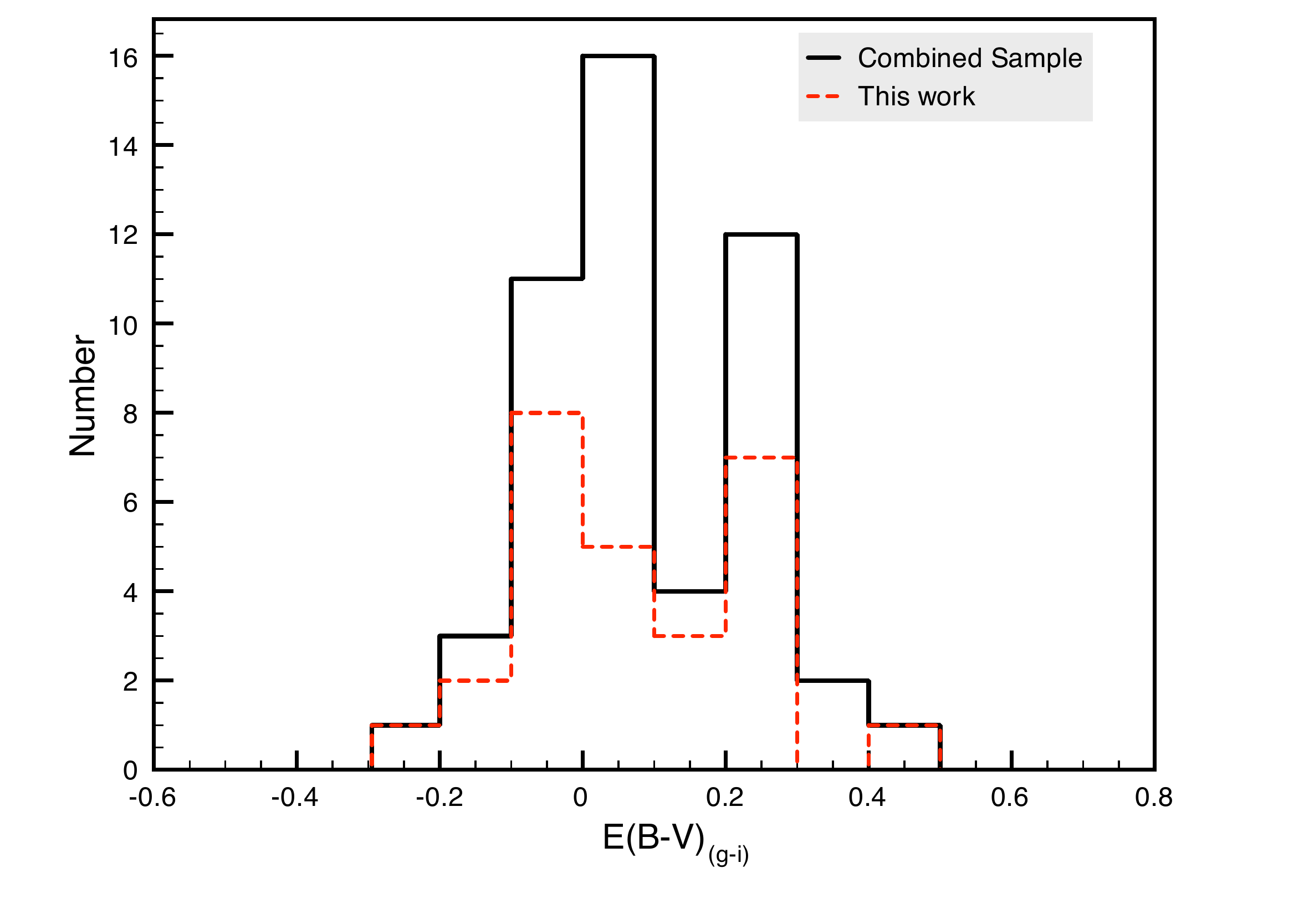}

\end{center}
\caption{The histogram of E(B-V) values by percentage in the combined sample is in black. The histogram of Sample II unique to this paper is in red. Values of E(B-V) $> 0.2$ may be considered as showing the QSOs are reddened by the foreground galaxy, as defined by York et al. (2006). \label{fig-ebv}}
\end{minipage}
\end{figure}

Figure~\ref{fig-ebv} shows an E(B-V)$_{(g-i)}$ histogram of the combined Sample I$+$II by number{ in addition to the histogram for Sample II alone (shown by the dashed line).  Statistically, the center of the distribution is offset from 0, indicating that on average QSOs with foreground galaxies are more reddened than the average QSO sample.  However, in any one case, we cannot separate systems with real extinction from error. Table 1 of York et al. (2006) quotes an average E(B-V)$_{(g-i)} = 0.01$ for their full sample of  807 QSOs. In the sample of QSOs without absorbers from York et al. (2006), the spread of E(B-V)$_{(g-i)}$ is $\sim 0.1$, compared to the spread of our sample of $\sim 0.2$. This broadening is most likely caused by the error introduced in the deconvolution of the QSO from the galaxy, which is not a factor in samples of QSOs without intervening galaxies. Intrinsic color variations have not been taken into account.}

We have performed a Kolmogorov-Smirnov (KS) test on the E(B-V)$_{(g-i)}$ values between our samples and the Mg II absorption-selected sample from York et al. (2006) in order to determine if the samples come from the same populations. Our H$\alpha$-selected Sample I compared to our multi-emission selected Sample II (numbering 23 and 27 data points respectively) return a discrepancy of 0.468, corresponding to a P-value of 0.855\%, the probability that the samples are from the same population. The KS test between only our multi-emission selected Sample II and the sample of York et al. (2006) (numbering 27 and 807 data points respectively) returns a discrepancy of 0.547 corresponding to a P-value of 0.00003\%. Similarly, when comparing the combined Sample I$+$II  with the sample from York et al. (2006), the test returns a discrepancy of 0.332 corresponding to a P-value of 0.006\%. This indicates that there is a very small chance the samples themselves are drawn from the same population, however, we caution that the two-sided KS test works best for sample sizes of 40 or more, and our Samples I and II each have $<40$ systems (though sample I$+$II has $>40$ systems). Individually and in combination, our samples are not drawn from the same population as the sample of York et al. (2006). In any case, the very small probability values still suggest that our sample is more reddened than the average of the York et al. sample, { which is in agreement with Figure~\ref{fig-ebv}}. 

Our targets with the highest reddening values from Sample II are 0.64, 0.36, and 0.29 (objects 5, 11, and 22 respectively). All three of these objects have Ca II absorption. Wild et al. (2006) studied a sample of 37 Ca II absorbers, and find E(B-V) values in the range -0.008 to 0.417, which is consistent with our measurements. Comparisons with other galaxy samples expected to have high reddening values include Ca II and DIB systems such as that of Ellison et al. (2008). This study searched nine Ca II-selected absorbers at redshifts up to $z=0.55$ for DIBs. They have one detection of a DIB ($\lambda$Gr5780) with E(B-V)$=0.23$ (towards J0013-0024), which is lower than our most reddened objects, higher than most of our sample, but also higher than expected for typical Ca II selected systems. 

Figure~\ref{fig-ebv_b}a is a plot of E(B-V)$_{(g-i)}$ vs. impact parameter. The points are represented the same as in previous figures. The Spearman correlation value for the combined sample (I$+$II) is r$_{s}$=-0.44, suggesting an anti-correlation at the $\alpha=0.05$ significance level. { Figure~\ref{fig-ebv_b}b is a plot of the same, but against a reduced-b parameter. This parameter is an estimate of the distance of the QSO from the center of the galaxy on a scale of 0 to 1 (0 being the galaxy center, 1 being the maximum galaxy radius measured in flux). Using this method, the slight anti-correlation found between E(B-V)$_{(g-i)}$ and b disappears.  }

Figure~\ref{fig-ebv_ur}c plots E(B-V)$_{(g-i)}$ vs. color (u-r). The points here are also represented the same as in previous figures. In Paper I of this series (Sample I), we found a slight correlation at the $\alpha=0.05$ significance level. However, after the addition of the sample new to this paper (Sample II),  we find no significant correlation for the combined sample.

\subsection{Measurements from galactic emission lines}

The Balmer decrement can also be used to determine extinction in a system and used to correct SFR. Values of H$\alpha$/H$\beta>$2.85 show extinction under the conditions normally assumed for galaxy H II regions (Case B recombination).  Table~\ref{tbl-SFR} reports the H$\alpha$/H$\beta$ values for each of our targets, which were used to correct the SFR for extinction.  Equation 4 is used to calculate the reddening values E(B-V) from H$\alpha$/H$\beta$. 

\begin{figure}
\begin{minipage}{1.0\linewidth}
\begin{center}

\subfigure[(a)]{\label{fig-ebv_b}
\includegraphics[trim = 0mm 0mm 0mm 0mm, clip, width=0.45\textwidth]{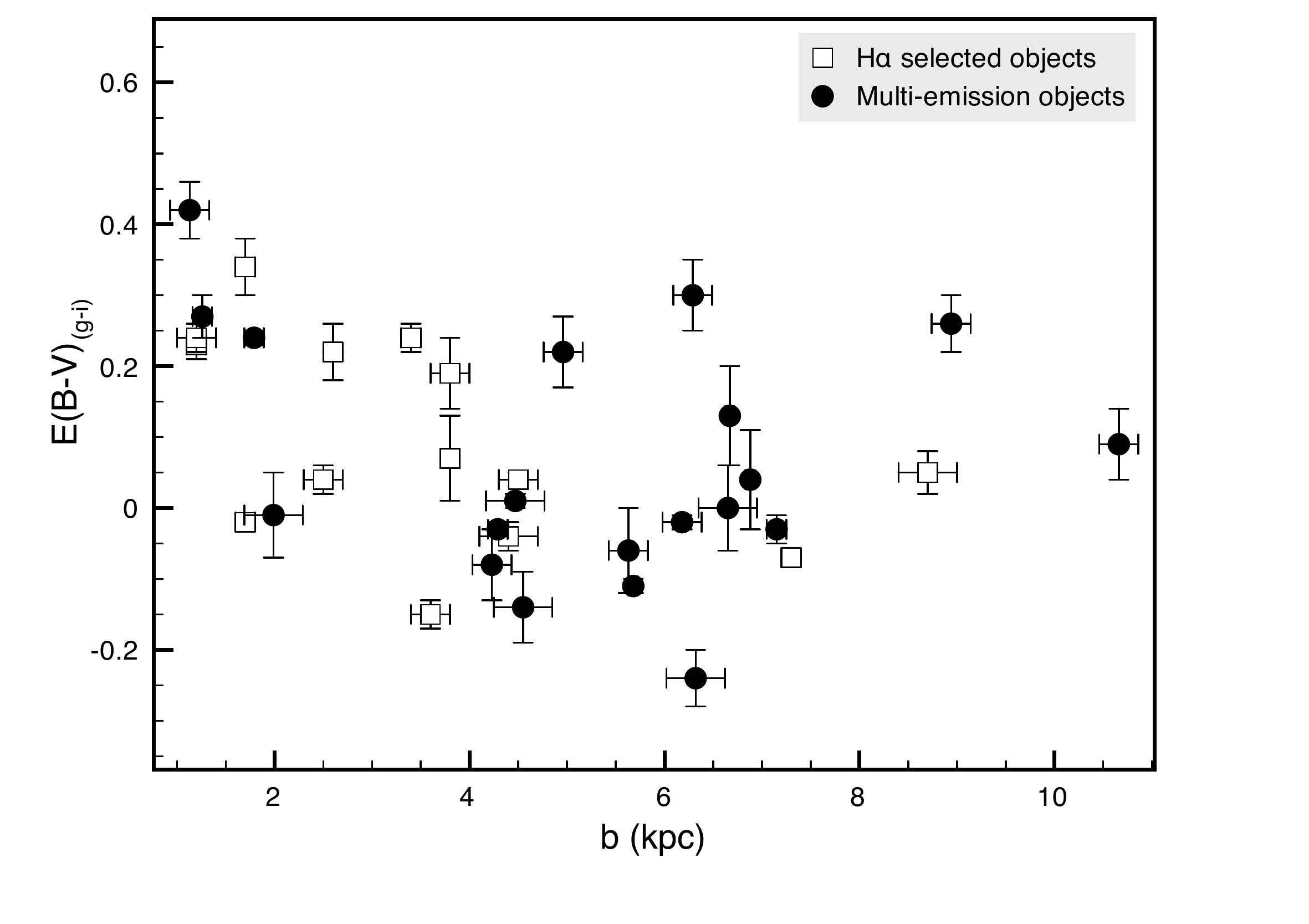}
}
\subfigure[(b)]{\label{fig-reduced_ebv}
\includegraphics[trim = 0mm 0mm 0mm 0mm, clip, width=0.45\textwidth]{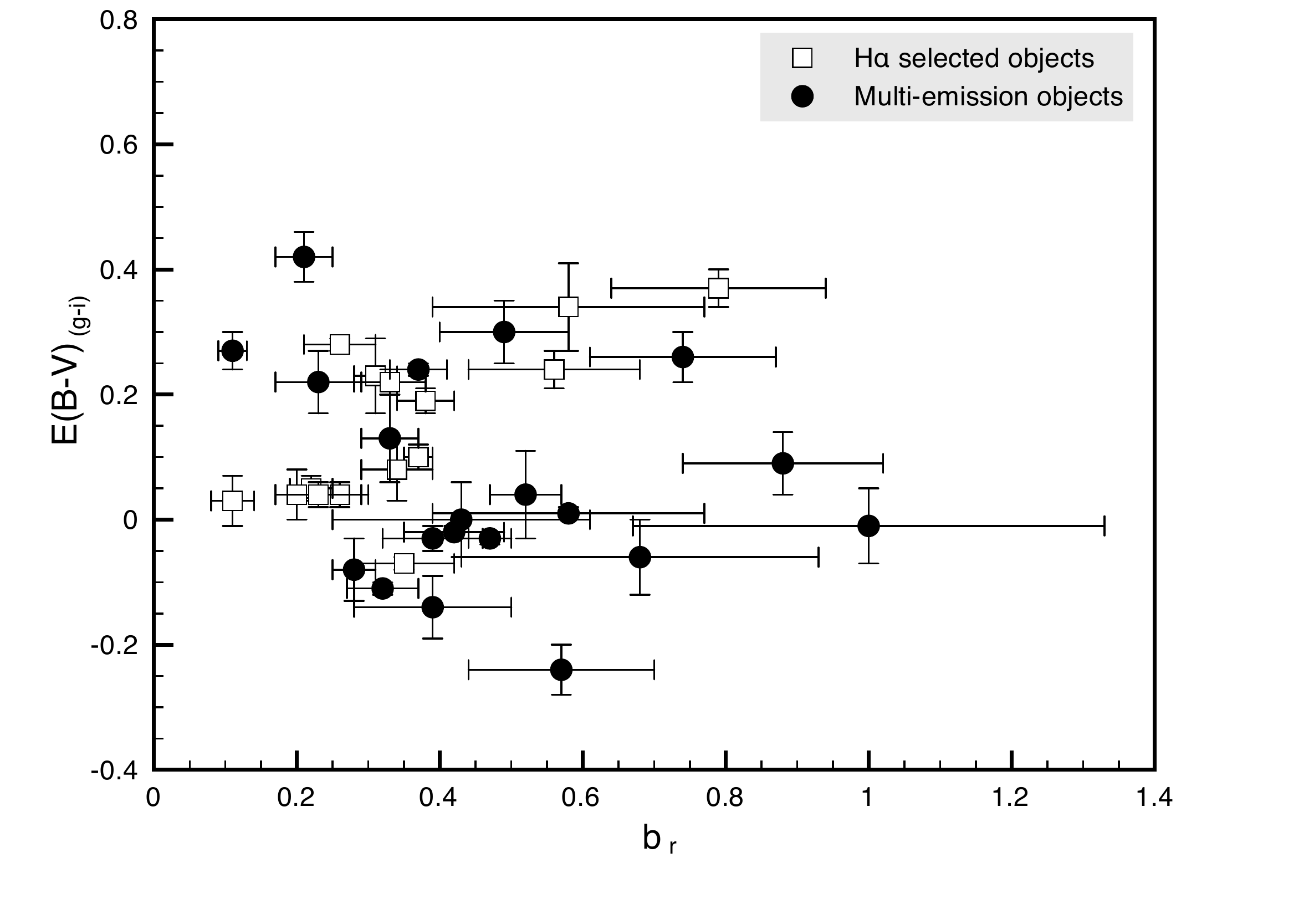}
}
\subfigure[(c)]{\label{fig-ebv_ur}
\includegraphics[trim = 0mm 0mm 0mm 0mm, clip, width=0.45\textwidth]{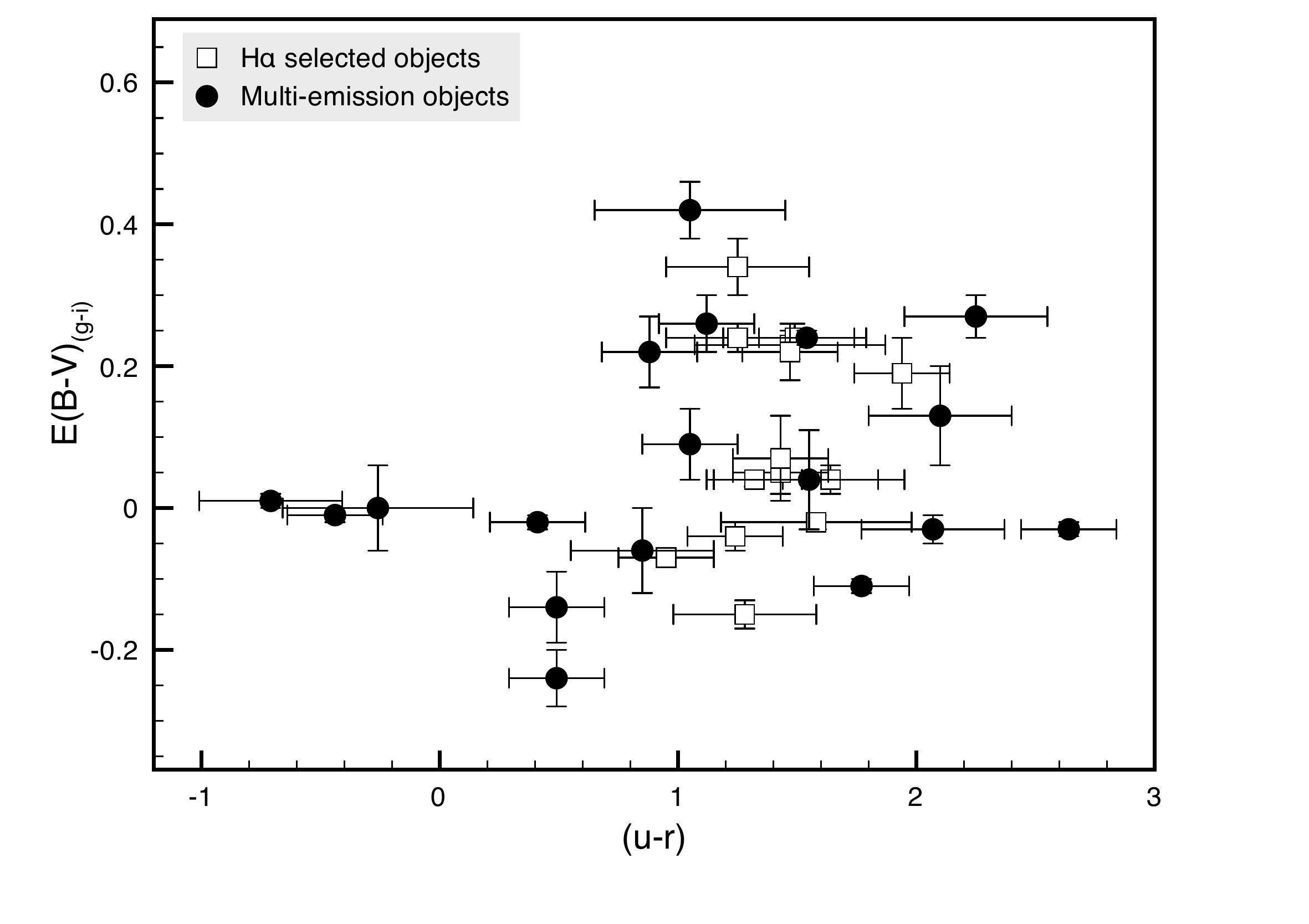}
}
\subfigure[(d)]{\label{fig-ebvgi_ebvhahb}
\includegraphics[trim = 0mm 0mm 0mm 0mm, clip, width=0.45\textwidth]{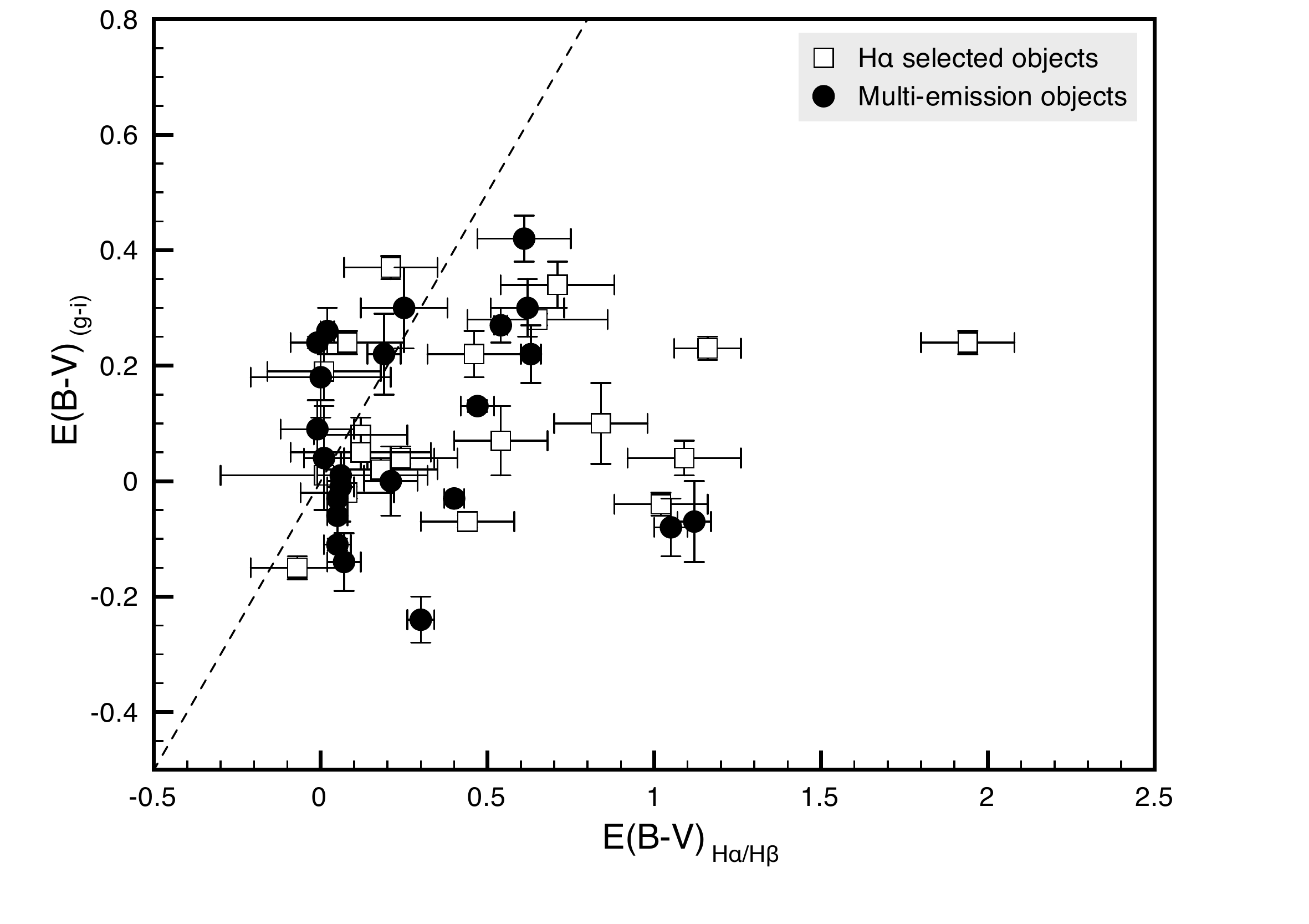}
}
\subfigure[(e)]{\label{fig-deltaebv_b}
\includegraphics[trim = 0mm 0mm 0mm 0mm, clip, width=0.45\textwidth]{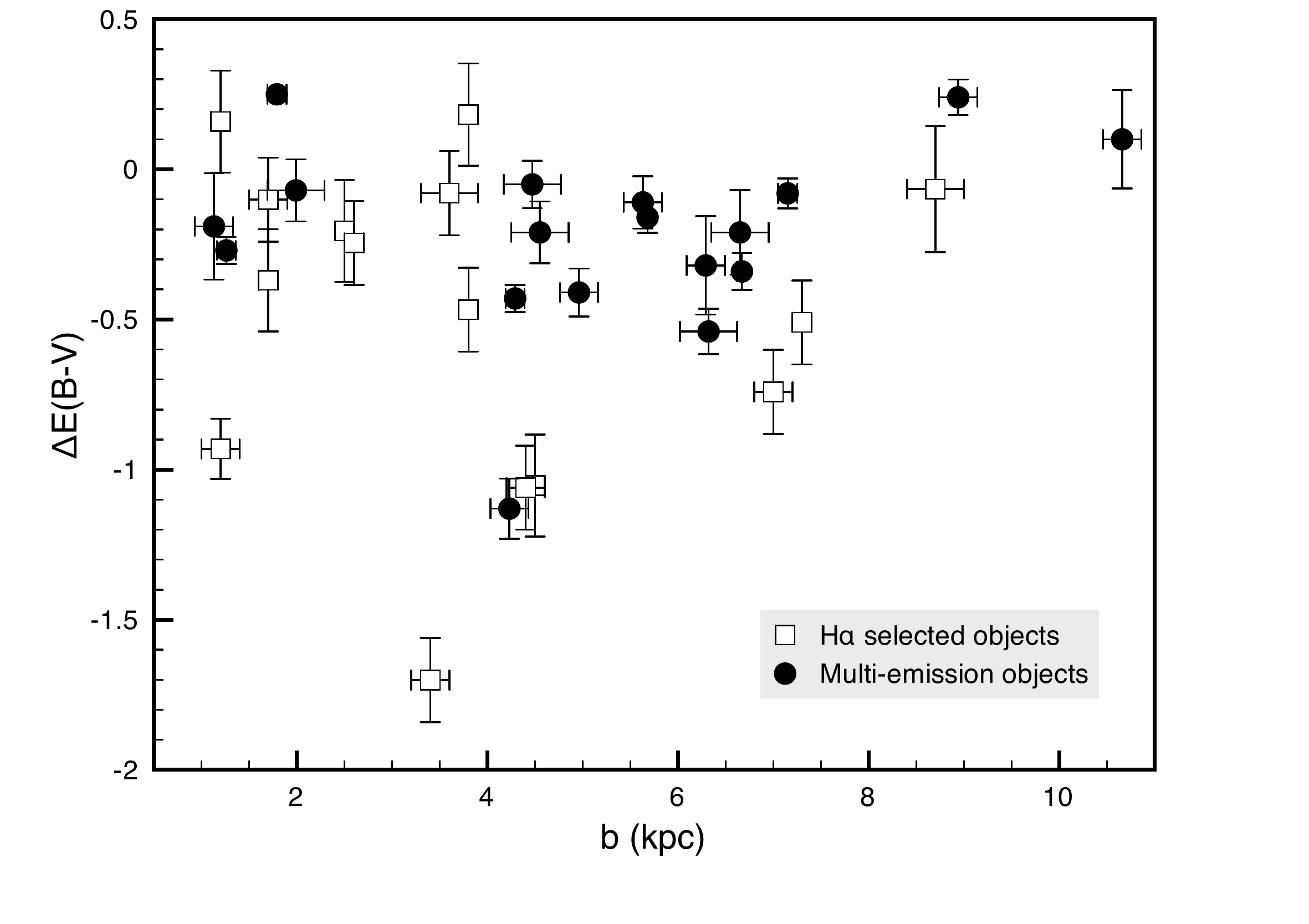}
}
\subfigure[(f)]{\label{fig-deltaebv_reducedb}
\includegraphics[trim = 0mm 0mm 0mm 0mm, clip, width=0.45\textwidth]{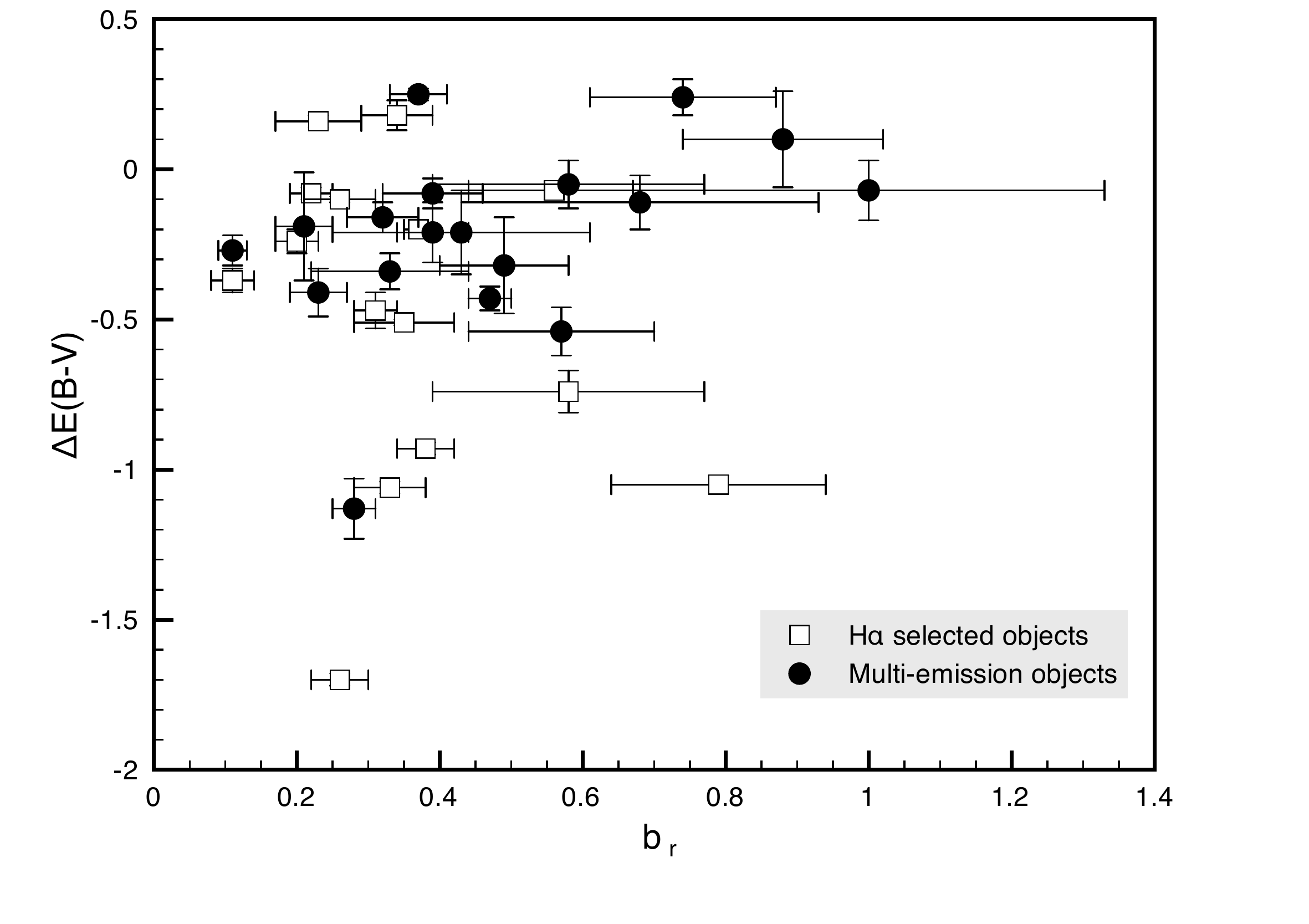}
}

\end{center}
\caption{ a) E(B-V)$_{(g-i)}$ vs. impact parameter. b) E(B-V)$_{(g-i)}$ vs. reduced impact parameter (a ratio of the impact parameter and the galaxy radius). c) E(B-V)$_{(g-i)}$ vs. color. c) Plot of E(B-V)$_{(g-i)}$ vs. E(B-V)$_{H\alpha/H\beta}$. The dotted line marks the one-to-one ratio between the two variables.  d) $\Delta$E(B-V) = E(B-V)$_{(g-i)}$ - E(B-V)$_{H\alpha/H\beta}$ vs. impact parameter. f) $\Delta$E(B-V) vs. reduced impact parameter. Intrinsic color variations have not been taken into account for the error bars on E(B-V)$_{(g-i)}$.}
\end{minipage}
\end{figure}

Figure~\ref{fig-ebvgi_ebvhahb}d plots the color excess of the QSO compared to that determined from the hydrogen emission lines for the combined sample. In general, the trend is that the gas in front of the QSO is less reddened than that in the regions of the galaxy producing emission lines. We find the mean E(B-V)$_{H\alpha/H\beta}$ to be 0.38 with a median of 0.23, compared to the median and mean values based on $\Delta$(g-i) of 0.06 and 0.09, respectively. Using a Spearman correlation test, we find no significant correlation at the $\alpha=0.05$ level or higher. 

Figure~\ref{fig-deltaebv_b}e plots the difference in our two E(B-V) values for each system against impact parameter. $\Delta$E(B-V) is the difference between the reddening in the QSO line of sight (from $\Delta$(g-i)) and the reddening in the rest of the galaxy (from the Balmer decrement). We find no significant correlation between these variables. This implies that the dust is patchy and found selectively near star forming regions that produce H$\alpha$ and H$\beta$ emission lines. 

{ We have again used a reduced b parameter to search for correlations between the reddening and the impact location of the QSO with the galaxy. Figure~\ref{fig-ebv_b}f shows $\Delta$E(B-V) vs. the reduced b parameter, as described for Figure~\ref{fig-ebv_b}b. As in Figure~\ref{fig-deltaebv_b}e, we find no correlation between $\Delta$E(B-V) and the QSO distance from galaxy center. }

The reddening-corrected and uncorrected SFR can be found in Table~\ref{tbl-SFR}. The geometric corrections to SFR for galaxy surface area falling outside the SDSS spectra fiber for each galaxy are in the range 0.48 - 0.99 (48\% - 99\% of these galaxies lie outside the fiber). All but one galaxy has $>50\%$ of its surface area outside the fiber. Correcting the SFR in this way produces an upper limit, as we do not expect star formation to be uniform across the entire galaxy. Taking this fact into account, our SFR most certainly agrees better with the large disk models at low redshifts.   

Our sample discussed in this paper has a mean extinction-corrected SFR of 2.81 M$_{\odot}$ yr$^{-1}$ with a median of 0.58 M$_{\odot}$ yr$^{-1}$.  The mean and median for the combined Sample I$+$II are 3.13 M$_{\odot}$ yr$^{-1}$ and 0.63 M$_{\odot}$ yr$^{-1}$, respectively. Comparing with field galaxies, the sample of low redshift field galaxies from Kewley et al. (2002) have a mean SFR$_{H\alpha}$ of 3.5 M$_{\odot}$ yr$^{-1}$ with a median value of 0.69 M$_{\odot}$ yr$^{-1}$, showing good agreement with our individual sample and combined samples both. Kulkarni et al. (2006) report low SFRs for galaxies known to contain metal-line absorbers, with SFRs $<5$ M$_{\odot}$ yr$^{-1}$. This is also consistent with our current measurements. However, as noted earlier, our measurements are lower limits with significant percentages of galaxy surface area falling outside the SDSS fiber. Followup spectroscopy is required to get a more complete view of the SFR in these systems. 

For targets with the appropriate available emission lines, we have calculated the emission line metallicity for several indices, namely R23, N2, and O3N2 from the extinction corrected fluxes. From these we were also able to calculate the $12+log(O/H)$ values by the analytic methods of McGaugh et al. (1999). Table~\ref{tbl-emission-metallicity-II} lists these values. For those systems with detected [O III]$\lambda5007$ and [N II]$\lambda5897$, we are able to determine which branch of R23 our values belong to (lower or upper), but for those systems without this ratio we have listed both the lower and upper branch values of R23. Based on the accepted solar value of 8.69 (Allende-Prieto et al. 2001), we find many of our systems are subsolar in metallicity. Those systems with [O~III]b/[N II]b $>2$ correspond to the lower, metal-poor branch of R23, and those with the ratio $<2$ correspond to the upper branch (Carollo et al. 2004). In the case where the branch cannot be determined, the corresponding values from N2 and O3N2 agree, indicating subsolar metallicites. This is in good agreement with the low SFRs we find. 

\begin{figure}
\begin{minipage}{1.0\linewidth}
\begin{center}

\includegraphics[trim = 0mm 0mm 0mm 0mm, clip, width=1.0\textwidth]{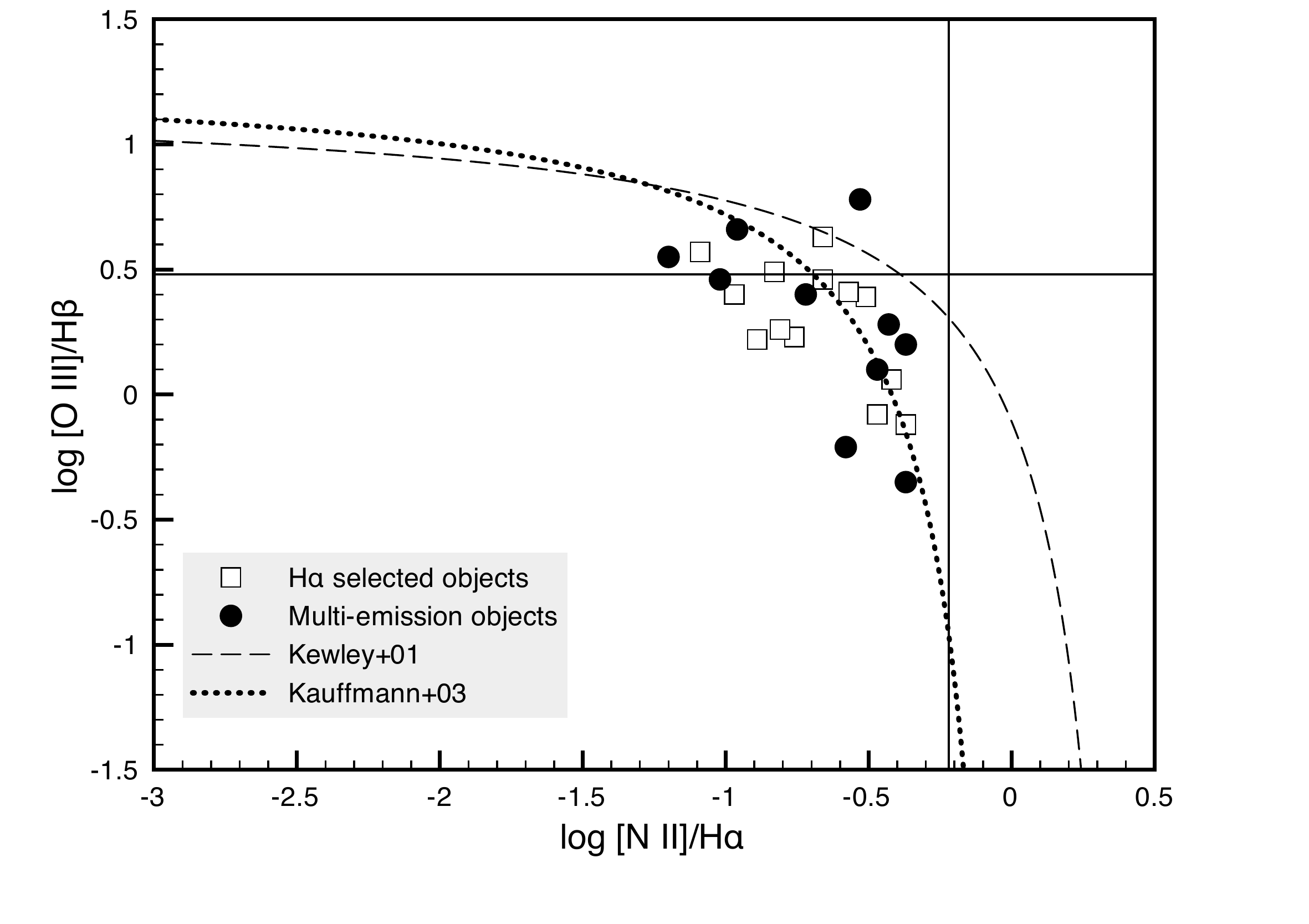}

\end{center}
\caption{BPT diagram (Baldwin et al. 1981) of H$\alpha$ selected objects from Paper I and multi-emission selected objects presented in this paper. The plot indicates that our sample of galaxies with these emission lines are normal galaxies. The dotted line follows the prescription of Kauffmann et al. (2003), while the dashed line is the prescription of Kewley et al. (2001) for the division between starburst galaxies and AGN. The crosshairs mark the division between Seyferts and LINERs. See text for details. These data points will benefit from followup spectroscopy of the galaxies, as the emission measured from SDSS spectra may not be the total emission given the 3$\arcsec$ diameter of the SDSS spectral fiber.  \label{fig-bpt-II}}
\end{minipage}
\end{figure}

Figure \ref{fig-bpt-II} shows the points from Papers I and II on a BPT (Baldwin-Phillips-Terlevich; Baldwin et al. 1981) diagram. This diagram has allowed us to get a preliminary idea of the nature of the galaxies we observe; i.e., whether they are HII regions, AGN, or composites. { This diagram can be used to help distinguish the ionization mechanism of the nebular gas we observe. The dashed curve represents the division between starburst galaxies and AGN as prescribed by Kewley et al. (2001). The dotted line represents the same division, but following the prescription of Kauffmann et al. (2003). Additionally, the general definition for Seyfert galaxies is log([O III]/H$\beta$)$>0.48$ and log([N II]/H$\alpha$)$>-0.22$, and the definition for LINERs (low-ionization nuclear emission regions) is log([O III]/H$\beta$)$<0.48$ and log([N II]/H$\alpha$)$>-0.22$. These limits are represented by the horizontal and vertical lines. 

According to these models and definitions, the majority of galaxies for those plotted fall within the normal HII and normal AGN regions. That is, the mechanism by which the gas we have detected in emission is ionized is from normal star forming regions, not Seyferts or LINERs. According to Kauffmann et al., $\sim$14 of our 23 galaxies fall within the starburst region with 4 on the border between starburst and AGN. However, according to Kewley et al., only one of our galaxies qualifies as a normal AGN with the rest being starburst galaxies. Followup spectroscopy will further constrain the positions of these galaxies on the BPT plot,  as it is possible a significant percentage of the galaxy surface area falls outside the SDSS spectral fiber, thus missing a significant percentage of the emission. }  

\subsection{Absorption lines}

\begin{figure}
\begin{minipage}{1.0\linewidth}
\begin{center}

\includegraphics[trim = 0mm 0mm 0mm 0mm, clip, width=1.0\textwidth]{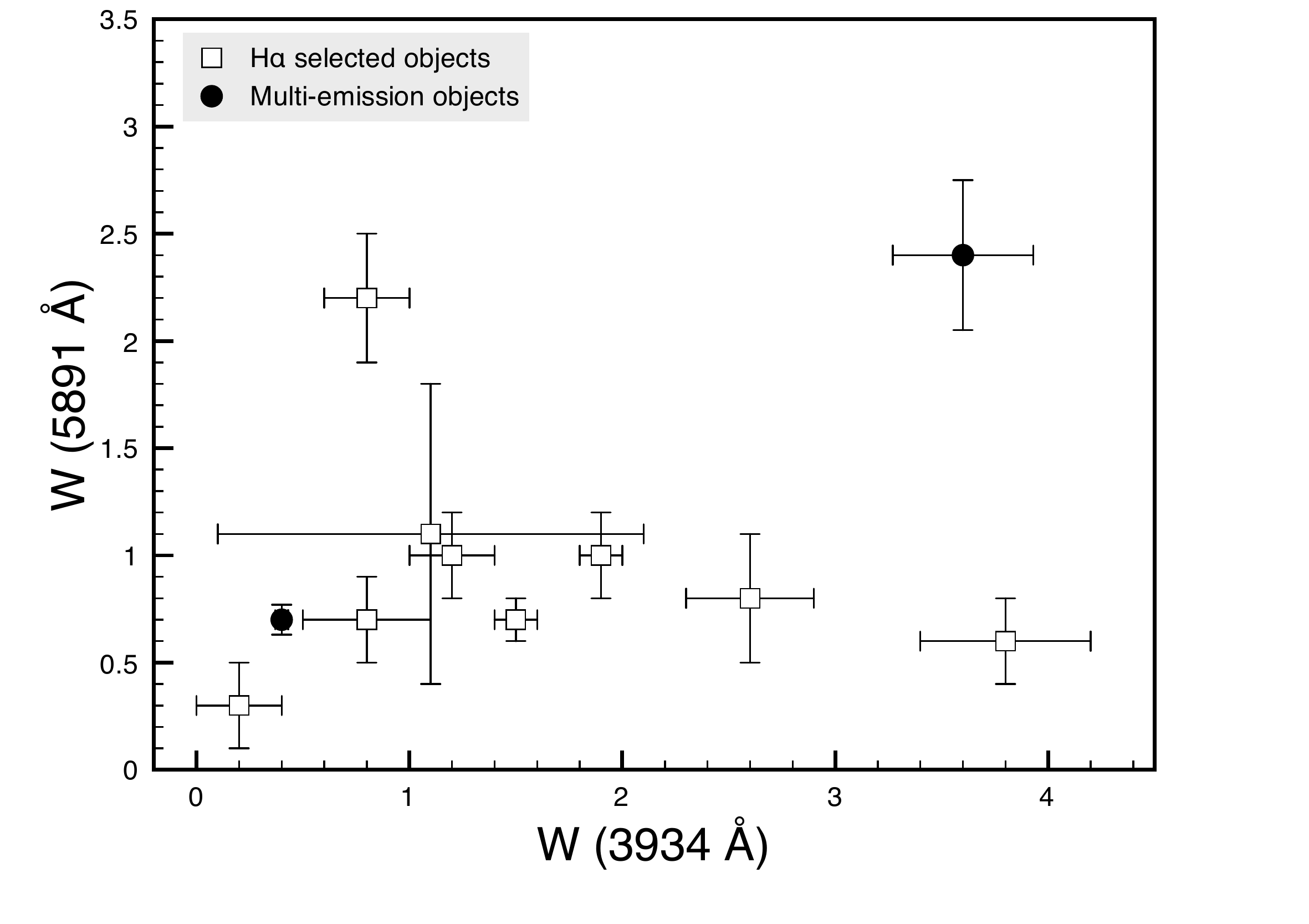}

\end{center}
\caption{Equivalent width of Na I (D2) vs. equivalent width of Ca II (K).  \label{cana-II}}
\end{minipage}
\end{figure}

Our H$\alpha$ selected sample from Paper I contributed 12 systems with Ca II (K) absorption features. Among Sample II, we find a further 12 systems with Ca II (K) absorption, listed in Table~\ref{tbl-ew} along with the equivalent width measurements of Ca II H $\lambda3969$, Na I D1 $\lambda5897$, and Na I D2 $\lambda5891$. { This table also shows the EW limits for those systems where absorption was not detected (all or in part). } Figure~\ref{cana-II} shows the plot of W$_{\lambda5891}$ vs. W$_{\lambda3934}$ from Paper I with our two new data points possessing both Ca II K and Na I D2 from Sample II added. As in Paper I of this series, we find no significant correlation between the two variables. 

\begin{figure}
\begin{minipage}{1.0\linewidth}
\begin{center}

\subfigure[(a)]
{\label{fig-ca_ebv}
\includegraphics[trim = 0mm 0mm 0mm 0mm, clip, width=0.45\textwidth]{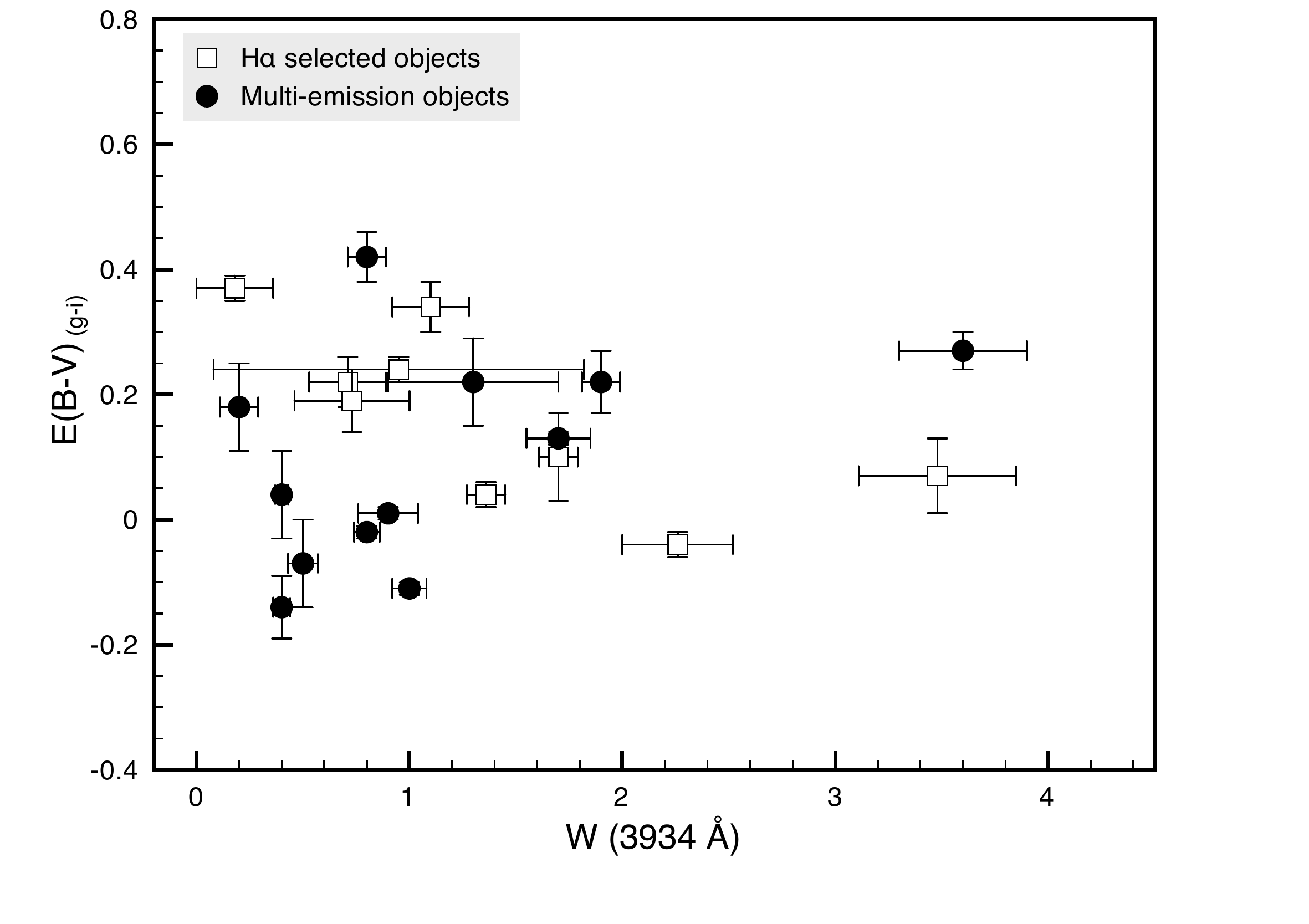}
}
\subfigure[(b)]
{\label{fig-na_ebv}
\includegraphics[trim = 0mm 0mm 0mm 0mm, clip, width=0.45\textwidth]{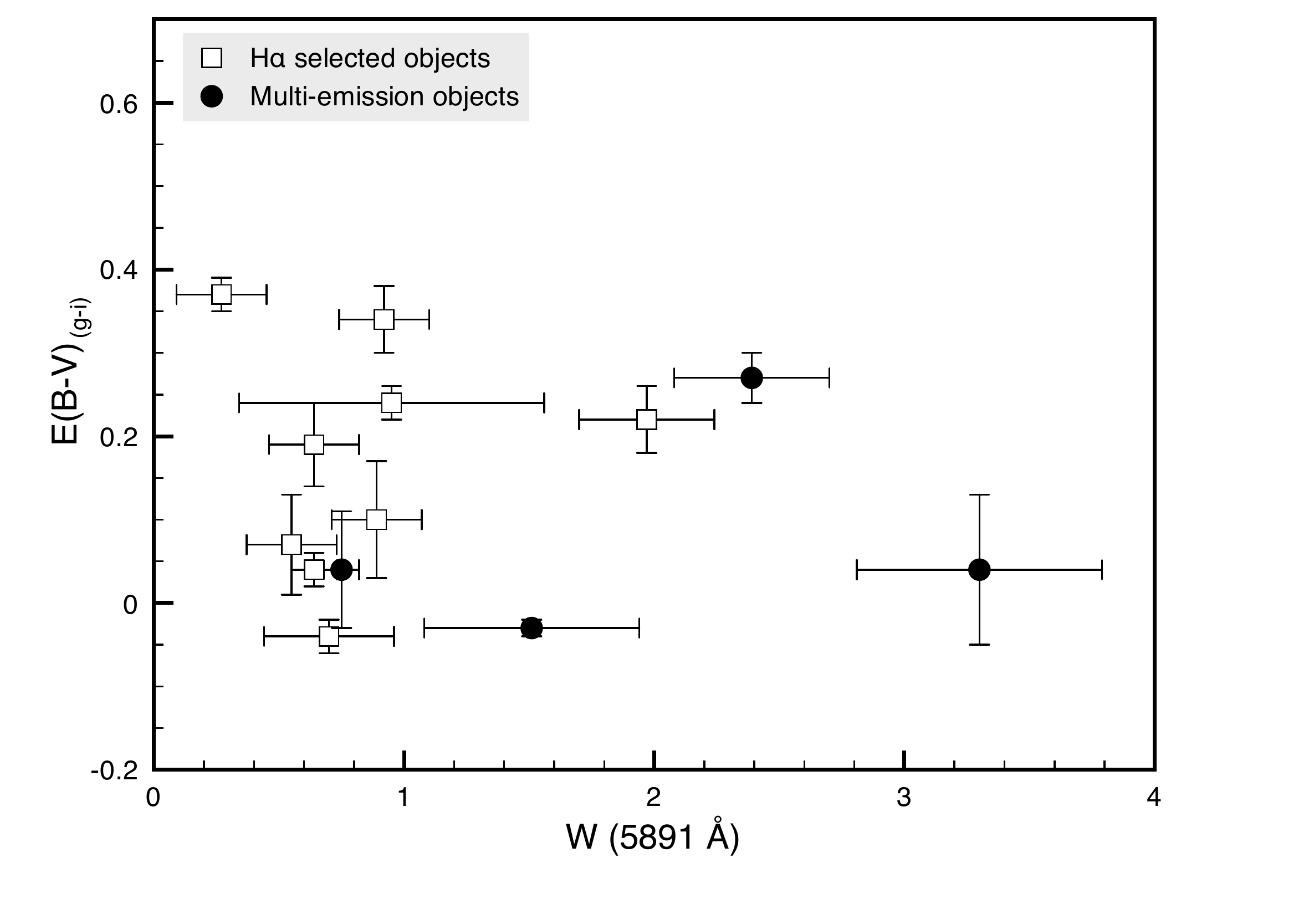}
}
\subfigure[(c)]
{\label{fig-ca_b}
\includegraphics[trim = 0mm 0mm 0mm 0mm, clip, width=0.45\textwidth]{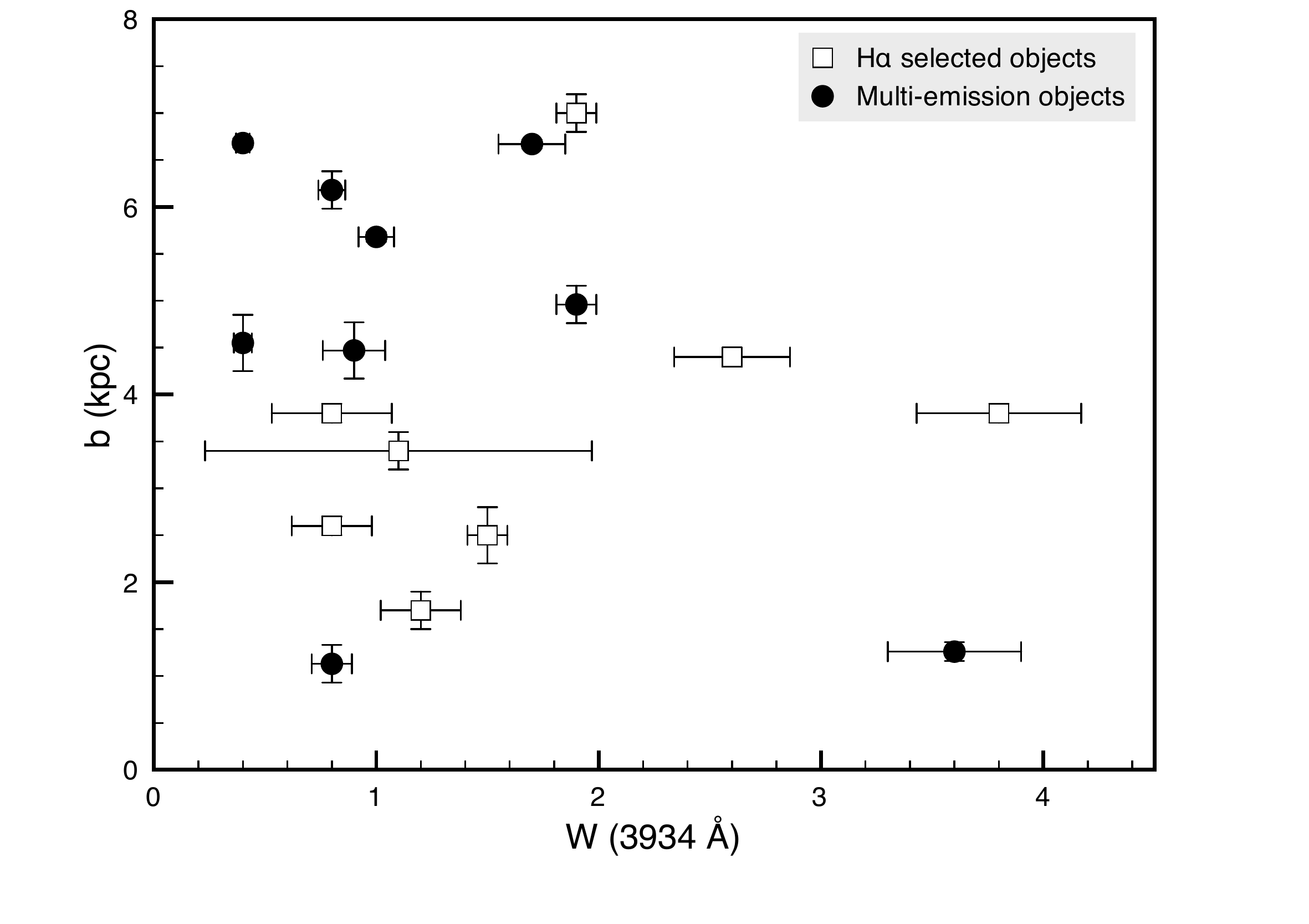}
}
\subfigure[(d)]
{\label{fig-reduced_ca}
\includegraphics[trim = 0mm 0mm 0mm 0mm, clip, width=0.45\textwidth]{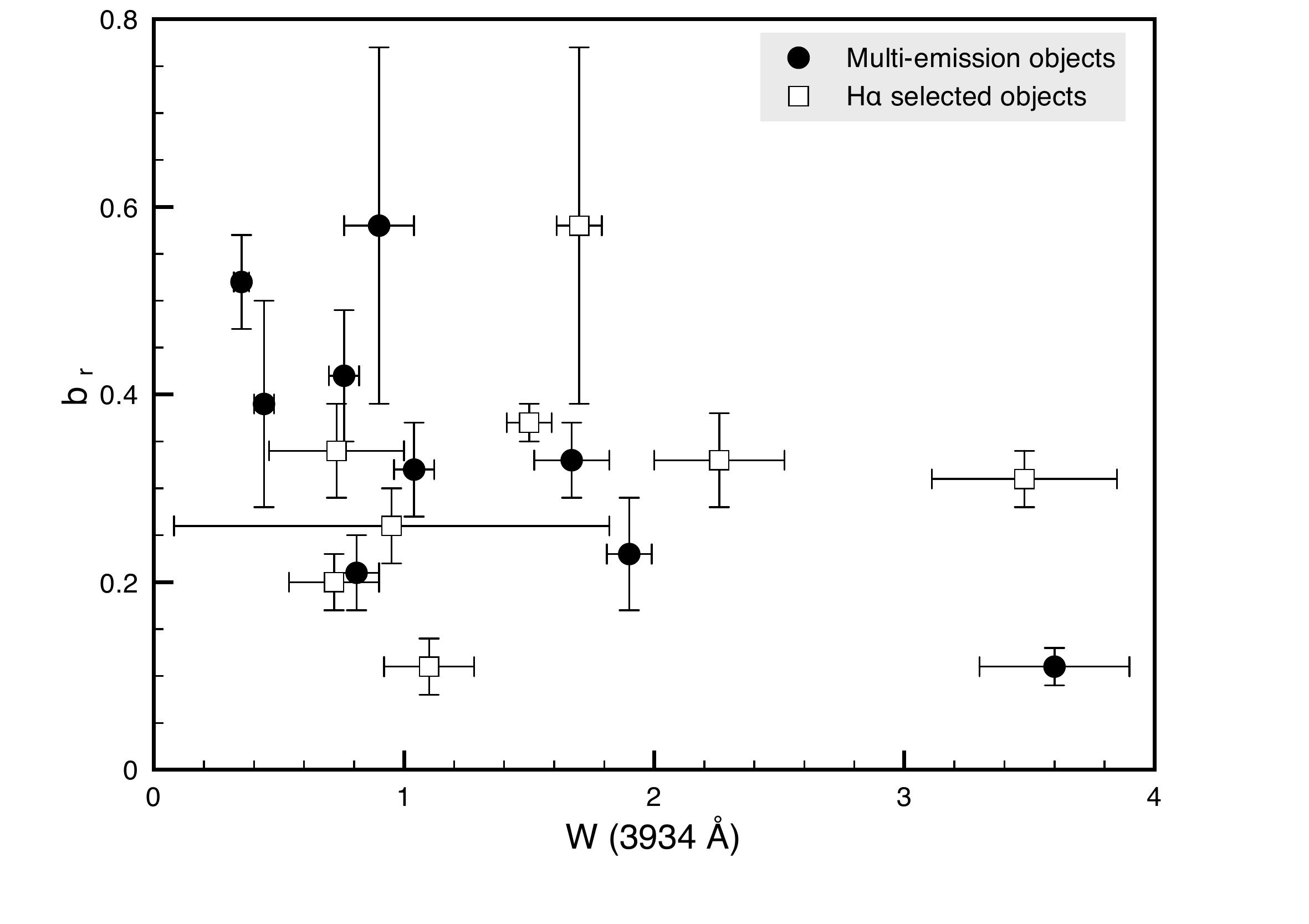}
}
\subfigure[(e)]
{\label{fig-na_b}
\includegraphics[trim = 0mm 0mm 0mm 0mm, clip, width=0.45\textwidth]{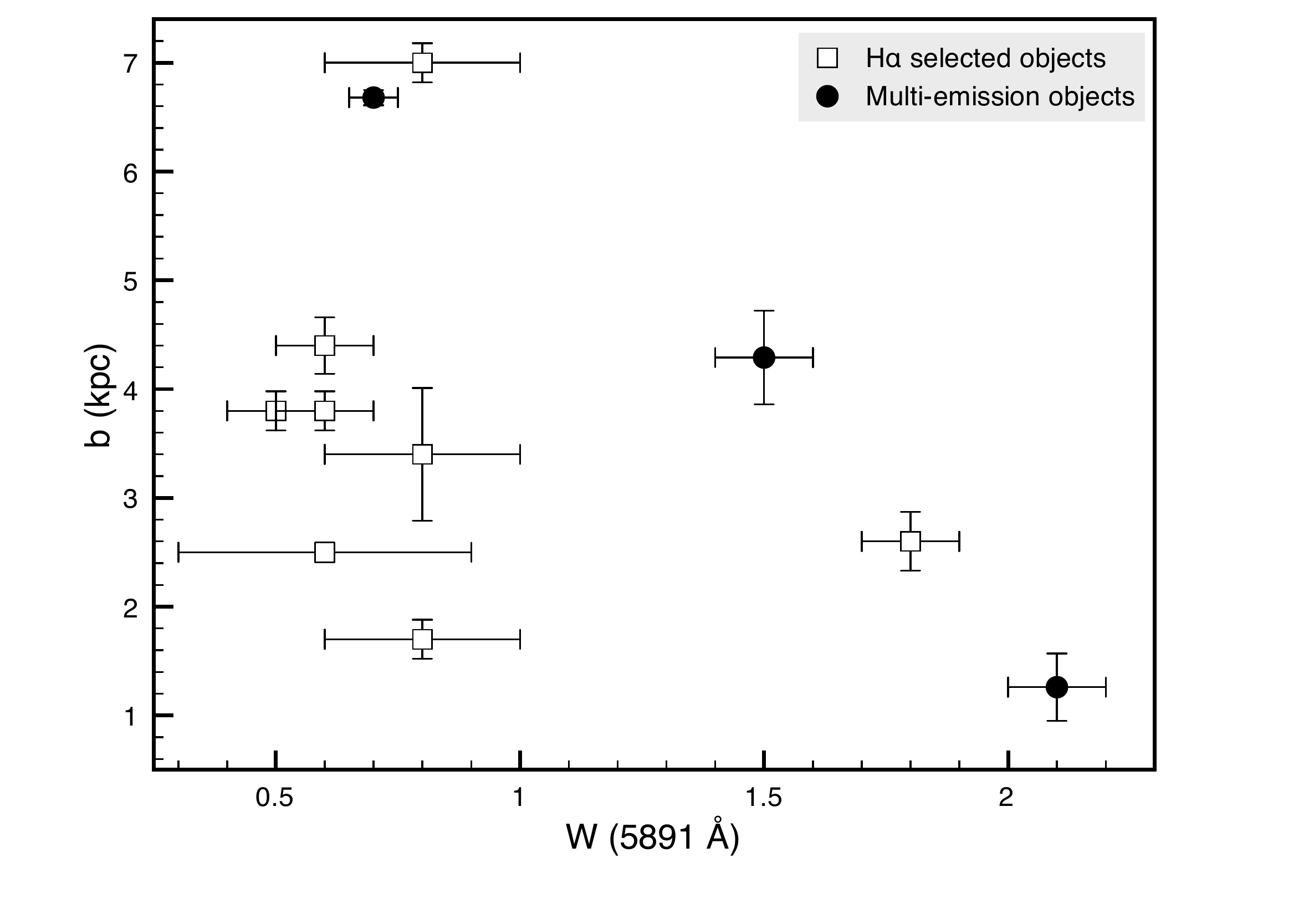}
}
\subfigure[(f)]
{\label{fig-reduced_na}
\includegraphics[trim = 0mm 0mm 0mm 0mm, clip, width=0.45\textwidth]{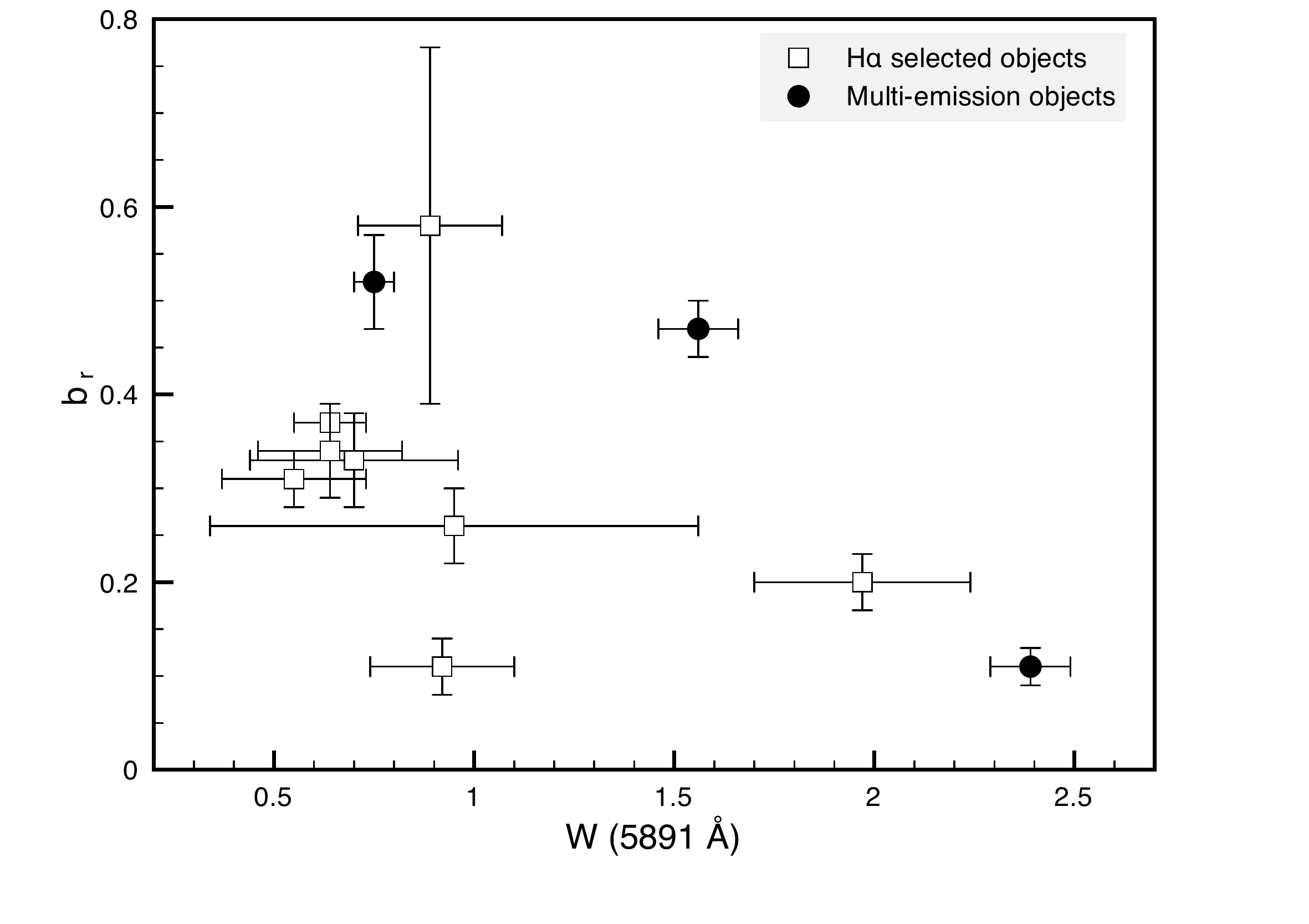}
}

\end{center}
\caption{a) E(B-V)$_{(g=i)}$ vs. equivalent width of Ca II (K). b) E(B-V)$_{(g-i)}$ vs equivalent width of Na I (D2). c) Impact parameter vs. equivalent width of Ca II (K). d) Reduced impact parameter vs. equivalent width of Ca II (K). e) Impact parameter vs. equivalent width of Na I (D2). f) Reduced impact parameter vs. equivalent width of Na I (D2). Please see text for details about these figures.}
\end{minipage}
\end{figure}

We have also added our new data points to the plot of impact parameter vs. W$_{\lambda3934}$, W$_{\lambda5891}$ (Figure~\ref{fig-ca_b}c and Figure~\ref{fig-na_b}e) and E(B-V)$_{(g-i)}$ vs. W$_{\lambda3934}$, W$_{\lambda5891}$ (Figure~\ref{fig-ca_ebv}a and Figure~\ref{fig-na_ebv}b). We find no significant correlation between impact parameter and  W$_{\lambda3934}$ or W$_{\lambda5891}$. In Paper I, we found a significant anti-correlation between E(B-V)$_{(g-i)}$ and W$_{\lambda3934}$ (r$_{s}$=-0.77, $\alpha=0.05$). However, with the addition of our new points, the anti-correlation disappears. { This lack of correlation between EW and impact parameter is in keeping with other surveys of QSOALS, where a correlation is sought between Mg II EW and impact parameter (e.g., Bowen \& Chelouche 2011, Hewett \& Wild 2007).}We similarly find no significant correlation between  E(B-V)$_{(g-i)}$ and W$_{\lambda5891}$. Figures~\ref{fig-ca_b}d and \ref{fig-ca_b}f show our absorption EW measurements against the reduced-b parameter. We find only a very weak anti-correlation for these variables at the $\alpha=$0.1 level, indicating a very weak correlation with the reddening and impact at varying distances from galaxy center. { This correlation or anti-correlation is difficult to ascertain due to our limited sample size. Enlarging the sample further is required to confirm this possible anti-correlation, or to refute it. Additionally, improving measurement errors for these absorption features may provide a stronger case for or against this relationship. Our current errors for our absorption measurements are in the range 0.2 - 0.4 \AA for Ca II and 0.2 - 0.6 \AA ~for Na I }

For these absorption measurements it should be noted that interstellar lines as strong as these are almost certainly saturated (especially Na I). Thus, they would be insensitive to correlations except in extreme circumstances (for instance, very low Ca II dust depletion in a region with highly disturbed gas, the Routly-Spitzer effect (Routly \& Spitzer 1952)). The outliers in any of our plots are worth further study. 

Of our sample, 19 targets are detected in both emission and absorption in either Ca II or Na I. Of these, 15 have a visible galaxy detected, leaving 4 with no visible galaxy, but emission and absorption detected in the QSO spectrum, perhaps indicating a dwarf or dark galaxy. In comparison, we have 5 targets with a visible galaxy and detected emission but no detected absorption. We also have 3 targets detected in emission but with no galaxy visible and with no absorption detected. No residuals were detected after PSF subtraction to indicate a possible galaxy presence. This could indicate a red galaxy with no gas left or with gas highly ionized, or perhaps the absorption gas for the species (particularly calcium) are depleted enough as to be undetectable. { These 8 systems with no absorption (target indices 1, 8, 9, 12, 13, 15, 26, and 27) have Ca II EW limits $\sim1.0$, with a few in the limit range $\sim0.5$. Na I EW limits for these targets are similar. These limits are not so stringent and higher resolution spectra are necessary to place a much stricter limit on these absorption features, or to detect any absorption that might be present. Additionally, these systems are at redshifts too low to detect strong absorption features in optical spectra such as Mg II, Fe II, and Ly-$\alpha$. For these, UV spectra are necessary. 

Studies have shown that for any survey of randomly selected QSO fields, there should be a non-negligible number of non-absorbing galaxies at small impact parameters (e.g., Charlton \& Churchill 1996). Recent studies such as Bowen \& Chelouche (2011) find a significant number of lines of sight that do not show absorption in Mg II. In our sample, 8 targets have no detected absorption ($\sim30\%$ of the sample), which could support this theory. Furthermore, suppressed star formation in systems could lead to lack of absorption in heavier metals (Ca II and Na I). }

\subsection{Dark Galaxy Connection}

{ We speculate that the targets in this sample that are not detected in imaging (indices 1, 2, 6, 13, 17, 21, and 27) are strong candidates for dark galaxies. Recent studies have shown that dark galaxies may be crucial in explaining evolutionary models of galaxies (Kuhlen et al. 2012, 2013). Models have long predicted an abundance of dwarf galaxies (dwarf halos) that are not seen in observations with low star formation rates (Kravtsov 2009). However, using a star formation prescription regulated by local H$_{2}$ abundance, Kuhlen et al. (2012, 2013) more accurately reproduce the observational qualities of dwarf galaxy halos (suppressed and quenched star formation) and predict a large population of gas-rich dark galaxies. They further show that dark galaxies have halo masses $<10^{10}$ M$_{\odot}$ and SFRs much less than 1 M$_{\odot}$ yr$^{-1}$. The former is in keeping with the well known fact that QSOALS can be produced in systems with as little gas mass as $10^{6}$ M$_{\odot}$. 

Observational detections of dark galaxies have recently been published supporting these theoretical claims. These dark galaxies have been detected in 21-cm emission and followed up with deep optical imaging and spectroscopy. Matsuoka et al. (2012) find no detectable galaxy in their image, placing a limit of R$_{AB} > 28$ mag arcsec$^{-2}$ in one of two regions of 21-cm emission. In the second region, they find a low surface brightness blue dwarf galaxy with mass $\sim10^{6}$ M$_{\odot}$.  Rhode et al. (2013) report the lowest mass system known that contains significant amounts of gas forming stars. Deep imaging of this object reveals gas-rich dwarf galaxy with mass $\sim10^{6}$ M$_{\odot}$, while spectroscopy shows a SFR of log (SFR) $= -4.27$. These systems are distinct in that they are gas-rich and metal-poor.  

Our own sample of GOTOQs has, on average, very low SFR (averaging $<1$ M$_{\odot}$ yr$^{-1}$), are gas-rich and metal poor, and many targets that have no imaging detections. These systems match the qualitative descriptions and quantitative definitions of dark galaxies and will require follow up observations (21-cm observations and deeper optical imaging). Including upper limits, we have 11 fields with galaxies $<0.1$L$^{\ast}_{r}$, the limit indicating a dwarf galaxy (7 of these are upper limits). These values can be found in Table~\ref{tbl-phot-galaxies}. Targets in this sample not detected in imaging or absorption (indices 1, 13, and 27) are candidates for dark galaxies and also the sample of non-absorbers predicted by Charlton \& Churchill (1996). The suppressed/quenched nature of SFR in dwarf and dark galaxies predicted by Kuhlen et al. (2012, 2013) could further support the lack of absorption features found in 8 of our systems (listed above).  }

\section{Conclusion}\label{section-conclusion}
 
 We have measured the properties of QSOs and foreground galaxies in 27 fields from the SDSS. These galaxies were detected via their narrow galactic emission lines overlaid on QSO spectra. Utilizing nine galactic emission lines, we required for this sample the presence of H$\alpha$ and at least two more lines with SL$>4$. We in turn inspected these results by eye to confirm that they were in fact GOTOQ and to search for visible galaxies. PSF subtraction allowed us to effectively separate the QSO and galaxy for photometric measurements without light contamination from either. 
 
The reddening values found for this sample are generally higher than those found in other samples, in particular the sample of Mg II absorbers from York et al. (2006). We find no trends relating E(B-V)$_{(g-i)}$ with (u-r), H$\alpha$/H$\beta$, or E(B-V)$_{H\alpha/H\beta}$; however, there appears to be a slight anti-correlation with b, the impact parameter. Similarly, we find no correlation between W$_{\lambda3934}$ and W$_{\lambda5891}$ or between each of these variables and impact parameter and E(B-V)$_{(g-i)}$. However, when comparing EW to the reduced-b parameter, we find a very weak anti-correlation. 

The increased sample size discussed in this study reaffirms the suggestion from Paper I that these galaxies are dusty and likely disk-dominated, according to comparisons to the studies of Strateva et al. (2001) and York et al. (2006) and the results of our KS tests. We find our measurements consistent with other work concerning Ca II absorbers, diffuse interstellar band (DIB) systems, and DLAs. The SFRs are generally low, agreeing with the SFRs found for DLAs and sub-DLAs in other studies. Even with our geometric correction, we find our SFR values to fall below the large disk scenario. Our emission line metallicities indicate sub-solar values, in agreement with the low SFRs we find. Future work on this sample includes searching these systems for DIBs in the dustiest of the systems, which can then also be  compared and correlated with other gas properties. { We hypothesize that a number of our targets are in fact dark galaxies and dwarf halos based on recent theoretical and observational publications. } 

In the future, it would also be of interest to obtain UV and IR spectra of the dustiest systems to study the 2175\AA ~bump from the carbonaceous dust and the 10$\mu m$ feature from the silicate dust (10$\mu m$ e.g., Kulkarni et al. 2011, Aller et al. 2012; { 2175 \AA: e.g., Motta et al. 2002, Wang et al. 2004, Noll \& Pierini 2005, York et al. 2006, Srianand et al. 2008, Conroy et al. 2010, Jiang et al. 2011). } We plan a further paper in this series for a sample of GOTOQ that were found by emission line detection of strong [O~III], rather than H$\alpha$, providing a higher redshift range (up to $z=0.83$). These additional targets will double the size of our current sample, allowing a higher degree of constraint on all of the variables discussed. The higher redshift range will allow us to extend our study over a wider range of lookback times, helping us determine if correlations exist at higher redshifts or over longer redshift ranges. 

\section*{Acknowledgements}

LAS acknowledges partial support from a South Carolina Space Grant Graduate Student Fellowship while at the University of South Carolina. LAS and VPK acknowledge partial funding from NSF grant AST/0908890 (PI Kulkarni). VPK also acknowledges partial funding from NSF grant AST/1108830. This material is based upon work supported by the National Science Foundation under Grant Number 0907743. The authors thank the anonymous referee for their valuable input in improving this paper. 

Funding for the creation and distribution of the SDSS Archive has been provided by the Alfred P. Sloan Foundation, the Participating Institutions, the National Aeronautics and Space Administration, the National Science Foundation, the U.S. Department of Energy, the Japanese Monbukagakusho, and the Max Planck Society. The SDSS Web site is http://www.sdss.org/.

\pagebreak

\clearpage

%Figure 1

\begin{figure}
\begin{minipage}{1.0\linewidth}
\begin{center}

\subfigure[7]{
\includegraphics[trim = 0mm 0mm 0mm 0mm, clip, width=0.3\textwidth]{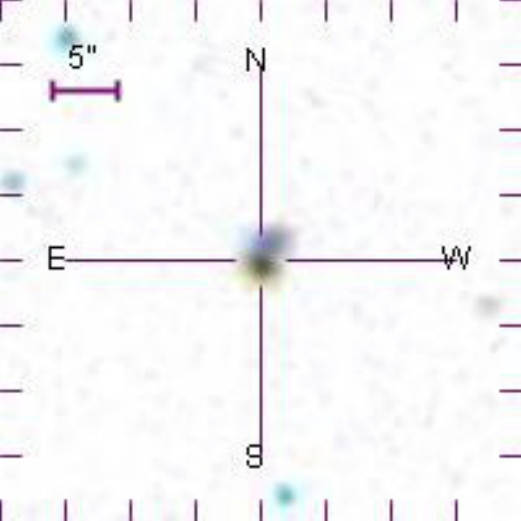}
}
\subfigure[$z_{gal}=0.2499$]{
\includegraphics[trim = 4.5mm 24mm 5mm 3mm, clip, width=0.6\textwidth]{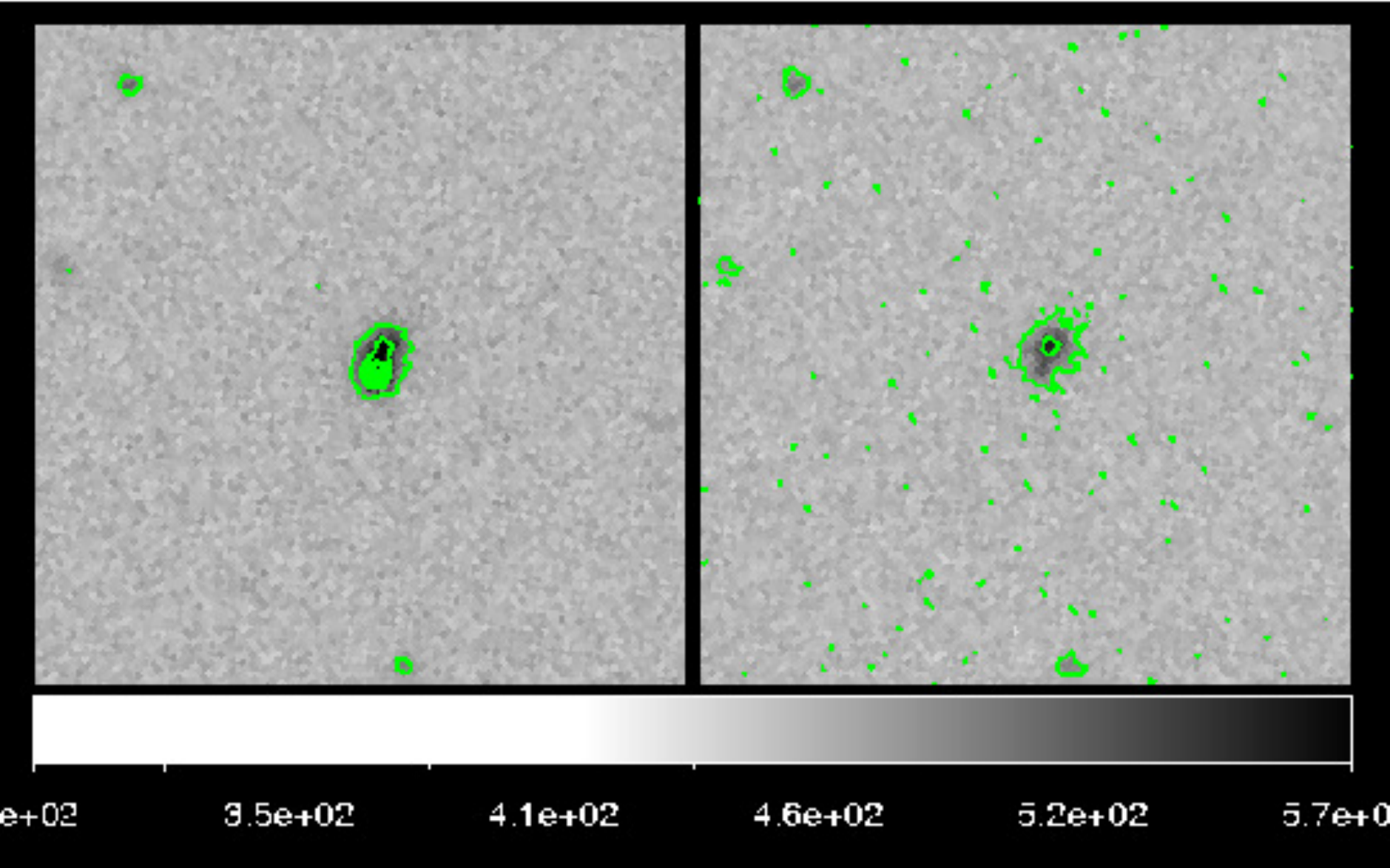}
}
\subfigure[8]{
\includegraphics[trim = 0mm 0mm 0mm 0mm, clip, width=0.3\textwidth]{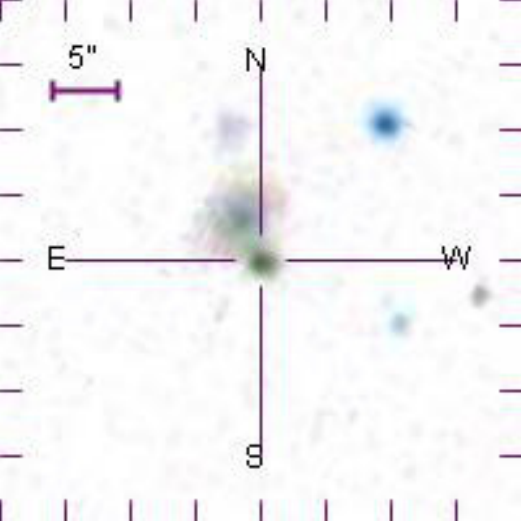}
}
\subfigure[$z_{gal}=0.1412$]{
\includegraphics[trim = 4.5mm 14mm 5mm 2.75mm, clip, width=0.6\textwidth]{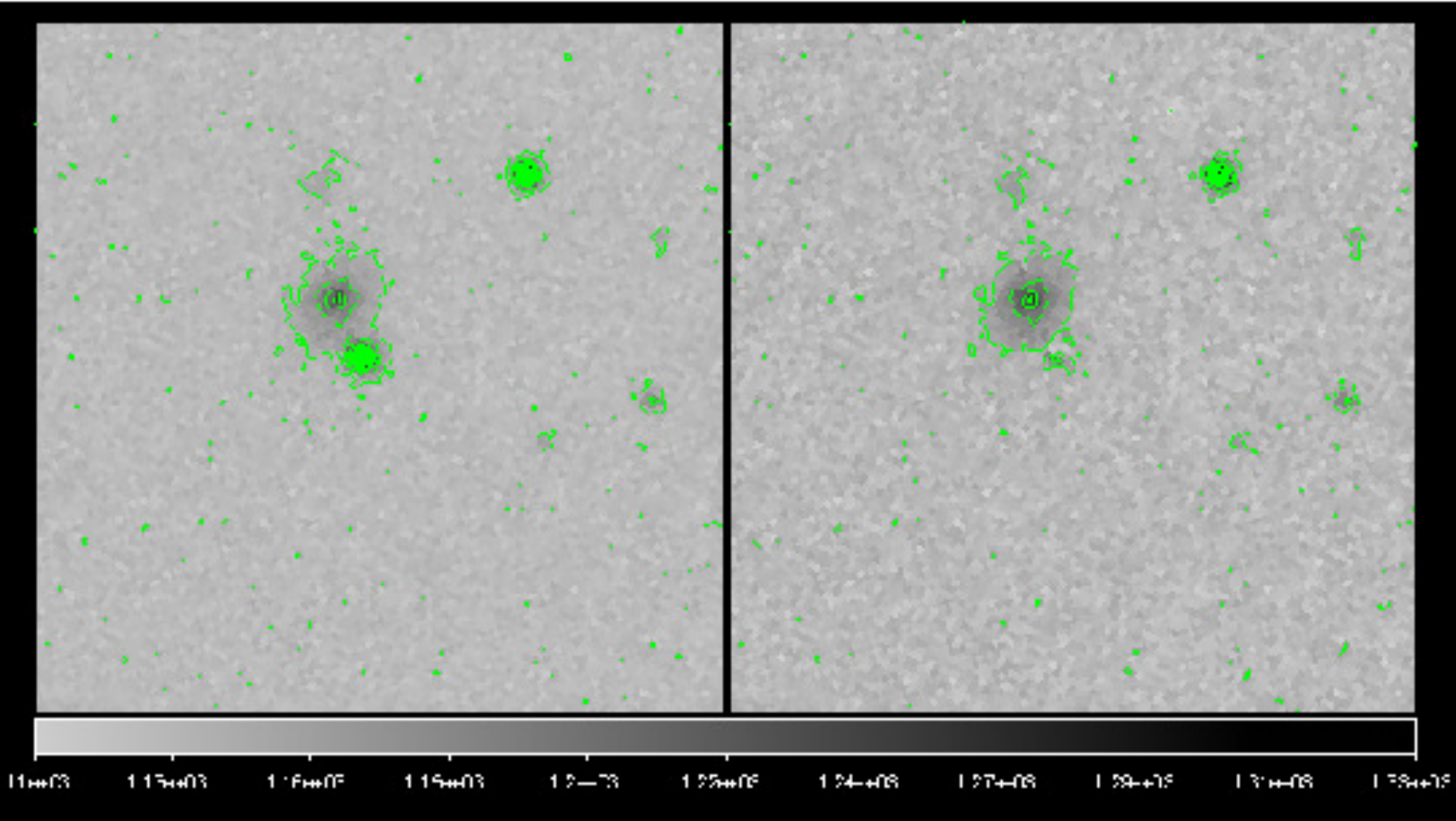}
}
\subfigure[9]{
\includegraphics[trim = 0mm 0mm 0mm 0mm, clip, width=0.3\textwidth]{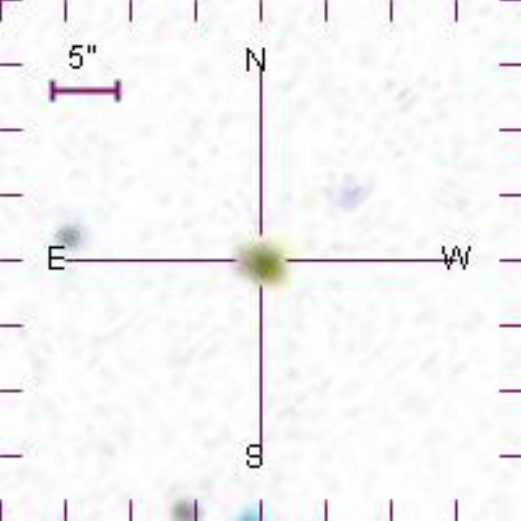}
}
\subfigure[$z_{gal}=0.3005$]{
\includegraphics[trim = 4.5mm 24mm 5mm 3mm, clip, width=0.6\textwidth]{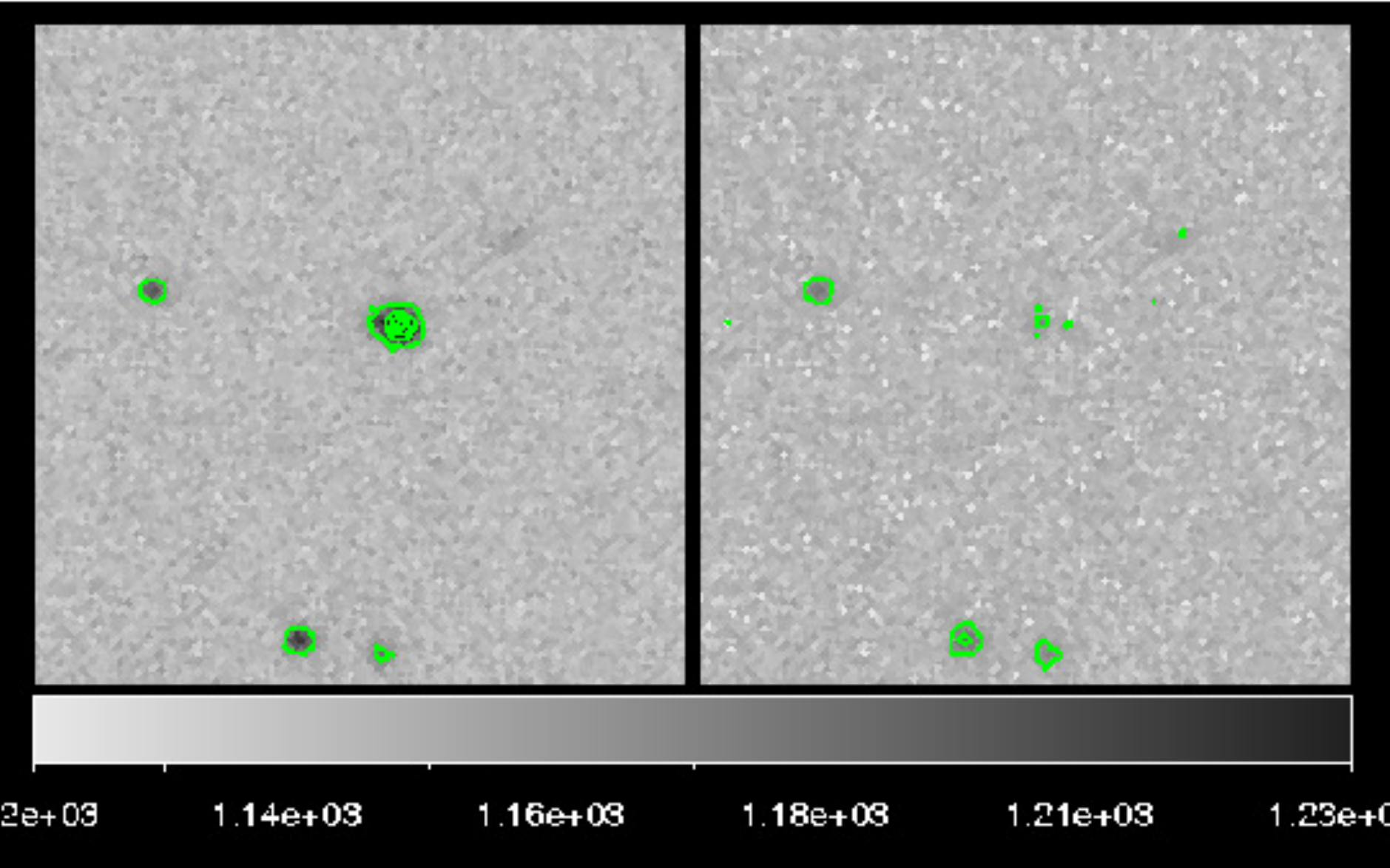}
}

\end{center}
\captcont{Continued. . }
\end{minipage}
\end{figure}

\begin{figure}
\begin{minipage}{1.0\linewidth}
\begin{center}

\subfigure[10]{
\includegraphics[trim = 0mm 0mm 0mm 0mm, clip, width=0.3\textwidth]{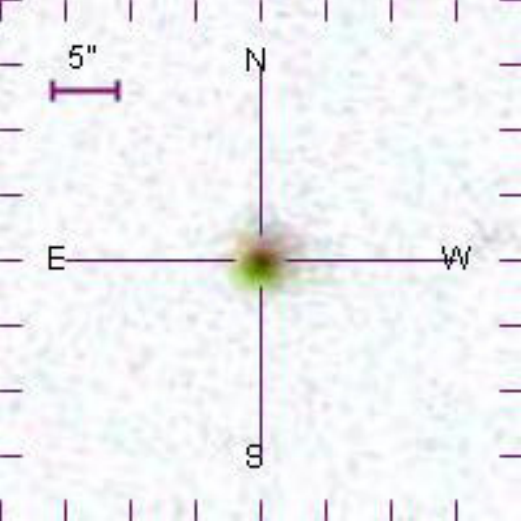}
}
\subfigure[$z_{gal}=0.1300$]{
\includegraphics[trim = 4.5mm 24mm 5mm 2.75mm, clip, width=0.6\textwidth]{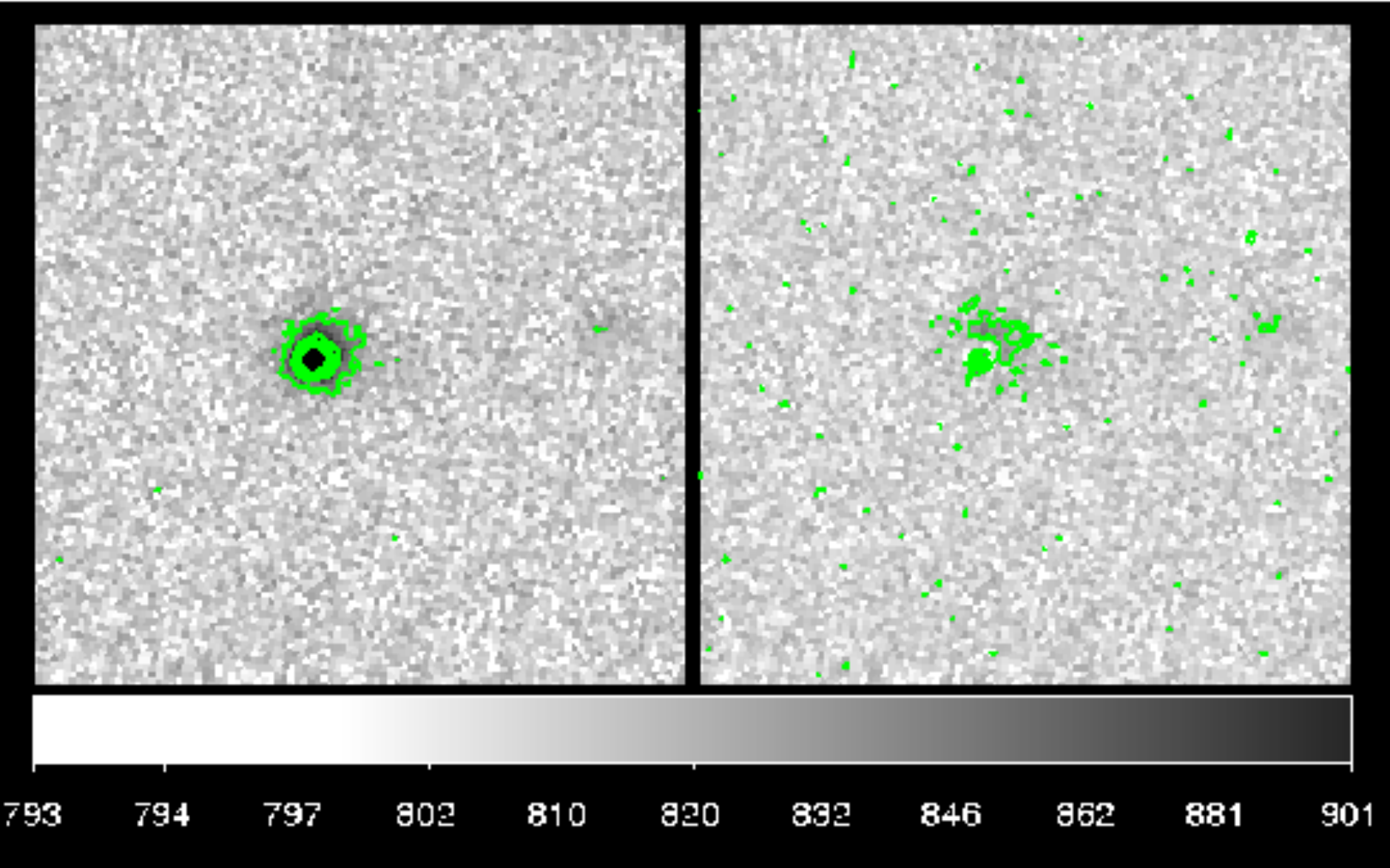}
}
\subfigure[11]{
\includegraphics[trim = 0mm 0mm 0mm 0mm, clip, width=0.3\textwidth]{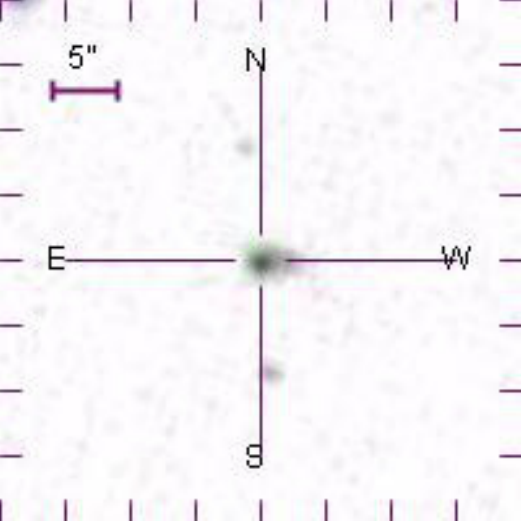}
}
\subfigure[$z_{gal}=0.2238$]{
\includegraphics[trim = 4.5mm 2.5mm 5mm 2.5mm, clip, width=0.6\textwidth]{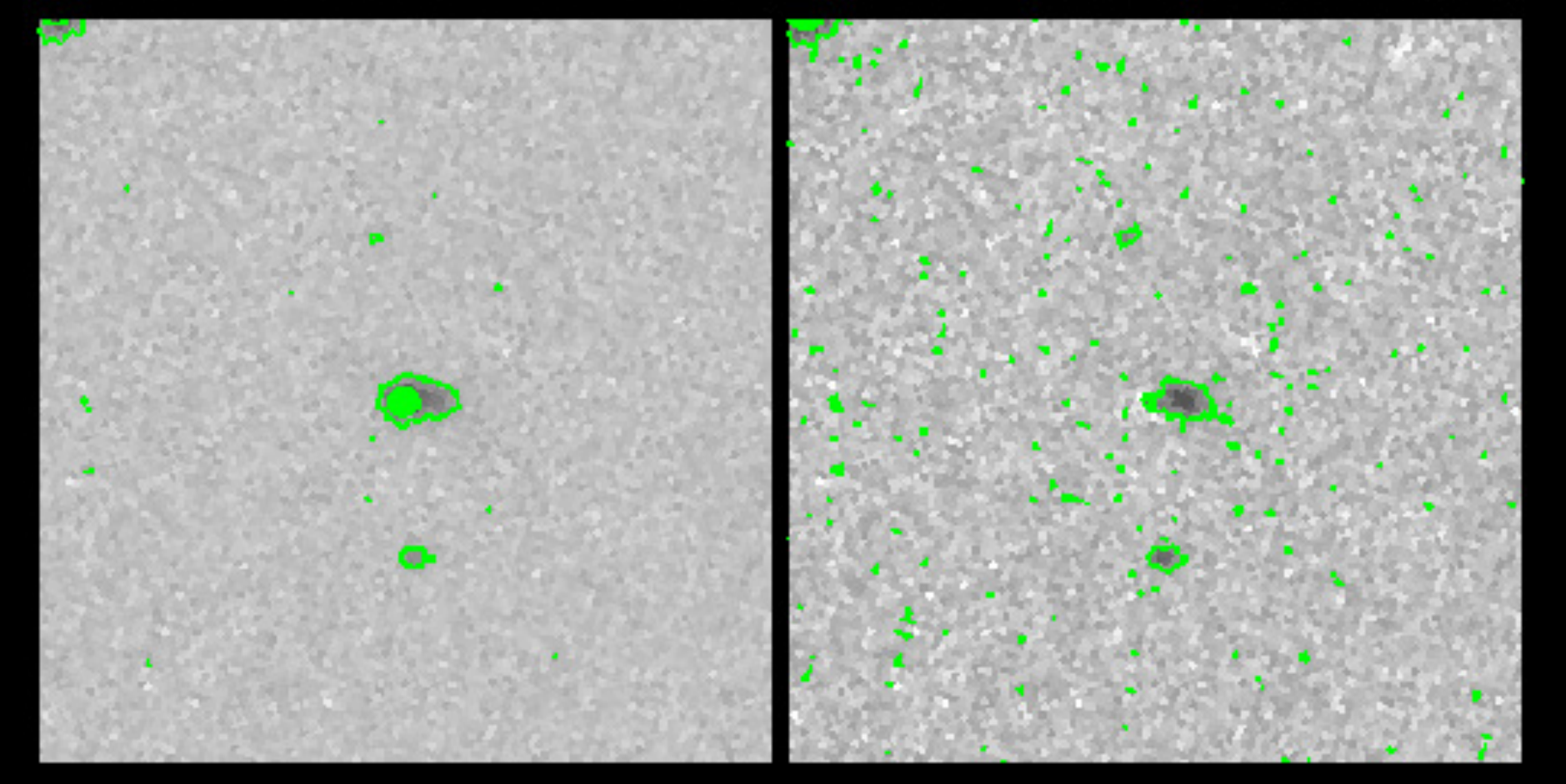}
}
\subfigure[12]{
\includegraphics[trim = 0mm 0mm 0mm 0mm, clip, width=0.3\textwidth]{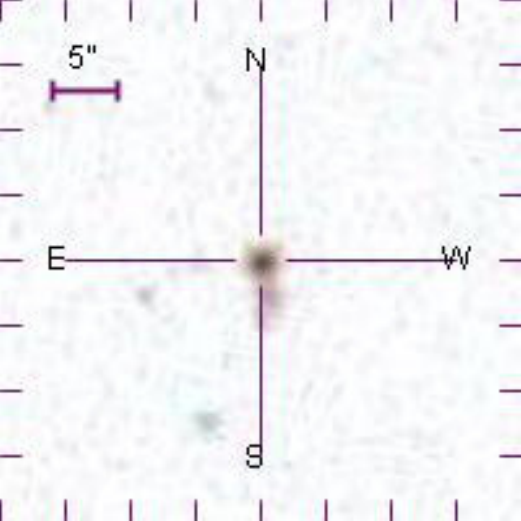}
}
\subfigure[$z_{gal}=0.2574$]{
\includegraphics[trim = 4.5mm 2.5mm 5mm 2.5mm, clip, width=0.6\textwidth]{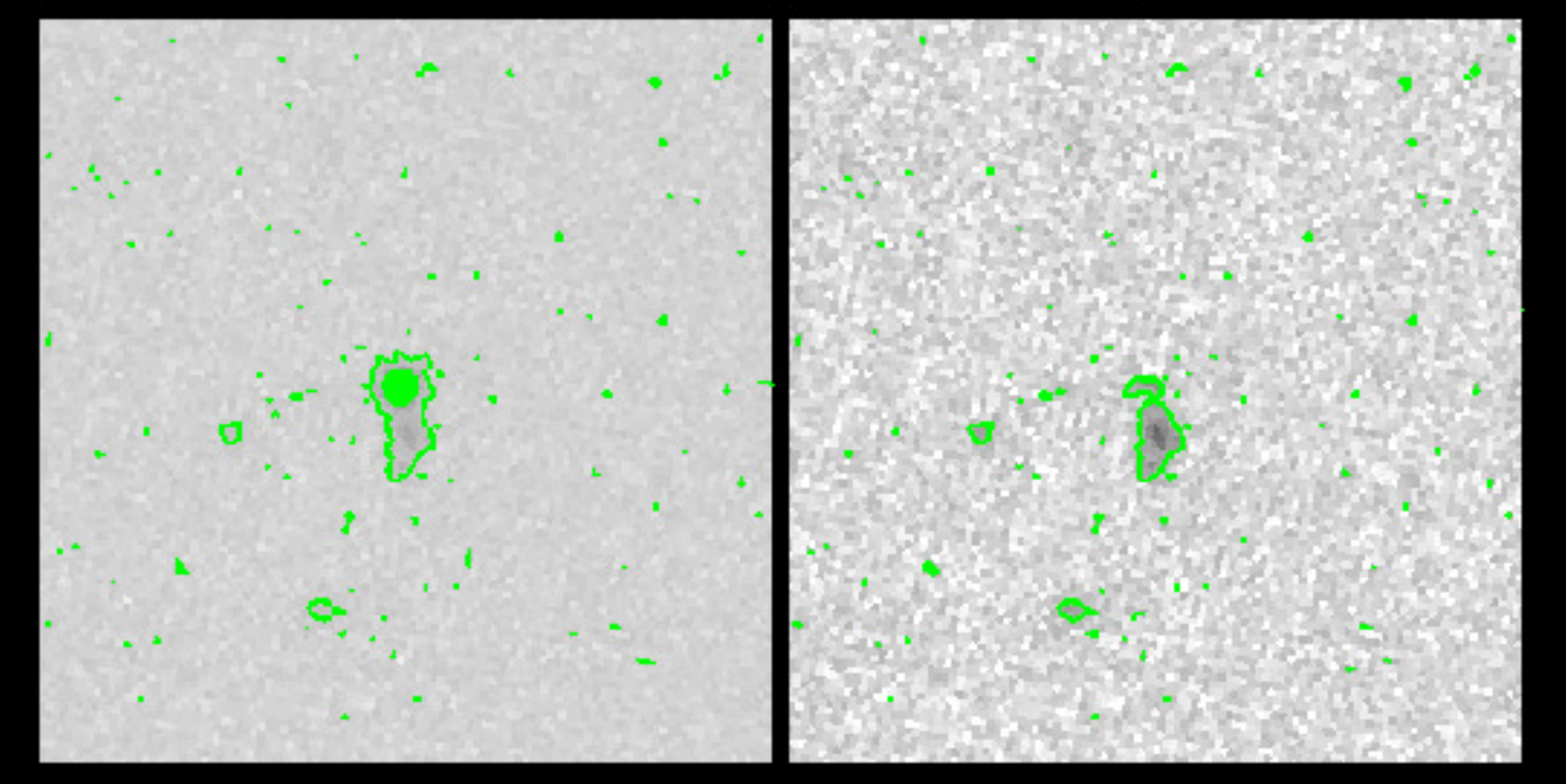}
}

\end{center}
\captcont{Continued.}
\end{minipage}
\end{figure}

\begin{figure}
\begin{minipage}{1.0\linewidth}
\begin{center}

\subfigure[14]{
\includegraphics[trim = 0mm 0mm 0mm 0mm, clip, width=0.3\textwidth]{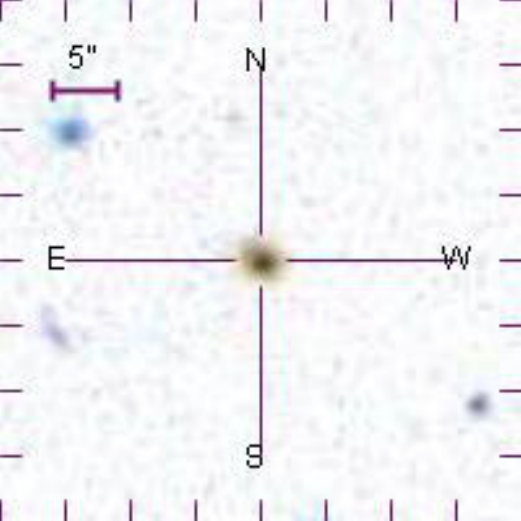}
}
\subfigure[$z_{gal}=0.2627$]{
\includegraphics[trim = 4.5mm 2.5mm 5mm 2.5mm, clip, width=0.6\textwidth]{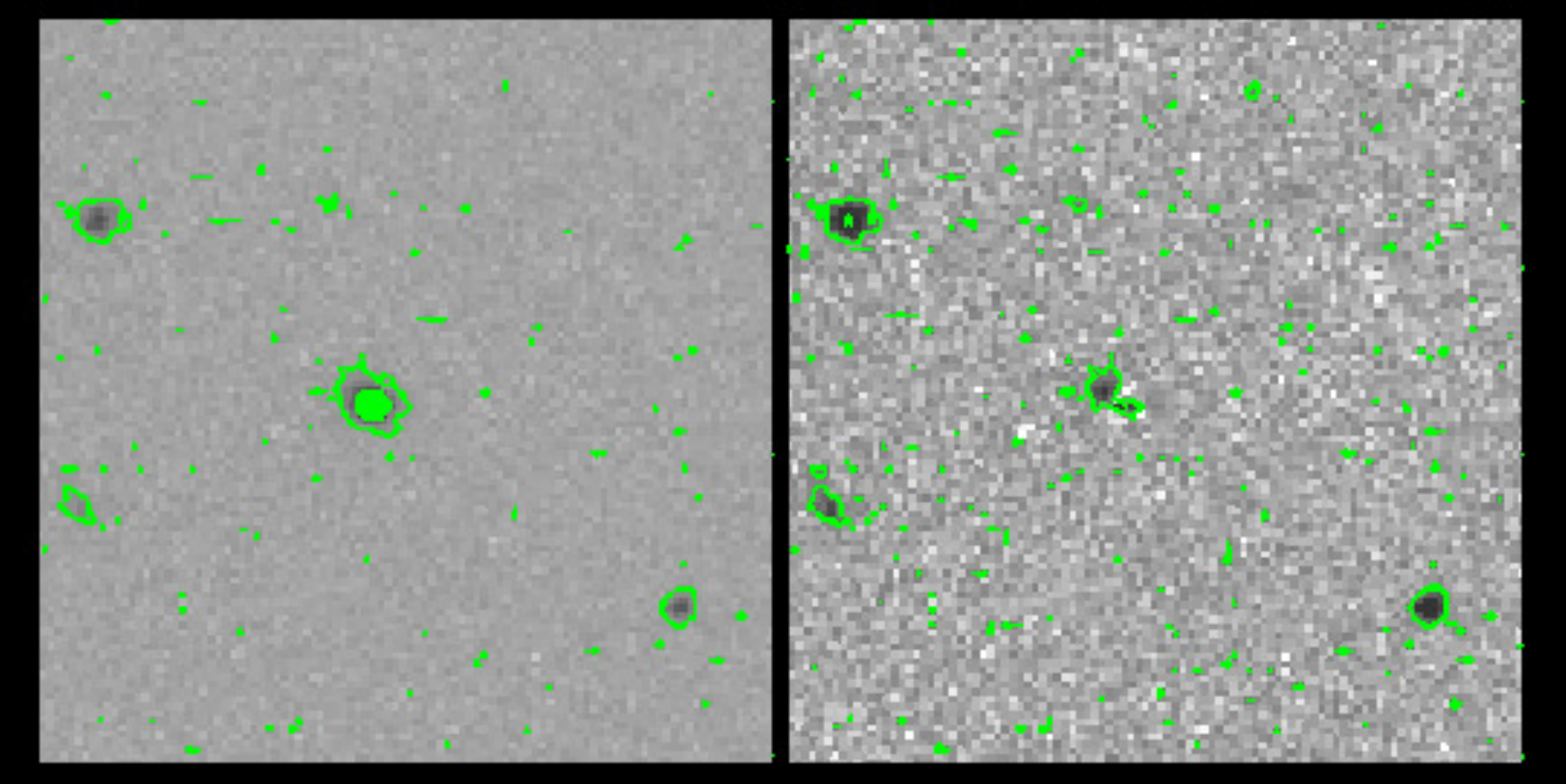}
}
\subfigure[15]{
\includegraphics[trim = 0mm 0mm 0mm 0mm, clip, width=0.3\textwidth]{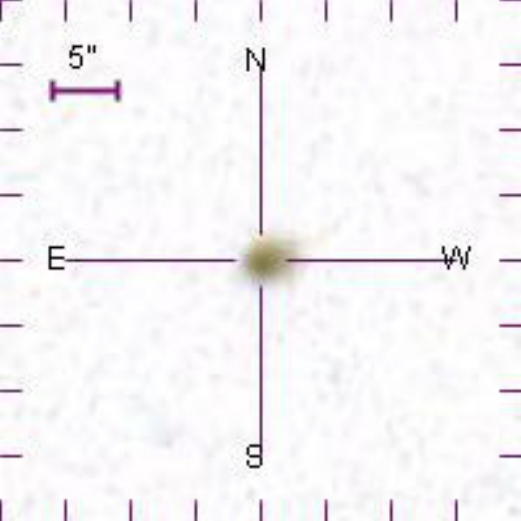}
}
\subfigure[$z_{gal}=0.0604$]{
\includegraphics[trim = 4.5mm 2.5mm 5mm 2.5mm, clip, width=0.6\textwidth]{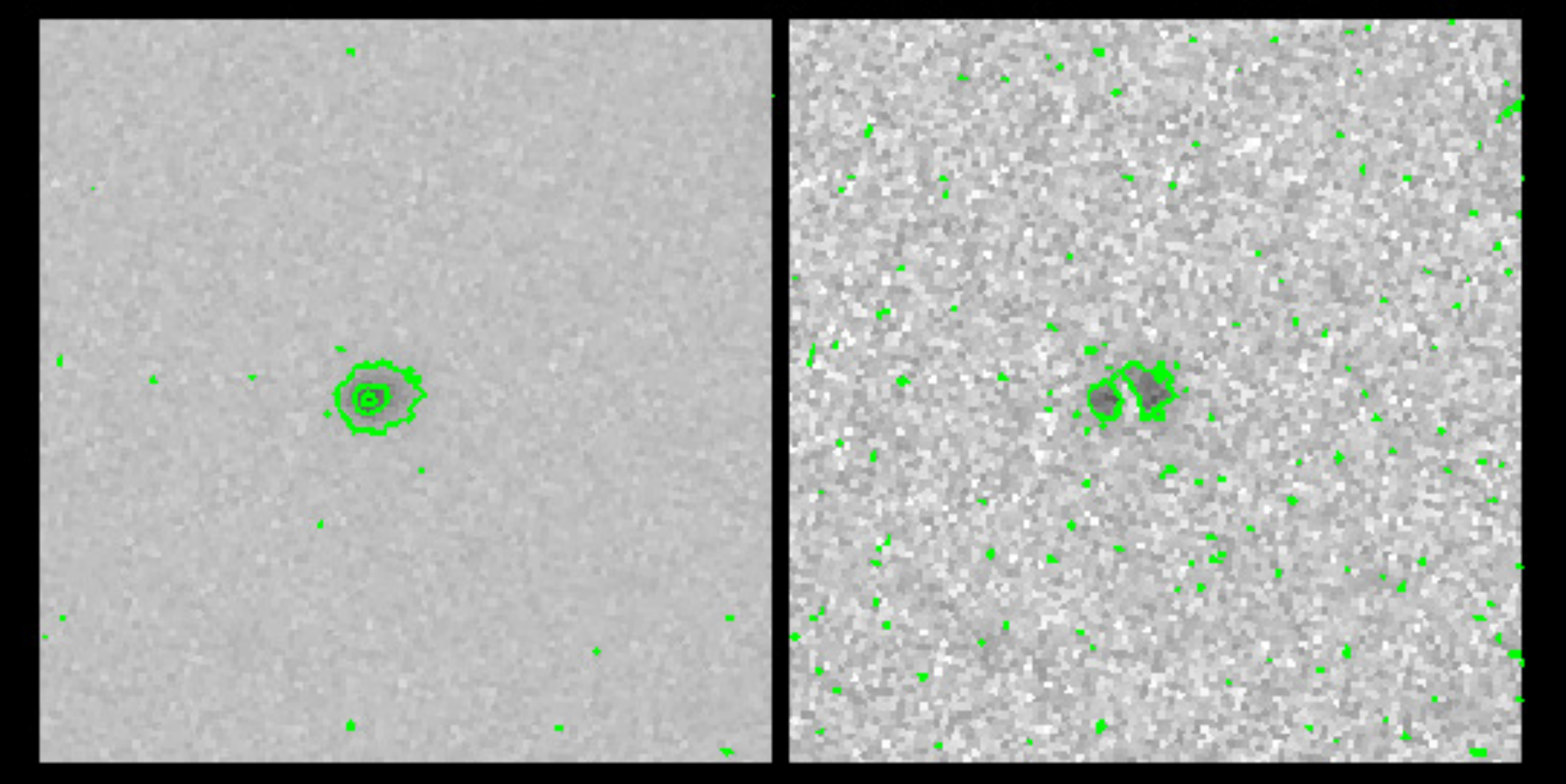}
}
\subfigure[16]{
\includegraphics[trim = 0mm 0mm 0mm 0mm, clip, width=0.3\textwidth]{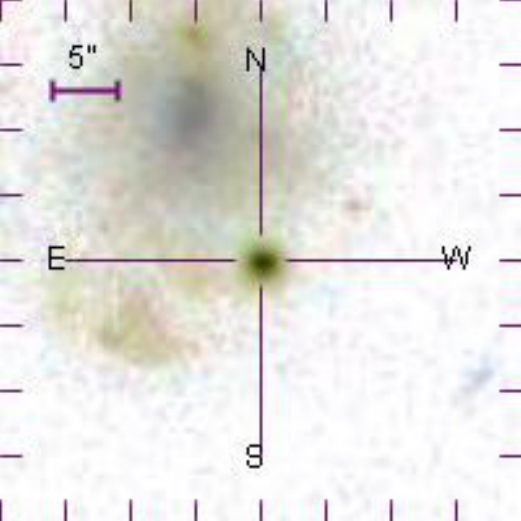}
}
\subfigure[$z_{gal}=0.0277$]{
\includegraphics[trim = 4.5mm 2.5mm 5mm 2.5mm, clip, width=0.6\textwidth]{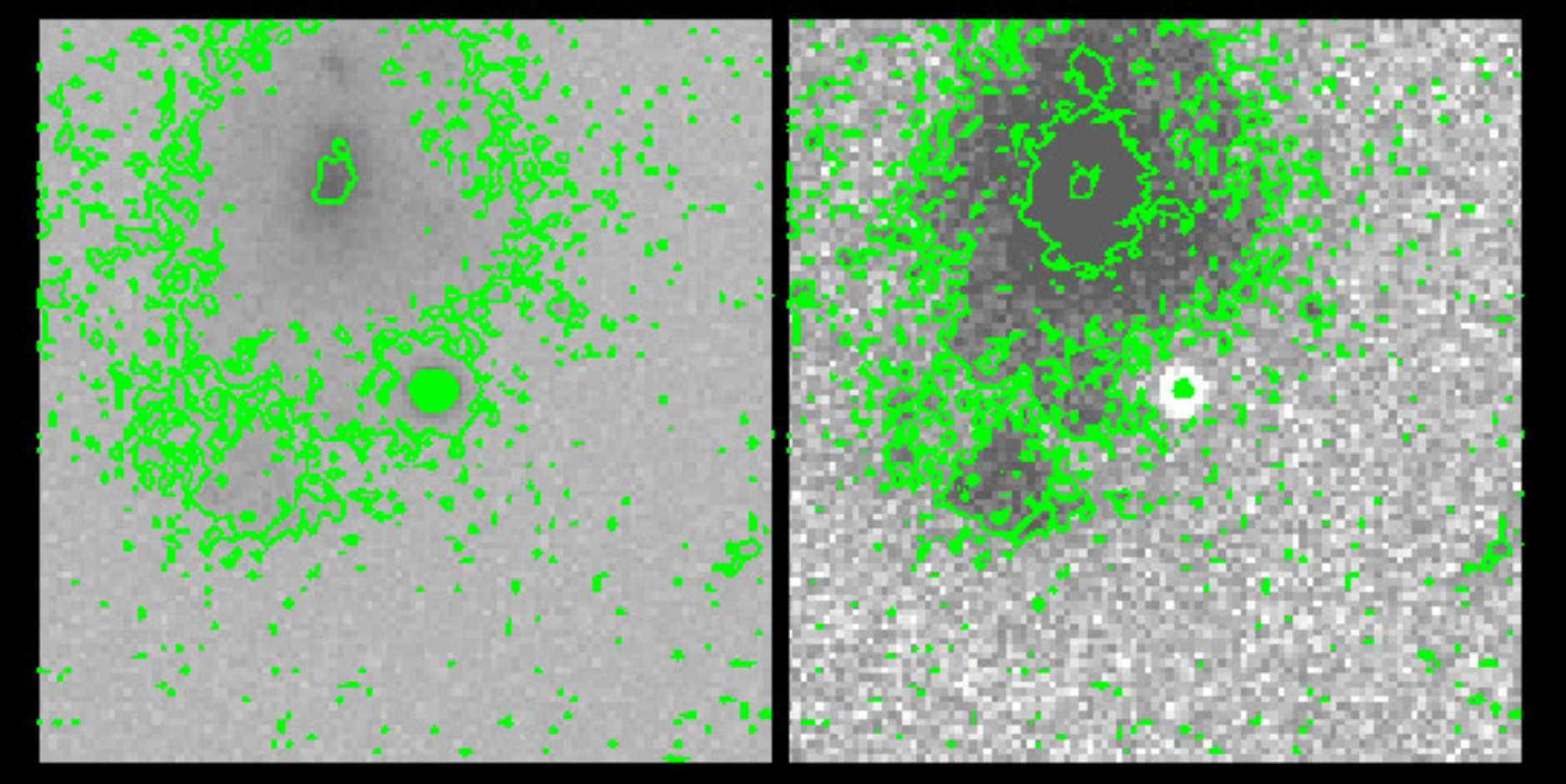}
}

\end{center}
\captcont{Continued.}
\end{minipage}
\end{figure}

\begin{figure}
\begin{minipage}{1.0\linewidth}
\begin{center}

\subfigure[18]{
\includegraphics[trim = 0mm 0mm 0mm 0mm, clip, width=0.3\textwidth]{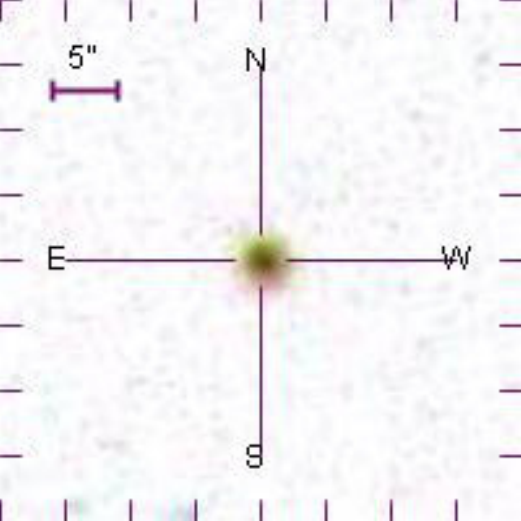}
}
\subfigure[$z_{gal}=0.1161$]{
\includegraphics[trim = 4.5mm 2.5mm 5mm 2.5mm, clip, width=0.6\textwidth]{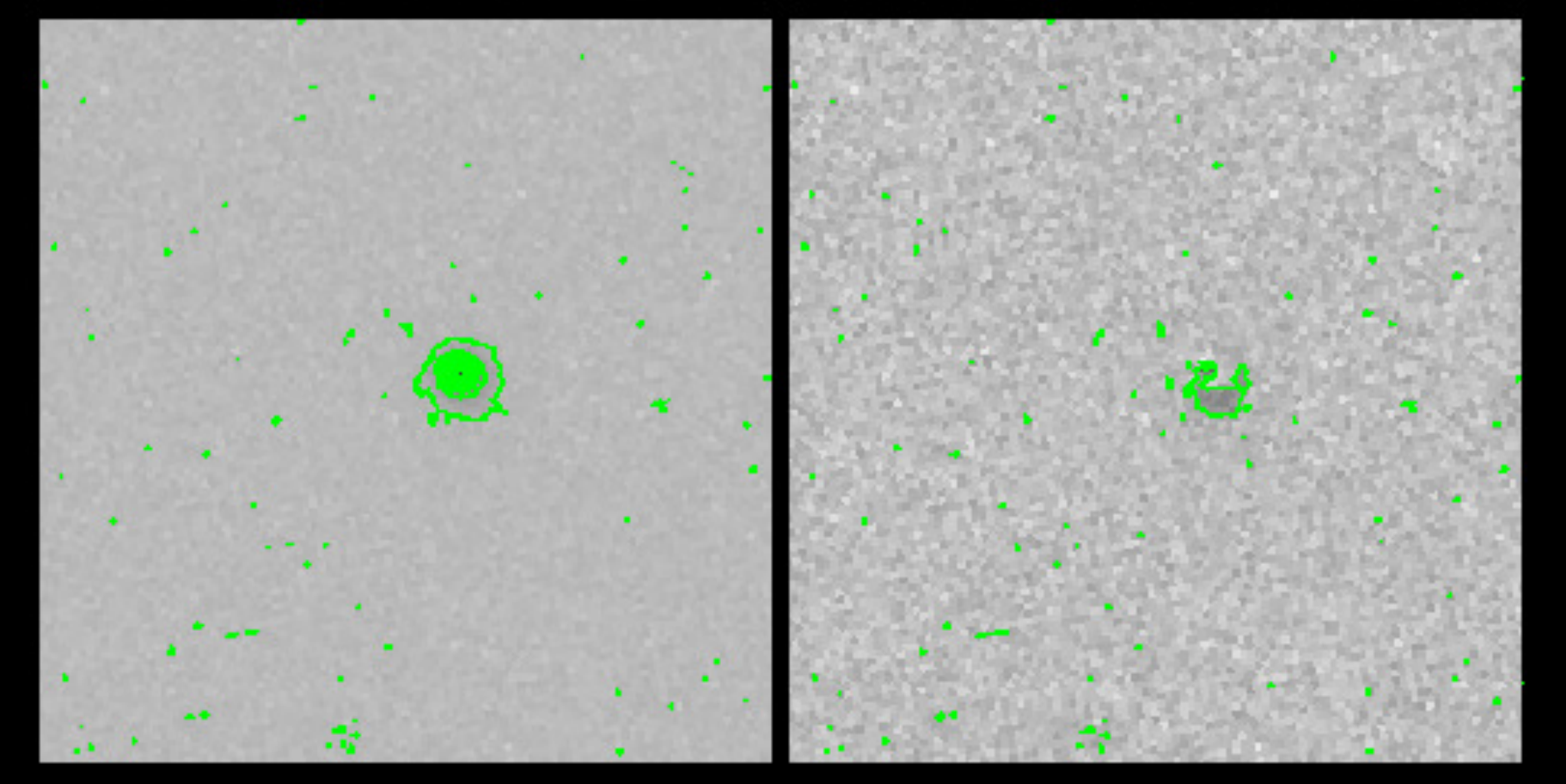}
}
\subfigure[19]{
\includegraphics[trim = 0mm 0mm 0mm 0mm, clip, width=0.3\textwidth]{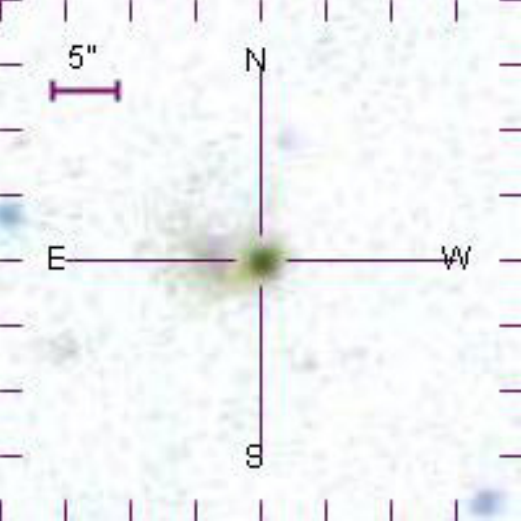}
}
\subfigure[$z_{gal}=0.0893$]{
\includegraphics[trim = 4.5mm 2.5mm 5mm 2.5mm, clip, width=0.6\textwidth]{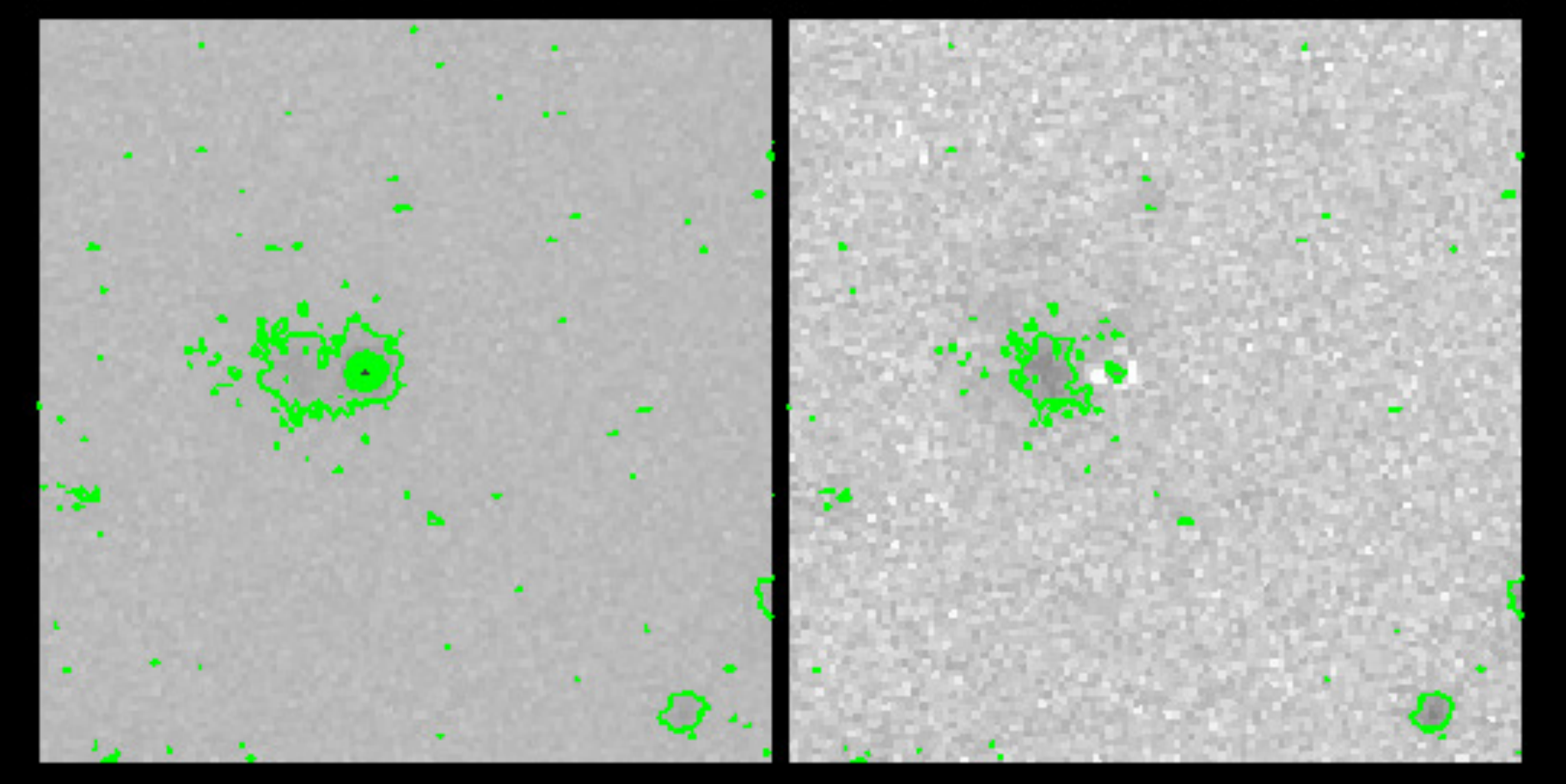}
}
\subfigure[20]{
\includegraphics[trim = 0mm 0mm 0mm 0mm, clip, width=0.3\textwidth]{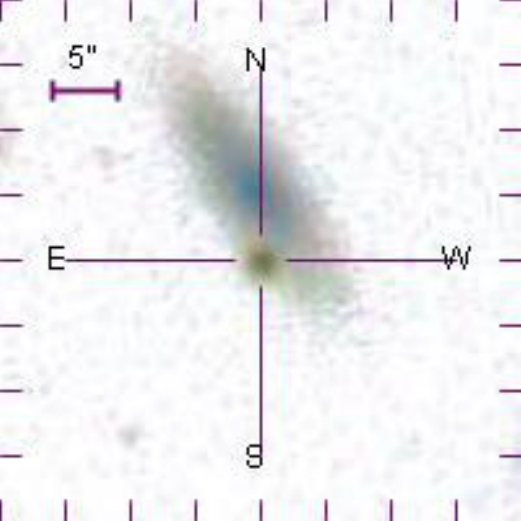}
}
\subfigure[$z_{gal}=0.0602$]{
\includegraphics[trim = 4.5mm 2.5mm 5mm 2.5mm, clip, width=0.6\textwidth]{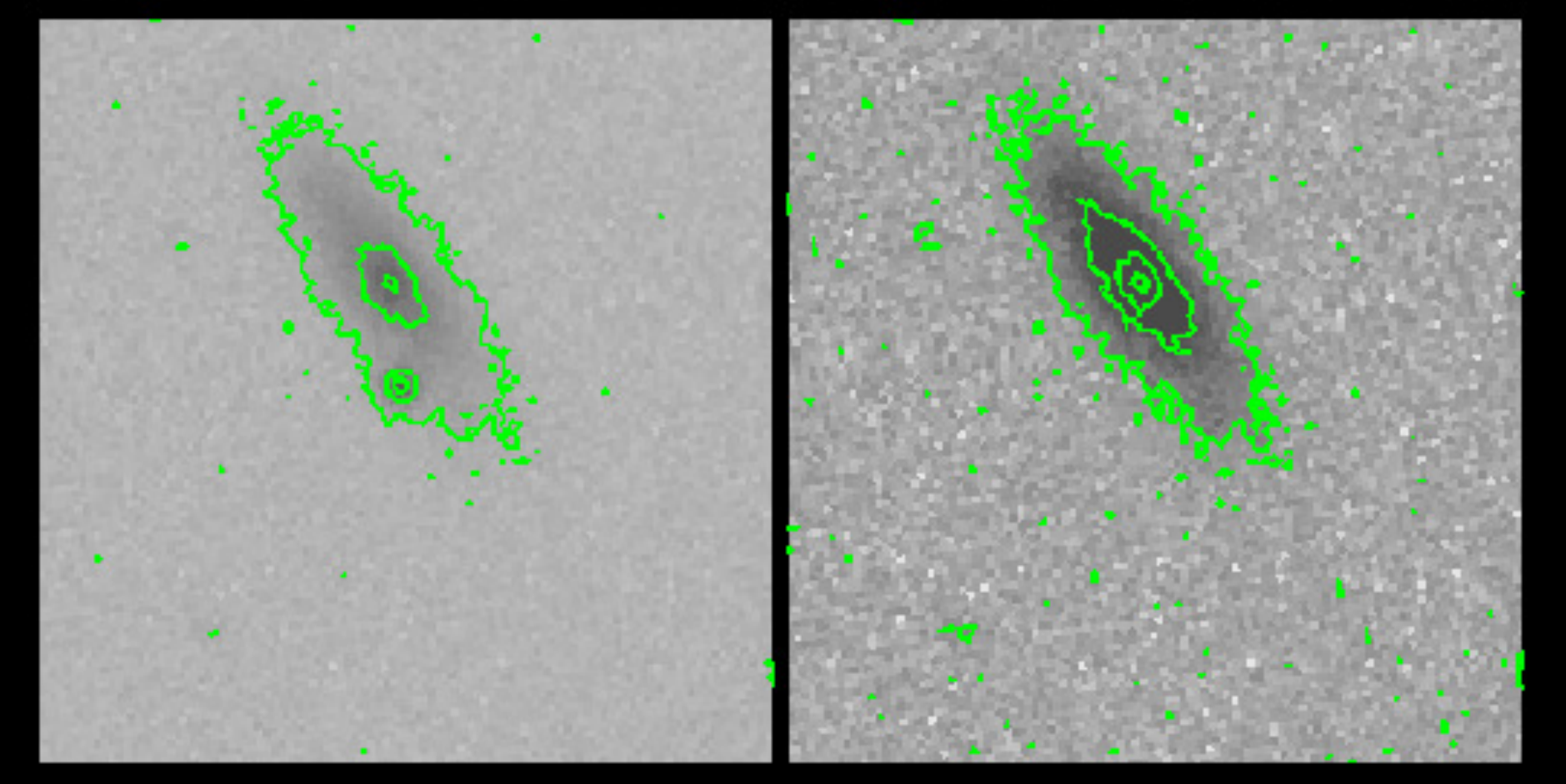}
}

\end{center}
\captcont{Continued.}
\end{minipage}
\end{figure}

\begin{figure}
\begin{minipage}{1.0\linewidth}
\begin{center}

\subfigure[22]{
\includegraphics[trim = 0mm 0mm 0mm 0mm, clip, width=0.3\textwidth]{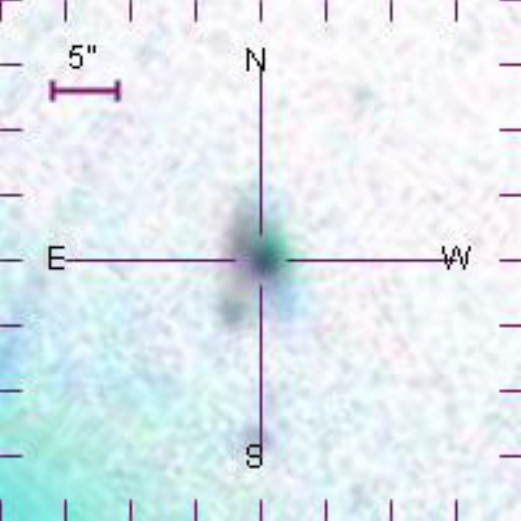}
}
\subfigure[$z_{gal}=0.0345$]{
\includegraphics[trim = 4.5mm 2.5mm 5mm 2.5mm, clip, width=0.6\textwidth]{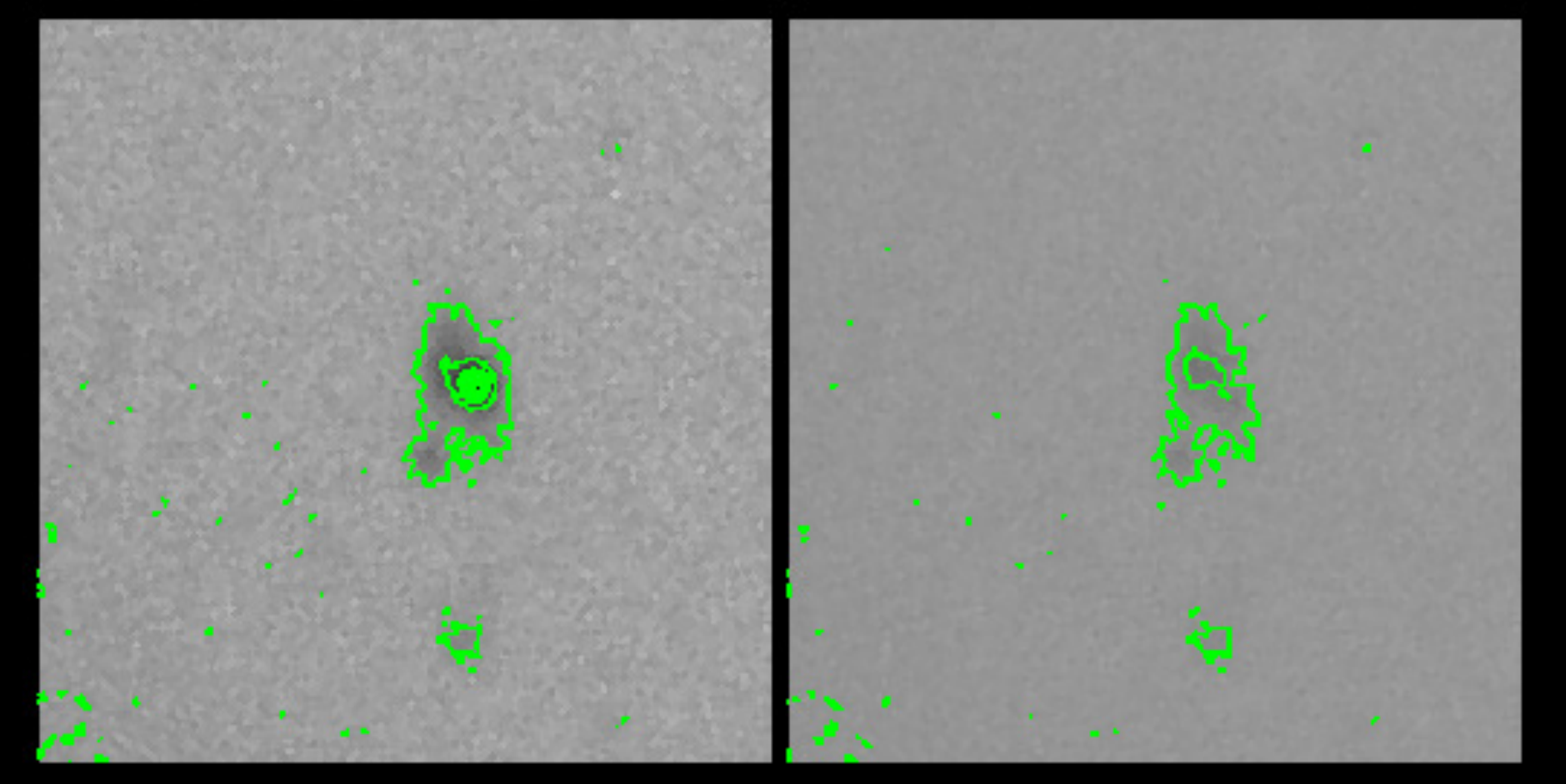}
}
\subfigure[23]{
\includegraphics[trim = 0mm 0mm 0mm 0mm, clip, width=0.3\textwidth]{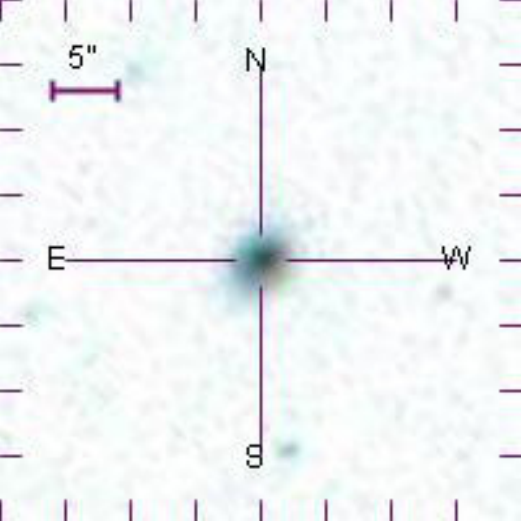}
}
\subfigure[$z_{gal}=0.2658$]{
\includegraphics[trim = 4.5mm 2.5mm 5mm 2.5mm, clip, width=0.6\textwidth]{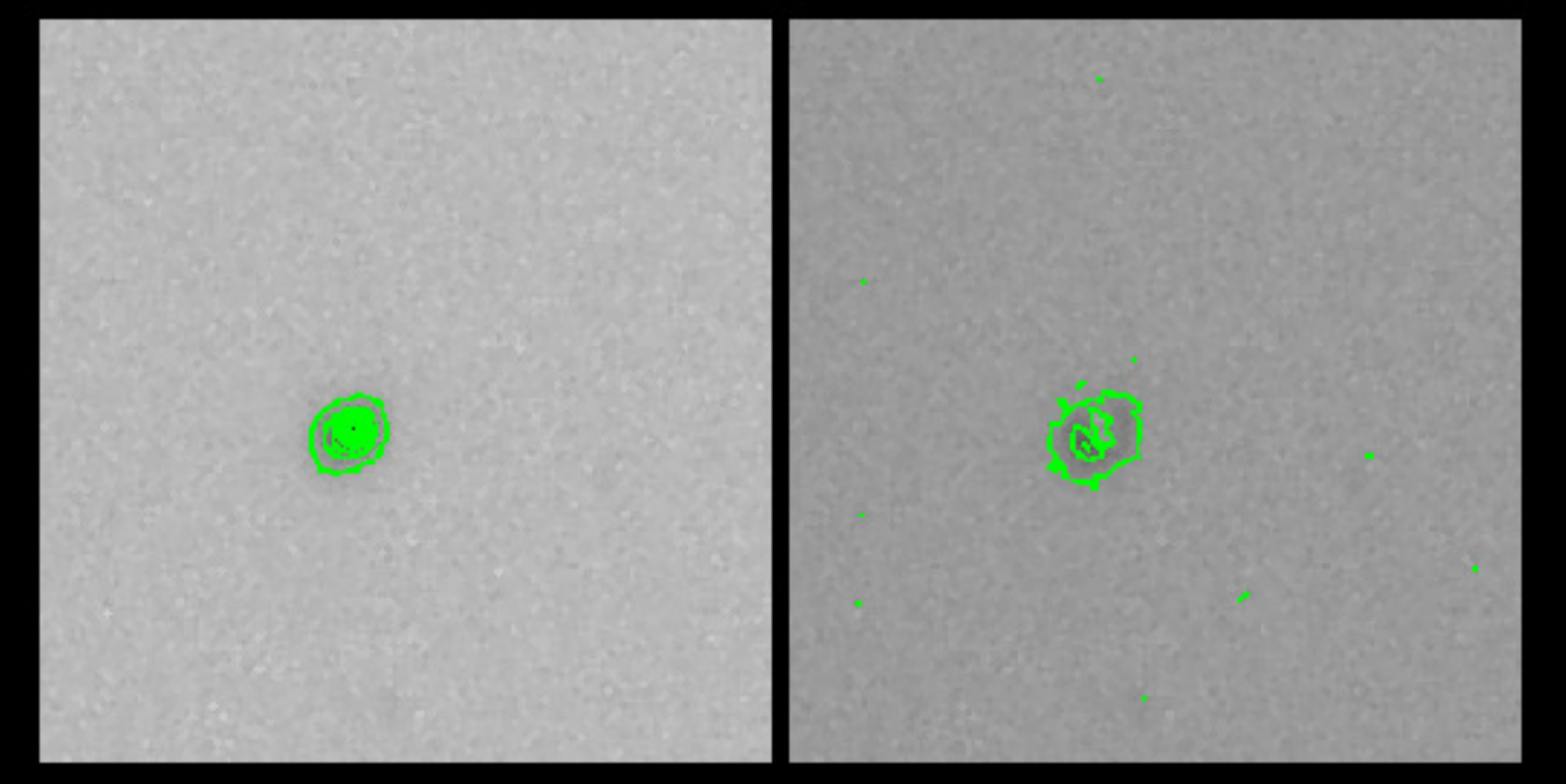}
}
\subfigure[24]{
\includegraphics[trim = 0mm 0mm 0mm 0mm, clip, width=0.3\textwidth]{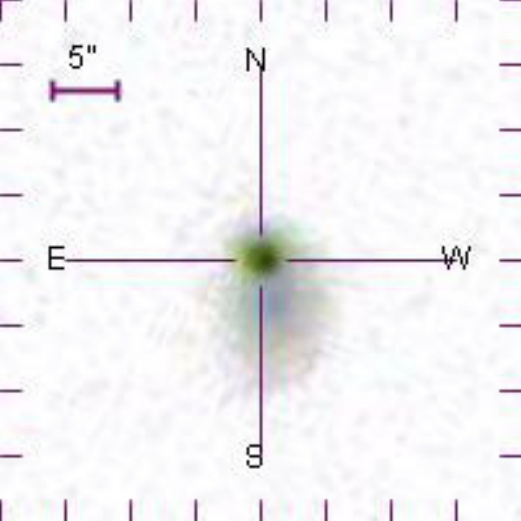}
}
\subfigure[$z_{gal}=0.1103$]{
\includegraphics[trim = 4.5mm 2.5mm 5mm 2.5mm, clip, width=0.6\textwidth]{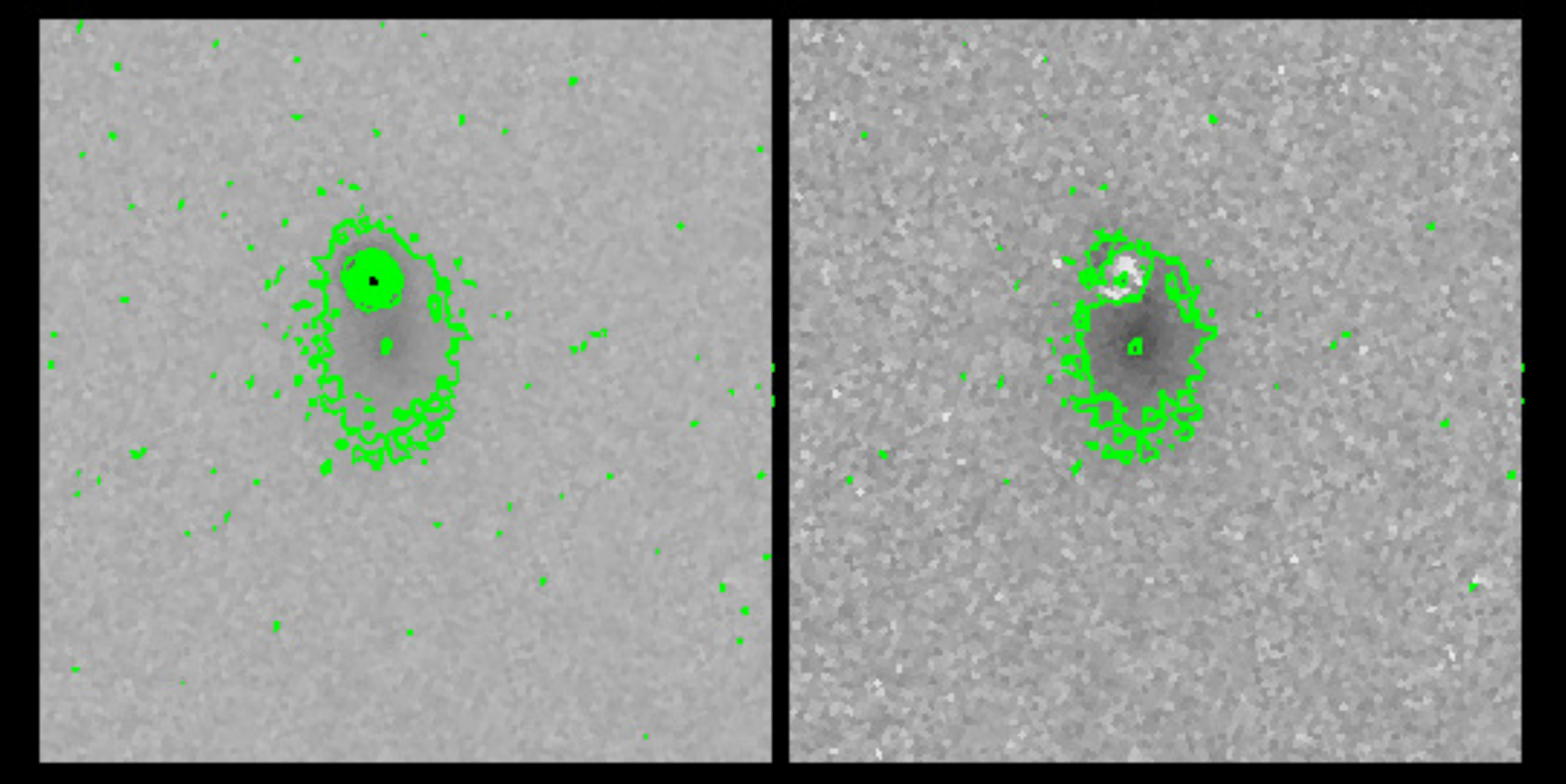}
}

\end{center}
\captcont{Continued.}
\end{minipage}
\end{figure}

\begin{figure}
\begin{minipage}{1.0\linewidth}
\begin{center}

\subfigure[25]{
\includegraphics[trim = 0mm 0mm 0mm 0mm, clip, width=0.3\textwidth]{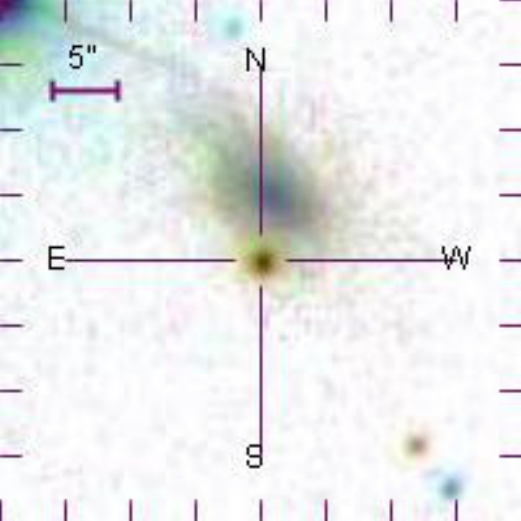}
}
\subfigure[$z_{gal}=0.0580$]{
\includegraphics[trim = 4.5mm 2.5mm 5mm 2.5mm, clip, width=0.6\textwidth]{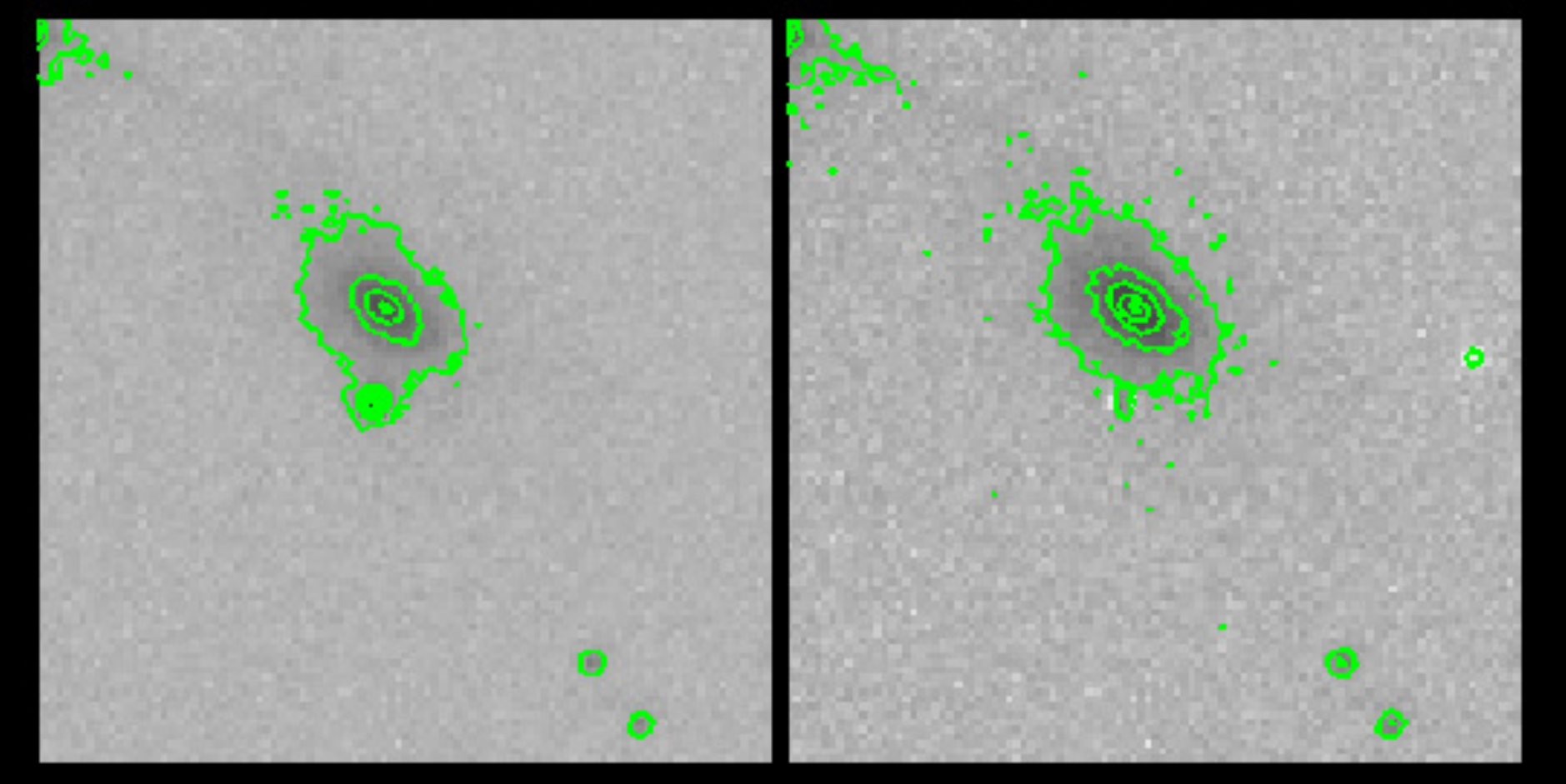}
}
\subfigure[26]{
\includegraphics[trim = 0mm 0mm 0mm 0mm, clip, width=0.3\textwidth]{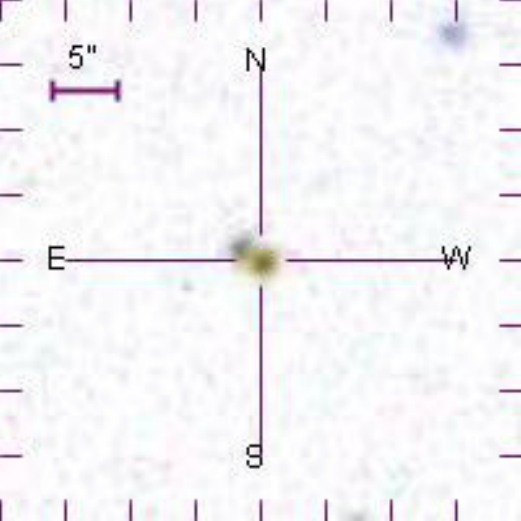}
}
\subfigure[$z_{gal}=0.1936$]{
\includegraphics[trim = 4.5mm 2.5mm 5mm 2.5mm, clip, width=0.6\textwidth]{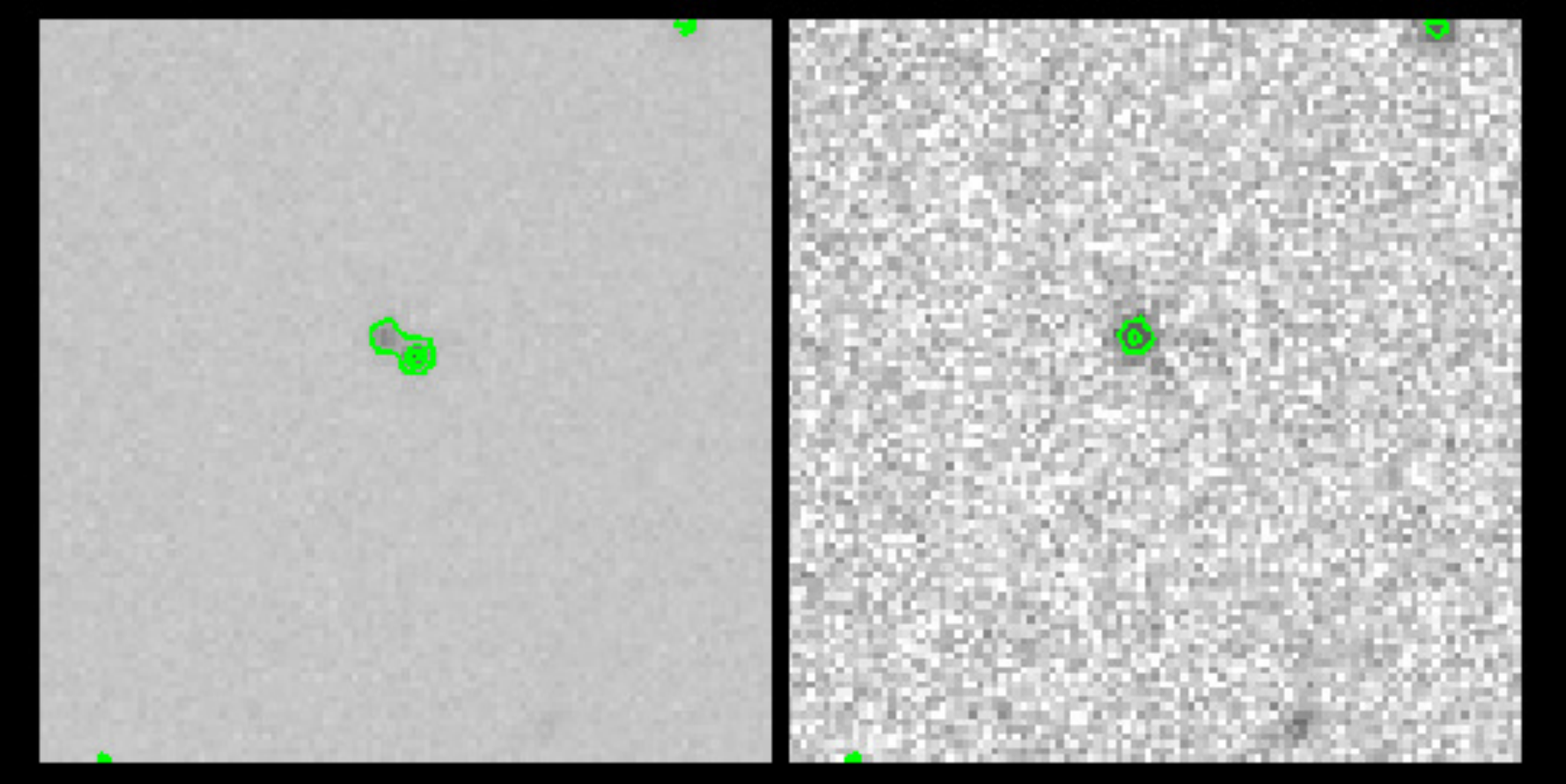}
}

\end{center}
\caption{Continued.}
\end{minipage}
\end{figure}

%Figure 3

\begin{figure}
\begin{minipage}{1.0\linewidth}
\begin{center}

\includegraphics[trim = 0mm 0mm 0mm 0mm, clip, width=0.7\textwidth]{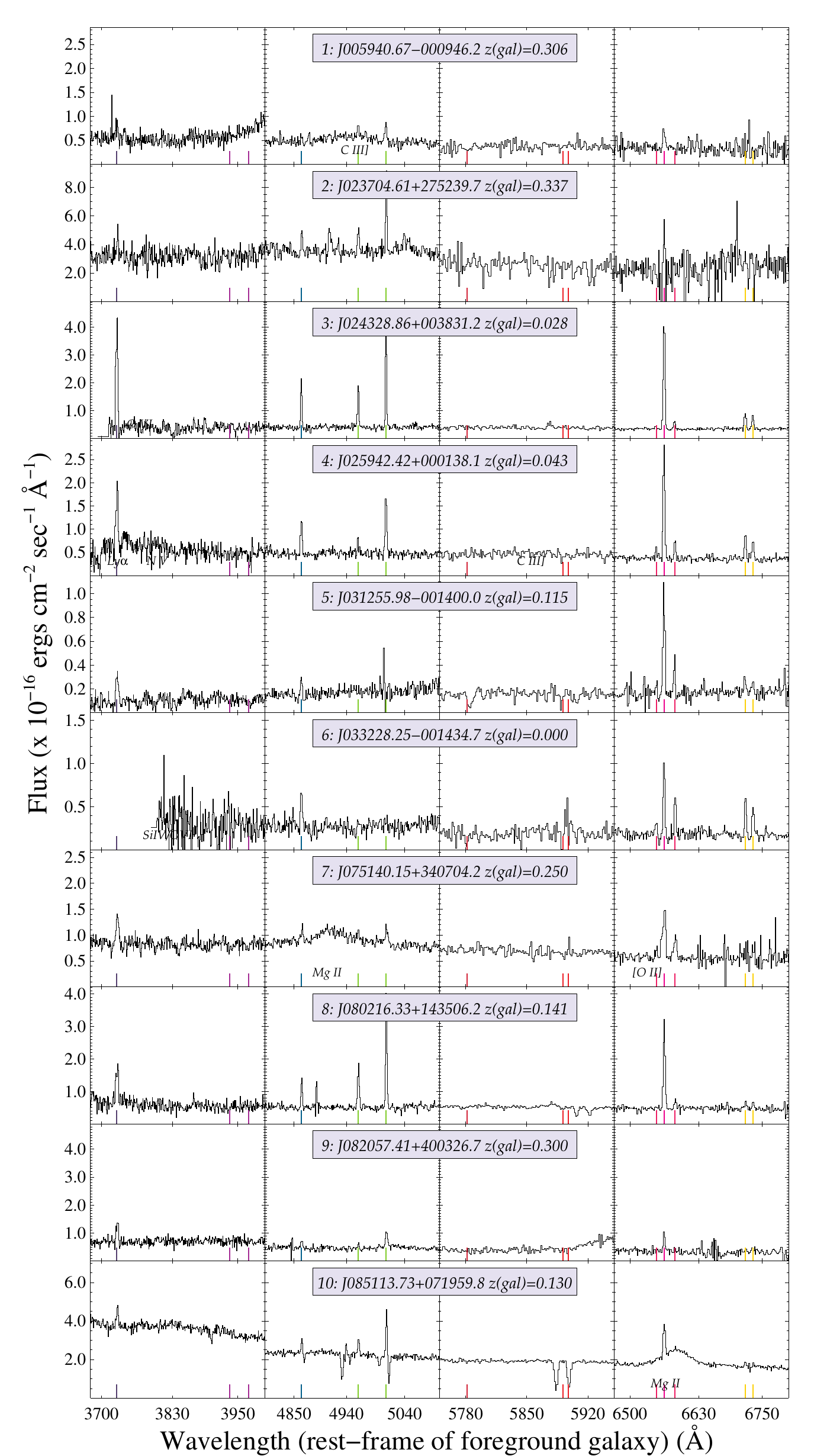}

\end{center}
\captcont{Spectra of targets 1-10 of our sample. The four panels highlight the relevant emission and absorption regions for our targets, with the relevant positions of the emission and absorption features marked by the lines at the bottom of each plot. The first shows the region of [O II] and Ca II, the second shows the region of H$\beta$ and [O III], the third shows the region of the 5780 \AA ~DIB and  Na II, and the fourth region shows H$\alpha$, [N II], and [S II]. Labels along the bottom of the spectra such as C III] and Mg II correspond to QSO emission line positions.}
\end{minipage}
\end{figure}

\begin{figure}
\begin{minipage}{1.0\linewidth}
\begin{center}

\includegraphics[trim = 0mm 0mm 0mm 0mm, clip, width=0.7\textwidth]{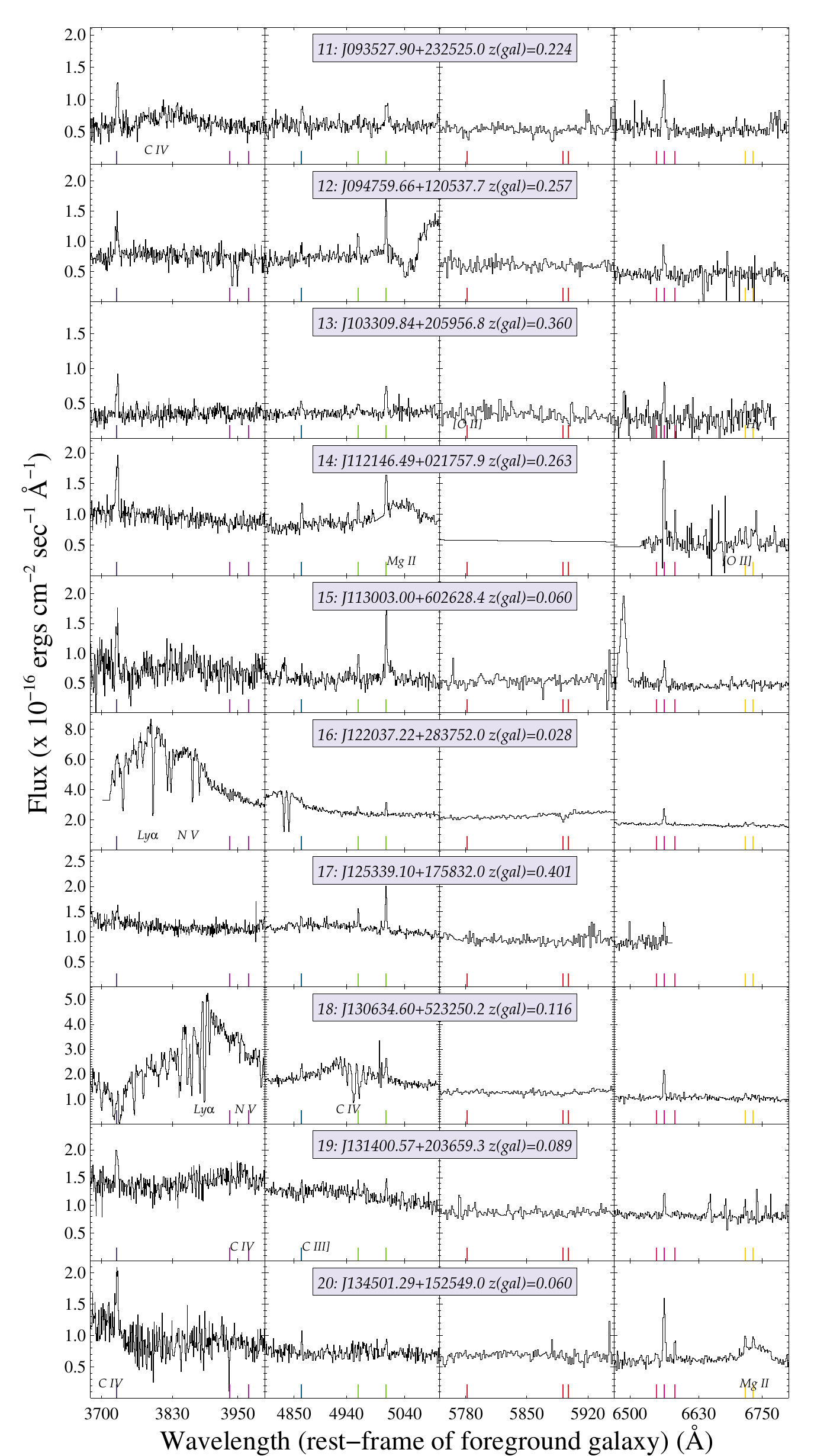}

\end{center}
\captcont{Spectra of targets 11-20 of our sample.}
\end{minipage}
\end{figure}

%Figure 4

%Figure 4

% Figure

%Figure 5

%Table 1

\begin{table}\footnotesize
\centering
\begin{minipage}{200mm}
\caption{QSO galaxy pairs in SDSS DR7 found from galactic emission superimposed on the QSO spectrum.\label{tbl-qgp}}
\begin{tabular}{lccccccccccc}

\hline\hline

Index & SDSS ID & Plate & MJD & Fiber & z$_{QSO}$ & z$_{gal}$ & Quality & $\Delta\theta^{e}$ & b  \\

\hline

1 & J005940.67-000946.2 & 1083 & 52520 & 105 & 2.39 & 0.3058 & B &--$^{f}$ &$< 4.69$ \\
2$^{d}$ & J023704.61+275239.7 & 2444 & 54082 & 194 & 0.97 & 0.3368 & B &-- &$<6.28$\\
3 & J024328.86+003831.2 & 807 & 52295 & 341 & 2.75 & 0.0279 & A & 3.18 & 1.79 \\
4 & J025942.42+000138.1 & 802 & 52289 & 244 & 2.20 & 0.0430 & A & 5.06 & 4.29\\
5 & J031255.98-001400.0 & 1179 & 52637 & 99 & 1.03 & 0.1147 & B & 0.60 & 1.26 \\
6 & J033228.25-001434.7 & 1063 & 52591 & 236 &1.73 & 0.0000 & B & -- & $< 0.09$ \\
7 & J075140.15+340704.2 & 756 & 52577 & 160 & 1.19 & 0.2499 & A & 1.08 & 4.23 \\
8$^{a,d}$ & J080216.33+143506.2 & 2266 & 53679 & 154 & 1.56 & 0.1412 & A & 3.60 & 8.94\\
9$^{a}$ & J082057.41+400326.7 & 760 & 52264 & 75 & 0.59 & 0.3005 & B & 1.49 & 6.65 \\
10$^{a}$ & J085113.73+071959.8 & 1299 & 52972 & 256 & 1.65 & 0.1300 & A & 2.43 & 5.63 \\
11$^{d}$ & J093527.90+232525.0 & 2293 & 53730 & 594 & 2.01 & 0.2238 & A & 1.75 & 6.29\\
12$^{a}$ & J094759.66+120537.7 & 1742 & 53053 & 624 & 1.29 & 0.2574 & A & 2.67 & 10.66  \\
13$^{a, d}$ & J103309.84+205956.8 & 2376 & 53770 & 427 & 1.11 & 0.3605 & B & -- & $<7.56$ \\
14$^{a}$ & J112146.49+021757.9 & 511 & 52636 & 175 & 1.27 & 0.2627 & B & 1.10 & 4.47 \\
15$^{a}$ & J113003.00+602628.4 & 952 & 52409 & 403 & 0.37 & 0.0604 & A & 1.69 & 1.99\\
16$^{b, d}$ & J122037.22+283752.0 & 2231 & 53816 & 567 & 2.20 & 0.0277 & A & 12.24 & 6.88 \\
17$^{a,c,d}$ & J125339.10+175832.0 & 2601 & 54144 & 625 & 0.50 & 0.4010 & B & -- & $<8.07$\\
18$^{b}$ & J130634.60+523250.2 & 887 & 52376 & 483 & 2.57 & 0.1161 & B & 1.44 & 4.55 \\
19$^{b,d}$ & J131400.57+203659.3 & 2618 & 54506 & 389 & 1.79 & 0.0893 & B & 3.72 & 6.18 \\
20$^{b,d}$ & J134501.29+152549.0 & 2742 & 54233 & 246 & 1.55 & 0.0602 &A & 5.75 & 6.67  \\
21$^{a}$ & J140103.31-005030.7 & 301 & 51942 & 52 & 0.93 & 0.3568 & A & -- & $<7.50 $\\
22 & J165643.35+254136.8 & 1693 & 53446 & 425 & 0.24 & 0.0345 & A & 1.67 & 1.13 \\
23$^{b}$ & J165743.05+221149.0 & 1415 & 52885 & 501 & 1.77 & 0.2658 & A & 1.21 & 4.96 \\
24$^{b}$ & J165958.93+620218.2 & 351 & 51780 & 476 & 0.23 & 0.1103 & A & 3.56 & 7.15 \\
25 & J211701.31-002638.8 & 1523 & 52937 & 65 & 1.14 & 0.0580 &A  & 5.06 & 5.68  \\
26 & J231148.54+004426.0 & 679 & 52177 & 383 & 2.20 & 0.1936 & A & 1.96 & 6.32\\
27$^{a}$ & J235621.28+002906.8 & 387 & 51791 & 343 & 1.05 & 0.3306 & B & -- & $< 7.14$ \\

\hline

\end{tabular}

\footnotetext[1]{System also found independently through different search criteria by Noterdaeme et al. (2010).}
\footnotetext[2]{System also found independently through different search criteria by Noterdaeme et al. (unpublished).}
\footnotetext[3]{FIRST radio source.}
\footnotetext[4]{Systems not present in SDSS DR5, and therefore could not be found with the search from York et al. (2012).}
\footnotetext[5]{Coordinate differences are measured relative to the QSO position.}
\footnotetext[6]{-- means that no offset could be determined.}

\end{minipage}
\end{table}

% TABLE 2

\begin{table}
\centering
\begin{minipage}{200mm}
\caption{Sample Criteria \label{tbl-sample}}
\begin{tabular}{lcccccc}

\hline\hline

Sample Name & Sample Source & SDSS & Size & Search Criteria \\

\hline

Sample I & York et al. (2012) & DR5 & 23 & H$\alpha$ FWHM $<5$ pixels, flux $> 5\times 10^{-17}$ ergs cm$^{-2}$ s$^{-1}$ \\
Sample II & This work & DR7 & 27 & three or more galactic emission lines with flux 4 times above background noise \\
ND10 & Noterdaeme et al. (2010)  & DR7 & 46 & [O III]$\lambda\lambda4959,5007$ better than 1$\sigma$ and 2$\sigma$ respectively \\

\hline

\end{tabular}

\end{minipage}
\end{table}

%Table 3

\begin{table}\footnotesize
\centering
\begin{minipage}{200mm}
\caption{Photometric data for QSOs\label{tbl-phot-QSOs}}
\begin{tabular}{lcccc|cccccc}

\hline\hline
\multicolumn{1}{c}{Index} & \multicolumn{2}{c}{SDSS measurements} & \multicolumn{4}{c}{Deconvolved apparent magnitudes$^{c}$} & \multicolumn{2}{c}{Reddening Estimates} \\
  & m$_{psf(r)}^{c}$ & Petrosian$^{d}$  & m$_{u}$ & m$_{g}$ & m$_{r}$ & m$_{i}$  & $\Delta$(g-i) & E(B-V)$_{(g-i)}$ \\

\hline

1  & 19.39 & 1.204  & 20.57 & 19.55 & 19.37 & 19.21  & -0.35 & 0.05\\
2  & 19.61 & 1.177 & $>23.5$ & 20.49 & 19.82 & 19.80 &  0.49 & 0.22 \\
3  & 20.17 & 6.840 &23.08 & 20.93 & 20.48 & 20.11 & 0.22 & 0.24 \\
4  & 20.18 & 7.517 & 20.52 & 20.26 & 20.06 & 19.82 & -0.21 & -0.03 \\
5$^d$  & 19.60 & 4.450 & 23.10 & 21.48 & 21.00 & 20.47 & 1.10 & 0.27\\
6  & 20.36 & 1.055 & 20.22 & 20.67 & 20.32 & 20.24 & -0.13 & 0.04 \\
7  & 19.14 & 2.718 & 20.00 & 19.60 &19.41 & 19.25 & -0.31 & -0.08 \\
8 & 19.60 & 1.218 & 20.71 & 20.16 & 19.55 & 19.10 & 0.35 & 0.26 \\ 
9  & 19.28 & 1.678 & 19.75 & 19.29 & 19.23 & 19.02 & -0.17 & 0.00 \\
10  & 17.97 & 1.528 & 18.34 & 18.09 & 18.00 & 17.63 &  -0.28 & -0.06  \\
11$^d$ & 19.33 & 1.729 & 20.39 & 20.11 & 19.46 & 18.97 & 0.54 & 0.30\\
12  & 19.15 & 1.167 & 20.05 & 19.64 & 19.14 & 19.03 & 0.03 & 0.09 \\
13  & 19.71 & 1.464 & 20.47 & 20.15 & 19.58 & 19.32 & 0.52 & 0.30 \\
14  & 18.79 & 1.286  & 19.63 & 19.17 & 18.81 & 18.74 & -0.12 & 0.01 \\
15  & 19.28 & 2.338 &  20.40 & 19.69 & 19.33 & 19.22 & -0.17 & -0.01 \\
16  & 17.91 & 1.133 & 18.78 & 18.19 & 17.87 & 17.74 & -0.11 & 0.04  \\
17 & 18.38 & 1.271 & 19.10 & 18.62 & 18.37 & 18.13 & 0.24 & 0.18 \\
18  & 18.31 & 1.699 & 19.23 & 18.50 & 18.32 & 18.21 & -0.39 & -0.14 \\
19  & 18.77 & 1.314 & 19.24 & 18.91 & 18.69 & 18.45 & -0.26 & -0.02 \\
20 & 19.35 & 1.502 & 19.77 & 19.64 & 19.28 & 18.97 & 0.04 & 0.13 \\
21  & 18.32 & 1.370 &  18.64 & 18.45 & 18.31 & 18.44 & -0.27 & -0.07 \\
22$^d$ & 18.16 & 2.791 & 20.71 & 19.74 & 19.01 & 18.65 &  0.46 & 0.42  \\
23  & 18.40 & 2.433  & 19.83 & 18.83 & 18.40 & 18.03 & 0.25 & 0.22  \\
24  & 17.80 & 1.543 & 17.89 & 17.87 & 17.78 & 17.37 & -0.26 & -0.03 \\
25  & 19.30 & 6.749 & 19.21 & 19.14 & 18.87 & 18.82 &  -0.37 & -0.11 \\
26 & 20.67 & 1.312 & 20.40 & 19.81 & 19.83 & 19.69 & -0.60 & -0.24  \\
27  & 18.68 & 1.342 & 19.56 & 19.12 & 18.70 & 18.67 & 0.10 & 0.12 \\

\hline

\end{tabular}

\footnotetext[1]{SDSS pipeline PSF fit.}
\footnotetext[2]{Petrosian radius measured in arcseconds, as determined by the SDSS pipeline.}
\footnotetext[3]{Deconvolved QSO apparent magnitudes as measured by the authors after PSF subtraction.}
\footnotetext[4]{Our three most reddened objects in this sample.}

\end{minipage}
\end{table}

%Table 4

\begin{table}\small
\centering
\begin{minipage}{200mm}
\caption{Photometric data for deconvolved galaxies\label{tbl-phot-galaxies}}
\begin{tabular}{ccccccccccccccc}

\hline\hline

Index & m$_{u}$ & L/L$_{\sun u}$ & m$_{g}$& L/L$_{\sun g}$ & m$_{r}$ & L/L$_{\sun r}$ & L$^{\ast}$/L$_{\sun r}$ & m$_{i}$ & L/L$_{\sun i}$ & (u-r) \\
 & & & & & & & & & &  \\

\hline

1 & $>23.66$ & $<9.00$ & $>24.48$ & $<8.18$ & $>24.00$ & $<8.20$ & $<0.01$ & $>23.57$ & $<8.31$ & --   \\
2 & $>23.50$ & $<9.65$ & $>24.54$ & $<8.74$ & $>24.03$ & $<8.77$ & $<0.04$ & $>23.60$ & $<8.88$  & -- \\
3 & 19.68 & 8.90 & 18.55 & 8.82 & 18.15 & 8.80  & 0.04 & 18.19 & 8.71   & 1.54  \\
4 & 20.56 & 8.97 & 18.53 & 9.23 & 17.92 & 9.28 & 0.12 & 17.86 & 9.24 & 2.64  \\
5 & 20.16 & 10.10 & 18.57 & 10.17 & 17.90 & 10.21  & 1.00 & 17.58 & 10.27  & 2.25  \\
6 & $>23.62$ & $<5.33$ & $> 24.48$ & $<4.49$ & $>24.11$ & $<4.46$ & $<0.01$  & $>23.68$ & $<4.57$ & --   \\
7 & $>23.84$ & $<9.22$ & 20.48 & 9.90 &19.37 & 9.88 & 1.70 & 19.13 & 9.32  & $>4.46$ \\
8 &  20.08 & 10.18 & 19.31 &  10.02 & 18.96 & 9.94 & 0.54 & 18.58  & 10.05  & 1.12 \\
9 &  20.93 & 9.80 & 21.90 & 9.86  & 21.20 & 9.85 & 0.44 & 21.11 & 9.77  & -0.26 \\
10 & 20.54 & 9.88 & 20.17 & 9.62 & 19.70 & 9.58 & 0.24 & 20.21 & 9.27  & 0.85 \\
11 & $>23.81$ & $<9.12$ & 20.90 & 9.90 & 20.31 & 9.88 & 0.47 & 20.97 & 9.32 & $>3.50$   \\
12 & 21.18 & 10.31  & 20.88 &  10.11 & 20.13 & 10.11 & 0.81 & 20.09 & 10.03  & 1.05   \\
13 & $>23.88$ & $<9.57$ & $>24.54$ & $<8.81$ &$>24.15$ & $<8.79$ & $<0.04$ & $>23.66$ & $<8.93$ & -- \\
14 &  19.88 & 9.86  & 20.61 & 10.06  & 20.59 & 9.74  & 0.34 & 20.74 & 9.52  & -0.71  \\
15 &  19.83 & 9.41  & 20.36 & 8.78  & 20.28 &  8.60 & 0.02  & 20.25 & 8.56  & -0.44  \\
16 & 19.47 & 8.98 & 18.36 & 8.90  & 17.92 & 8.89 & 0.05 & 17.69 & 8.93  & 1.55 \\
17 & $> 23.80$ & $<9.71$ & $>24.56$ & $<8.91$ & $>24.14$ & $<8.91$ & $<0.05$ & $23.52$ & $<9.09$  & --  \\
18 & 20.83 & 9.97 & 20.51 & 9.79 & 20.34 &  9.64 & 0.27 & 21.04 & 9.27  & 0.49 \\
19 & 19.96 & 9.74 & 19.92 &  9.34 & 19.56 & 9.28  & 0.12 & 19.28 & 9.33  & 0.41 \\
20 & 19.24 & 9.81 & 17.64 & 9.89 & 17.15 & 9.89 & 0.48 & 16.84 & 9.95  & 2.10 \\
21 &  $>23.85$ & $<9.57$& $> 24.44$ & $<8.84$& $>24.16$ & $<8.78$  & $<0.04$ & $>23.54$ & $<8.97$ & --  \\
22 & 19.69 & 9.06 & 19.43 & 8.67 & 18.65 & 8.78 & 0.04 & 18.23 & 8.89  & 1.05 \\
23 & 19.74 & 10.86  & 19.73 & 10.65 & 18.86 & 10.67 & 2.92 & 18.53 & 10.72 & 0.88  \\
24 & 20.61 & 9.86 & 18.96 & 9.93 & 18.53 & 9.89 & 0.48 & 18.33 & 9.91  & 2.07 \\
25 & 19.09 & 9.82 & 17.78 & 9.80 & 17.32 & 9.79  & 0.38 & 17.06 & 9.83  & 1.77 \\
26 & 21.05 & 10.04 & 20.47 & 9.87 & 20.09 & 9.78 & 0.38 & 20.02 & 9.73   & 0.49  \\
27 & $>23.61$ & $<9.59$ & $>24.39$ & $<8.79$ & $> 24.03$ & $<8.75$ & $<0.04$ & $>23.44$ & $<8.93$ & --   \\

\hline

\end{tabular}
\end{minipage}
\end{table}

%Table 6

\begin{table}\small
\centering
\begin{minipage}{200mm}
\caption[Emission line strengths for galaxies in front of QSOs.]{Emission line strengths for galaxies in front of QSOs.}\label{tbl-emission}
\begin{tabular}{lccccccccc}

\hline\hline

Index & $f_{H\alpha}^{a}$  & $f_{H\beta}$  & $f_{[O II]}$ & $f_{[O III]a}$  & $f_{[O III]b}$ & $f_{N IIa}$  & $f_{N IIb}$  & $f_{[S II]a}$  & $f_{[S II]b}$  \\

\hline

1 	& 35.15  	& -- $^{b}$ 	& 11.15 	& 8.51  	& 29.79 	& -- 		 & --  & --  & --\\
2  	& 38.46  	& 11.28 	& 7.91 	 	& 9.38 		& 18.35			& -- 		 & --& -- & -- \\
3  	& 170.37  	& 59.57  	 & 46.87 		& 41.34 	 	& 131.18	 			& -- 		 & 16.12  & 29.93   & 21.51  \\
4 	& 106.52 	 	& 25.62 	 & 79.18 		& 10.00	 	& 54.70 		& 6.69 	 & 20.24   & 24.79  & 17.47  \\
5  	& 201.56 		& 42.70 	& 120.47				& -- 					& --  			& 35.37   & 82.00   & 39.33  & 31.49  \\
6  	& 45.76 	 	& 15.78	 				& -- 				& -- 	 		& -- 			 & 13.50  	 & 18.26  & 21.06  & 11.78  \\
7   	& 78.50  	 	& 10.47	 & 51.72 				& -- 		 	& 16.74 	 	& 6.99  	 & 33.43  & --  & -- \\
8  	& 132.12 	 	& 44.93 	 & 69.50 		& 58.48  		& 147.35			& --  		 & 14.35  & 10.00  & 15.04  \\
9 	& 36.04		& 10.32 	&  68.08 		& 15.23 	& 28.61			& -- 		 & -- & --  & --\\
10  	& 87.10 	 	& 28.78 	 & 52.81  		& 25.20 	 	& 20.83 	 			& -- 		 & 5.45    & 9.48  & --  \\
11  	& 70.48 			& 13.96 	 & 33.94				& --   			& 19.13 				& --  		 & --& --  & --\\
12  	& 36.29		& 12.77 	 & 45.97		& 22.84 		& 43.52 				& --   		 & --  & --  & --\\
13  	& 37.58 		& 10.43	& 33.09		& 6.95 		& 25.53 			& --  		 & --  & --  & -- \\
14  	& 63.13 		& 20.74 	& 51.45 		& 13.31 	 	& 26.57 				& --  		 & 23.58  & --  & --\\
15 	& 25.31 		& 8.35  	 & 50.83 			& 13.05 		& 45.29	 			& --  		 & --  & --  & --\\
16  	& 17.07 	 			& -- 		 & 109.87 	 	& 9.03 	 	& 30.57				& -- 		 & --  & 15.72  & 15.45 \\
17 	& 26.55 		& 9.24	 & 14.66    	& 13.34 		& 38.16 			& -- 		 & -- & --  & --\\
18  	& 52.40 		& 17.02 			& --  				& -- 		 	& 36.87 				& --  		 & --   & 14.82 & -- \\
19  	& 16.11  			& --  		& 39.62 		& 6.62	 	& 14.70				& --  		& --  & 8.24  & -- \\
20  	& 47.22 		& 10.69 	 & 30.80	 			& -- 			& 6.66				& --  		 & 12.54  & 8.63  & 8.25 \\
21  	& 38.76 	 	& 4.85 	  & 13.83  	& 25.64 	 	& 49.01 	 			& --  		 & -- & -- & -- \\
22 	& 34.72 		& 6.92	 & 27.94   	& 15.85  	 	& 25.72	 			 & -- 		 & 10.32  & --  & --\\
23  	& 272.73 	 	& 53.55 	& 58.02 				& --  			& 23.82		 & 25.74  	 & 115.14  & 29.23  & 24.15  \\
24 	& 95.75 		& 31.85 	& 86.85				& --  			& 42.47			& --  		 & --  & 14.98 & --\\
25 	& 57.59 	 	& 19.13 	 & 74.21  				& --  		 	& 24.07		& 8.71 	& 19.38 & 15.78  & 8.86 \\
26  	& 92.04 	 	& 24.30 	 & 61.85  		& 25.96  	 	& 57.42	 & -- & --  & -- & --\\
27  	& 36.35 	  			& -- 			 & 50.14  	 	& 11.82 	 	& 39.76	 & --  & --  & --  & -- \\

\hline

\end{tabular}

\footnotetext[1]{All fluxes measured in 10$^{-17}$ ergs cm$^{-2}$ s$^{-1}$. }
\footnotetext[2]{Indicates no emission line detected or emission line is redshifted out of range of the spectrograph.}

\end{minipage}
\end{table}

% Table 7

\begin{table}
\centering
\begin{minipage}{200mm}
\caption{Star formation rates\label{tbl-SFR}}
\begin{tabular}{lccccccc}

\hline\hline

Index & H$\alpha$/H$\beta$  & E(B-V)$_{H\alpha/H\beta}$ & SFR$_{H\alpha}^{a}$ & SFR$_{[O II]}^{a}$ &SFR$_{H\alpha}^{b}$ &  SFR$_{[O II]}^{b}$ \\
& & & (M$_{\sun}$ yr$^{-1}$)& (M$_{\sun}$ yr$^{-1}$)& (M$_{\sun}$ yr$^{-1}$)& (M$_{\sun}$ yr$^{-1}$)\\

\hline

1 	& -- 	& --		& 0.83	& 0.47 	& --	& --  	\\
2	& 3.41	& 0.19		& 1.15	& 0.42	& 1.74	& 0.91 	\\
3	& 2.86	& -0.01		& 0.02	& 0.01 	& 0.02	& 0.01 	\\
4	& 4.16	& 0.40		& 0.04	& 0.05 	& 0.09	& 0.26 	 \\
5	& 4.72	& 0.54		& 0.54	& 0.58 	& 1.84	& 5.48   \\
6	& 2.90	& 0.01		& 0.00	& -- 	& 0.00	& -- 		\\
7	& 7.50	& 1.05		& 1.18	& 1.39 	& 12.46	& 107.39 \\
8	& 2.94	& 0.02		& 0.55	& 0.52 	& 0.58	& 0.57 \\
9	& 3.49	& 0.21		& 0.83 	& 2.79 	& 1.33	& 6.70  \\
10	& 3.03	& 0.05		& 0.31	& 0.33 	& 0.35	& 0.41 \\
11	& 5.05	& 0.62		& 0.83 	& 0.71 	& 3.29	& 9.09 	\\
12	& 2.84	& -0.01		& 0.58 	& 1.31 	& 0.56	& 1.23 \\
13	& 3.60	& 0.25		& 1.32	& 2.07 	& 2.29 	& 5.73 	\\
14	& 3.04	& 0.06		& 1.06	& 1.55 	& 1.22	& 2.00 \\
15	& 3.03	& 0.06		& 0.02 	& 0.06 	& 0.02	& 0.08  \\
16	& --	& --		& 0.01 	& 0.03 	& --	& -- 	\\
17	& 2.87	& 0.00		& 1.19 	& 1.17 	& 1.18	& 1.16 	\\
18	& 3.08 	& 0.07		& 0.14 	& --	& 0.17 	& -- 	 \\
19	& --	& --		& 0.03 	& 0.11 	& --	& -- 	 \\
20	& 4.42 	& 0.47		& 0.03	& 0.04 	& 0.09	& 0.26 	\\
21	& 7.99 	& 1.12		& 1.32 	& 0.84 	& 16.31	& 86.78 \\
22	& 5.02	& 0.61		& 0.01	& 0.01	& 0.03 	& 0.13  \\
23	& 5.09 	& 0.63		& 4.72	& 1.79 	&19.19	& 23.89 \\
24	& 3.01	& 0.05		& 0.24	& 0.38 	& 0.26	& 0.46 \\
25	& 3.01 	& 0.05		& 0.04	& 0.08 	& 0.04 	& 0.10 	\\
26	& 3.79 	& 0.30		& 0.78 	& 0.94	& 1.53	& 3.26 \\
27	& -- 	& --		& 1.04 	& 1.20 	& -- 	& -- 	 \\

\hline

\end{tabular}
\footnotetext[1]{SFR uncorrected for extinction.}
\footnotetext[2]{SFR corrected for extinction using H$\alpha$/H$\beta$.}

\end{minipage}
\end{table}

 \begin{table}
 \centering
 \begin{minipage}{200mm}
 \caption{Small number statistics for Samples I and II \label{tbl-small-II}}
 \begin{tabular}{lcccccccc}
 
 \hline\hline
 
 \multicolumn{1}{c}{} & \multicolumn{2}{c}{Sample I$^{a}$} & \multicolumn{2}{c}{Sample II$^{b}$} & \multicolumn{2}{c}{Combined Sample$^{c}$} \\
 Measurement & Mean & Median & Mean & Median & Mean & Median \\
 \hline
 
 (u-r) & 1.41 & 1.43 & 1.07 & 1.05 & 1.21 & 1.30 \\
 $\Delta$(g-i) & 0.17 & 0.08 & 0.15 & 0.07 & 0.16 & 0.08 \\
 E(B-V)$_{(g-i)}$ & 0.10 & 0.05 & 0.08 & 0.04 &  0.09 & 0.05 \\
 H$\alpha$/H$\beta$ & 5.04 & 3.63 & 3.95 & 3.41 & 4.47 & 3.54 \\
 E(B-V)$_{H\alpha/H\beta}$ & 0.48 & 0.24 & 0.29 & 0.19  & 0.38 & 0.23  \\
 SFR (H$\alpha$)$^{d}$ & 3.98 & 0.67 & 2.81 & 0.58 & 3.13 & 0.63  \\

 \hline
 
 \end{tabular}
 \footnotetext[1]{Sample from York et al. (2012).}
 \footnotetext[2]{Sample from this paper.}
 \footnotetext[3]{Total sample from samples I and II.}
 \footnotetext[4]{Measured in M$_{\odot}$ yr$^{-1}$.}
  
 \end{minipage}
 \end{table}

% Table 9

\begin{table}
\centering
\begin{minipage}{200mm}
\caption{Emission line metallicities\label{tbl-emission-metallicity-II}}
\begin{tabular}{lccccccccc}

\hline\hline

\multicolumn{1}{c}{} & \multicolumn{4}{c}{Metallicity Indices} & \multicolumn{2}{c}{$12+log(O/H)$} & \multicolumn{2}{c}{$12+log(O/H)$} \\
\cline{2-5}
Index & R23 & [O III]b/[N II]b & N2 & O3N2 & R23$_{l}$ & R23$_{u}$ & N2 & O3N2 \\
\hline

1 & 0.15	&--	&--	&-- & 6.04 & 9.03 & -- & -- \\
2&-0.03	&--	&--	&-- & 6.05 & 9.05 & -- & --\\
3&0.11	&8.14	&-1.02	&1.37 & 5.96 & 9.03 & 8.32 & 8.29 \\
4&0.13	&2.7	&-0.72	&1.05 & 6.90 & 9.03 & 8.49 & 8.39 \\
5&--	&--	&-0.39	&-- & -- & -- & 8.68 & --\\
6&--	&--	&-0.4	&--  & -- & -- & 8.67 & --\\
7&--	&0.5	&-0.37	&0.57 & -- & -- & 8.69 & 8.55\\
8&0.32	&10.27	&-0.96	&1.48 & 6.29 & 9.01 & 8.35 & 8.26  \\
9&0.49	&--	&--	&-- & 7.34 & 8.89  & -- & -- \\
10&0.25	&14.03	&-1.2	&1.63 & 6.61 & 9.01 & 8.21 & 8.21  \\
11&--	&--	&--	&-- & -- & -- & -- & -- \\
12&0.49	&--	&--	&-- & 7.04 & 8.93 & -- & --\\
13&0.24	&--	&--	&-- & 6.93 & 9.00 & -- & --\\
14&0.16	&1.13	&-0.43	&0.54 & 6.94 & 9.02 & 8.66 & 8.56  \\
15&0.63	&--	&--	&-- & 7.36 & 8.84 & -- & -- \\
16&0.51	&--	&--	&-- & 7.47 & 8.87 & -- & --\\
17&0.4	&--	&--	&-- & 6.16 & 9.01 & -- & -- \\
18&--	&--	&--	&-- & -- & -- & -- & -- \\
19&0.58	&--	&--	&-- & 7.50 & 8.84 & -- & --\\
20&--	&0.53	&-0.58	&0.37 & -- & -- & 8.57 & 8.61 \\
21&0.36	&--	&--	&-- & 5.43 & 9.05 & -- & --\\
22&0.3	&2.49	&-0.53	&1.1 & 6.81 & 8.99 & 8.60 & 8.38  \\
23&--	&0.21	&-0.37	&0.02 & -- & -- & 8.69 & 8.72\\
24&--	&--	&--	&-- & -- & -- & -- & --\\
25&--	&1.24	&-0.47	&0.57 & -- & -- & 8.63 & 8.55 \\
26&0.2	&--	&--	&-- & 6.77 & 9.02 & -- & --\\
27&0.45	&--	&--	&-- & 7.14 & 8.93 & -- & -- \\

\hline

\end{tabular}
\end{minipage}
\end{table}

%Table 10

\begin{table}\footnotesize
\centering
\begin{minipage}{200mm}
\caption{Rest-frame equivalent widths (W) for Ca II and Na I interstellar lines in foreground galaxies.}\label{tbl-ew}
\begin{tabular}{ccccccccc}

\hline\hline

Index & $\lambda_{obs}$ & W$_{\lambda3934.78}$  & $\lambda_{obs}$ & W$_{\lambda3969.59}$  & $\lambda_{obs}$ & W$_{\lambda5891.58}$  & $\lambda_{obs}$ &W$_{\lambda5897.56}$ \\
& & Ca II K & & Ca II H & & Na I D2 & & Na I D1 \\

\hline

1& 5138.04 & $<0.9^{a}$ & 5183.41 & $<0.9$ & 7693.23 & $<1.1$ & 7701.03 & $<0.8$  \\
2 & 5261.31 & 1.3 & 5306.55 & $<0.7$ & 7875.86 & $<0.8$ & 7883.86 & $<1.1$ \\
3 & 4044.56 & $< 1.2$ & 4079.68 & $<1.6$  & 6055.96 & $<0.6$ & 6062.01 & 1.1 \\
4 & 4103.24 & $<1.1$ & 4140.28 & $< 1.4$  &6144.92 & 1.5 &  6151.00 & $<0.7$ \\
5 & 4385.08 & 3.6 & 4424.75 & --$^{d}$ & 6568.28 & 2.4$^{c}$ & 6574.01 & 2.4$^{c}$ \\
6 & 3934.78 & $<2.8$ & 3969.59 & $<0.2$  & 5890.23 & 3.3 & 5898.01 & $<1.6$ \\
7 & 4917.64 & 0.5 & 4960.23 & $<0.4$  & 7363.89 & $<0.5$ & 7372.09 & $<0.5$ \\
8 & 4490.37 & $< 1.1$ & 4530.10 & $<1.0$ & 6723.47 & $<0.4$ & 6730.30 & $<0.4$ \\
9 & 5117.44 & $<0.9$ & 5162.45 & $<0.7$ & 7662.00 & $<1.0$ & 7669.78 & $<1.0$ \\
10 & 4446.30 & $<0.2$ & 4485.64 & $<0.2$ & 6657.49 & 3.1$^{c}$ & 6665.35 & 3.1$^{c}$ \\
11 & 4815.38 & 0.8 & 4856.18 & --$^{d}$ & 7210.12 & $<0.4$ & 7217.43 & $<0.4$ \\
12 & 4947.35 & $<0.5$ & 4991.36 & $<0.4$ & 7408.07 & $<0.7$ & 7415.59 & $<0.7$ \\
13 &  5353.27 & $< 0.6$ & 5400.63 & $<0.5$ & 8015.49 & $<1.7$ & 8023.63 & $<1.7$ \\
14 &  4967.22 & 0.9 & 5012.76 & $<0.6$ & 7439.30 & --$^{b}$ & 7446.85 & --$^{b}$ \\
15 & 4172.44 & $<1.1$ & 4209.35 & $<0.6$  & 6247.43 & $<0.7$ & 6253.77 & $<0.7$ \\
16 & 4043.90 & 0.4 & 4079.55 & $<0.4$  & 6054.78 & 0.8 & 6060.11 & 0.3 \\
17 & 5512.69 & 0.2 & 5561.40 & $<0.7$ & 8254.1 & $<0.5$ & 8262.48 & $<0.5$ \\
18 & 4391.67 & 0.4 & 4430.46 & $<0.3$ & 6575.59 & $<0.2$ & 6582.27 & $<0.2$ \\
19 & 4286.94 & 0.8 & 4324.52 & 0.4 & 6417.70 & $<0.3$ & 6424.21 & $<0.3$ \\
20 & 4171.00 & 1.8 & 4209.24 & $<0.9$  & 6246.25 & $<0.6$ & 6252.59 & $<0.6$ \\
21 & 5337.73 & 0.5 & 5385.94 & $<0.3$ & 7993.70 & $<1.6$ & 8001.81 & $<1.6$ \\
22 & 4070.14 & 0.8 & 4106.54 & $<0.3$  & 6094.84 & $<0.3$ & 6101.03 & $<0.3$ \\
23 & 4980.15 & 1.9 & 5023.73 & $<4.6$ & 7457.56 & 4.4$^{c}$ & 7460.89 & 4.4$^{c}$ \\
24 &  4368.79 & 0.7 & 4407.63 & 0.3 & 6541.42 & $<0.3$ & 6548.06 & $<0.3$ \\
25 & 4163.10 & 1.0 & 4199.83 & $<0.7$ & 6233.29 & $<0.4$ & 6239.62 & $<0.4$ \\
26 & 4969.55 & $<0.9$ & 4738.10 & $<1.1$  & 7032.19 & $<0.7$ & 7039.33 & $<0.7$  \\
27 & 5235.62 & $<0.4$ & 5281.94 & $<0.4$ & 7839.34 & $<0.8$ & 7847.29 & $<0.8 $ \\

\hline

\end{tabular}
\footnotetext[1]{Limits are 3$\sigma$ measured in the noise at the location of the expected absorption line. All detections are 4$\sigma$ or better. }
\footnotetext[2]{No spectral coverage in this region for this target.}
\footnotetext[3]{Blended Na I D1 and Na I D2 absorption.}
\footnotetext[4]{Line blended with another absorption feature. }
\end{minipage}
\end{table}

\bsp

\label{lastpage}


\begin{thebibliography}{}

\bibitem{allende01} Allende-Prieto, C., Lambert, D., Asplund, M. 2001, ApJ, 556, L63

\bibitem{aller12} Aller, M., Kulkarni, V., York, D., Vladilo, G., Welty, D., Som, D. 2012, ApJ, 748, 19

\bibitem[Arp et al. (1967)]{arp67} Arp, H., Bolton, J., Kinman, T. 1967, ApJ, 147, 840

\bibitem{baldwin81} Baldwin, J., Phillips, M., Terlevich, R. 1981, PASP, 93, 5

\bibitem[Bell et al. (2004)]{bell04} Bell, E., Wolf, C., Meisenheimer, K., Rix, H-W., Borch, A., Dye, S., Kleinheinrich, M., Wisotzki, L., McIntosh, D. 2004, ApJ, 608, 752

\bibitem[Bergeron \& Boisse (1991)]{bergeron91} Bergeron, J. \& Boisse, P. 1991, A\&A, 234, 344

\bibitem{bouche12} Bouch\'e, N., Murphy, M., P\'eroux, C., Contini, T., Martin, C., Forster Schreiber, N., Genzel, R., Lutz, D. et al. 2012, MNRAS, 419, 2

\bibitem{bowen11} Bowen, D. \& Chelouche, D. 2011, ApJ, 727, 47

\bibitem[Burbidge et al. (1966)]{burbidge66} Burbidge, E., Lynds, C., Burbidge, G. 1966, ApJ, 144, 447

\bibitem{charlton96} Charlton, J. \& Churchill, C. 1996, ApJ, 465, 631

\bibitem{conroy10} Conroy, C., Schiminovich, D., Blanton, M. 2010, ApJ, 718, 184

\bibitem{ellison08} Ellison, S., York, B., Murphy, M., Zych, B., Smith, A., Sarre, P.  2008, MNRAS, 383, L30

\bibitem{fynbo10} Fynbo, J., Laursen, P., Ledoux, C., Moller, P., Durgapal, A., Goldoni, P., Gullberg, B. et al. 2010, MNRAS, 408, 2128

\bibitem{fynbo11} Fynbo, J., Ledoux, C. Noterdaeme, P., Christensen, L., Moller, P., Durgapal, A., Goldoni, P. et al. 2011, MNRAS, 413, 2481

\bibitem{fynbo13} Fynbo, J., Geier, S., Christensen, L., Gallazzi, A., Krogager, J., Kruhler, T., Ledoux, C., Maund, J. et al. 2013, MNRAS, accepted, arXiv:1306.2940

\bibitem{hewett07} Hewett, P. \& Wild, V. 2007, MNRAS, 379, 738

\bibitem{hewitt89} Hewitt, A. \& Burbidge, G. 1989, ApJS, 69, 1

\bibitem{Jaffe et al. (2011)} Jaffe, Y., Aragon-Salamanca, A., De Lucia, G., Jablonka, P., Rudnick, G., Saglia, R., Zaritsky, D. 2011, MNRAS, 410, 280

\bibitem{jiang11} Jiang, P., Ge, J., Zhou, H., Wang, J., Wang, T. 2011, ApJ, 732, 110

\bibitem{kauffmann03} Kauffmann, G., Heckman, T., Tremonti, C., Brinchmann, J., Charlot, S., White, S., Ridgway, S., Brinkmann, J. et al. 2003, MNRAS, 346, 1055

\bibitem[Kennicutt (1998)]{kennicutt98} Kennicutt, R.C. 1998, ARAA, 36, 189

\bibitem{kewley01} Kewley, L., Dopita, M., Sutherland, R., Heisler, C., Trevena, J. 2001, ApJ, 556, 121

\bibitem{kewley02} Kewley, L., Geller, M., Jansen, R., Dopita, M. 2002, AJ, 124, 3135

\bibitem{khare12} Khare, P., Vanden Berk, D., York, D., Lundgren, B., Kularni, V. 2012, MNRAS, 419, 1028

\bibitem{kravtsov09} Kravtsov, A. 2009, ASPC, 419, 283

\bibitem{kuhlen12} Kuhlen, M., Krumholz, M., Madau, P., Smith, B., Wise, J. 2012, ApJ, 749, 36

\bibitem{kuhlen13} Kuhlen, M., Madau, P., Krumholz, M. 2013, submitted, arXiv:1305.5538v1

\bibitem{kulkarni06} Kulkarni, V., Woodgate, B., York, D., Thatte, D., Meiring, J., Palunas, P., Wassel, E. 2006, ApJ, 636, 30

\bibitem{kulkarni11} Kulkarni, V., Torres-Garcia, L., Som, D., York, D., Welty, D., Vladilo, G. 2011, ApJ, 726, 14

\bibitem{kulkarni12} Kulkarni, V., Meiring, J., som, D., P\'eroux, C., York, D., Khare, P., Lauroesch, J. 2012, ApJ, 749, 176

\bibitem{lundgren09} Lundgren, B., Brunner, R., York, D., Ross, A., Quashnock, J., Myers, A., Schneider, D., Al Sayyad, Y., Bahcall, N. 2009, ApJ, 698, 819

\bibitem[Masters et al. (2011)]{masters11} Masters, K., Mosleh, M., Romer, A., Nichol, R., Bamford, S., Schawinski, K., Lintott, C., Andreescu, D., Campbell, H., Crowcroft, B., Doyle, I., Edmondson, E., Murray, P., Raddick, M., Slosar, A., Szalay, A., Vandenberg, J. 2011, MNRAS, 405, 783

\bibitem{matsuoka12} Matsuoka, Y., Ienaka, N., Oyabu, S., Wada, K., Takino, S. 2012, AJ, 144, 159

\bibitem{moller04} Moller, P., Fynbo, J., Fall, S. 2004, A\&A, 422, L33

\bibitem{motta02} Motta, V. et al. 2002, ApJ, 574, 719

\bibitem{noll05} Noll, S. \& Pierini, D. 2005, A\&A, 444, 137

\bibitem{noterdaeme10} Noterdaeme, P., Srianand, R., \& Mohan, V. 2010, MNRAS, 403, 906

\bibitem{noterdaeme12} Noterdaeme, P., Laursen, P., Petitjean, P., Vergani, S., Maureira, M., Ledoux, C., Fynbo, J., L\'opez, S., Srianand, R. 2012, A\&A, 540, 63

\bibitem[Pei (1992)]{pei92} Pei, Y. 1992, ApJ, 395, 130

\bibitem{peroux11} P\'eroux, C., Bouch\'e, N., Kulkarni, V., York, D., Vladilo, G. 2011, MNRAS, 410, 2237

\bibitem{peroux12} P\'eroux, C., Bouch\'e, N., Kulkarni, V., York, D., Vladilo, G. 2012, MNRAS, 419, 3060

\bibitem{rhode13} Rhode, K., Salzer, J., Haurber, N., Van Sistine, A., Young, M., Haynes, M., Giovanelli, R., Cannon, J. et al. 2013, AJ, 145, 149 

\bibitem{routly52} Routly, P. and Spitzer, L. 1952, ApJ, 115, 227

\bibitem[Sandage (1965)]{sandage65} Sandage A. 1965, ApJ, 141, 1560

\bibitem[Schneider et al. (2007)]{schneider07} Schneider, D., Hall, P., Richards, G., Strauss, M., Vanden Berk, D., Anderson, S., Brandt, W., Fan, X., Jester, S., Gray, J. et al. 2007, AJ, 134, 102

\bibitem{schneider10} Schneider, D., Richards, G., Hall, P., Strauss, M., Anderson, S., Boroson, T., Ross, N., Shen, Y., Brandt, W., Fan, X. et al. 2010, AJ, 139, 2360

\bibitem{srianand08} Srianand, R., Gupta, N., Petitjean, P., Noterdaeme, P., Saikia, D. 2008, MNRAS, 391, L69

\bibitem{stobie06} Stobie, E. \& Ferro, A. 2006, ASPC, 351, 540

\bibitem{stoughton02} Stoughton, C., Lupton, R., Bernardi, M., Blanton, M., Burles, S., Castander, F., Connolly, A., Eisenstein, D., Frieman, J., Hennessy, G. et al. 2002, AJ, 123, 485

\bibitem{strateva01} Strateva, I., Ivezic, Z., Knapp, G., Narayanan, V., Strauss, M., Gunn, J., Lupton, R., Schlegel, D. et al. 2001, AJ, 122, 1861

\bibitem{vandenberk08} Vanden Berk, D., Khare, P., York, D., Richards, G., Lundgren, B., Alsayyad, Y., Kulkarni, V., SubbaRao, M. et al. 2008, ApJ, 679, 239

\bibitem{wang04} Wang, J., Hall, P., Ge, J., Li, A., Schneider, D. 2004, ApJ, 609, 589

\bibitem{wild06} Wild, V., Hewett, P., Pettini, M. 2006, MNRAS, 367, 211

\bibitem[Womble (1993)]{womble93} Womble, D. 1993, PASP, 691, 1043

\bibitem[Yanny \& York (1992)]{yanny92} Yanny, B. \& York, D. 1992, ApJ, 391, 569

\bibitem{york00} York, D., Adelman, J., Anderson, J., Anderson, S., Annis, J., Bahcall, N., Bakken, J., Barkhouser, R., Bastian, S., Berman, E. et al.  2000, 120, 1579

\bibitem[York et al. (2006)]{york06} York, D. G., Khare, P., Vanden Berk, D., Kulkarni, V.P., Crotts, A.P.S., Lauroesch, J.T., Richards, G.T., Schneider, D.P. et al. 2006, MNRAS, 367, 945

\bibitem[York et al. (2012)]{york12} York, D. G., Straka, L.A., Bishof, M., Kuttruff, S., Quashnock, J., Bowen, D., Subbarao, M., Richards, G.T. et al. 2012, MNRAS, 423, 3692

\bibitem{zych07} Zych, B., Murphy, M., Pettini, M., Hewett, P., Ryan-Weber, E., Ellison, S. 2007, MNRAS, 379, 1409

\end{thebibliography}
\end{document}